\documentclass[12pt,a4paper]{article}
\pdfoutput=1
\usepackage{color}
\usepackage{amssymb,amsmath,bm,bbold}
\usepackage{epsf}
\usepackage{epsfig}
\usepackage{relsize}
\usepackage[dvipsnames]{xcolor}
\usepackage[linktoc=page,bookmarks=false,colorlinks=false,linkbordercolor=RoyalBlue,citebordercolor=ForestGreen,urlbordercolor=CornflowerBlue]{hyperref}
\usepackage{latexsym,mathrsfs,dsfont}
\usepackage[normalem]{ulem} % for strikeout \sout
\usepackage[compress]{cite}
\usepackage{graphicx}
\usepackage{url}
\usepackage{booktabs}
\usepackage{float}
\usepackage{multirow}
\usepackage{changepage}
\usepackage[hypcap]{caption, subcaption}

\usepackage[titles]{tocloft}

\usepackage{tabularx,colortbl}

\usepackage{fancyhdr}
\advance \headheight by 3.0truept       % for 12pt mandatory...
\allowdisplaybreaks[1]

\definecolor{red}{cmyk}{0,1,1,0.4}
\definecolor{darkgreen}{rgb}{0.0,0.6,0.0}
\definecolor{cDarkGrey}{RGB}{91,91,91}
\definecolor{cGrey}{RGB}{245,243,238}
\definecolor{cBlue}{RGB}{0,110,191}
\definecolor{cLightBlue}{RGB}{214,237,252}
\definecolor{cRed}{RGB}{196,0,100}
\definecolor{cLightRed}{RGB}{254,222,237}
\definecolor{cGreen}{RGB}{0,166,80}
\definecolor{cLightGreen}{RGB}{254,222,237}
\definecolor{cOrange}{RGB}{221,74,44}
\definecolor{cLightOrange}{RGB}{255,215,210}
\definecolor{cPurple}{RGB}{93,35,125}
\definecolor{cLightPurple}{RGB}{241,230,252}
\definecolor{cYellow}{RGB}{252,191,10}
\definecolor{cISSRBlue}{RGB}{0,111,174}
\definecolor{cISSRGrey}{RGB}{167,169,172}

\newcommand{\beq}{\begin{equation}}
\newcommand{\eeq}{\end{equation}}
\newcommand{\be}{\begin{equation}}
\newcommand{\ee}{\end{equation}}
\newcommand{\bi}{\begin{itemize}}
\newcommand{\ei}{\end{itemize}}
\newcommand{\ba}{\begin{array}}
\newcommand{\ea}{\end{array}}
\newcommand{\beqa}{\begin{eqnarray}}
\newcommand{\eeqa}{\end{eqnarray}}
\newcommand{\bea}{\begin{eqnarray}}
\newcommand{\eea}{\end{eqnarray}}
\newcommand{\beqn}{\begin{eqnarray}}
\newcommand{\eeqn}{\end{eqnarray}}

\newcounter{TODO}

\newcommand{\oL}[1]{\overline{#1}}

\newcommand{\ord}{{\cal O}}

\newcommand{\cB}{\mathcal{B}}

\newcommand{\mev}{\text{MeV}}
\newcommand{\GeV}{\,\text{GeV}}

\newcommand{\vcb}{|V_{cb}|}
\newcommand{\vtd}{|V_{td}|}
\newcommand{\vub}{|V_{ub}|}
\newcommand{\vts}{|V_{ts}|}
\newcommand{\vus}{|V_{us}|}

\def\kpn{K^+\rightarrow\pi^+\nu\bar\nu}
\def\klpn{K_{L}\rightarrow\pi^0\nu\bar\nu}
\def\ksm{K_S\to\mu^+\mu^-}
\def\klm{K_L\to\mu^+\mu^-}

\newcommand{\IM}{\rm{Im}}
\newcommand{\RE}{\rm{Re}}

\newcommand{\eps}{\epsilon}

\newcommand{\BR}{{\cal B}}

\newcommand\cdt[1]{\cdot10^{#1}}

\begin{document}

\begin{flushleft}
%{\em Version of \today}
\end{flushleft}

\vspace{-14mm}
\begin{flushright}
  AJB-21-7\\
  TUM-HEP-1364/21
\end{flushright}

\medskip

\begin{center}
{\large\bf\boldmath
  Searching for New Physics in Rare $K$ and $B$
  Decays \\ without $\vcb$ and $\vub$
  Uncertainties
}
\\[1.0cm]
{\bf
    Andrzej~J.~Buras$^{a,b}$ and
  Elena Venturini$^{b}$
}\\[0.3cm]

{\small
$^a$TUM Institute for Advanced Study,
    Lichtenbergstr. 2a, D-85747 Garching, Germany \\[0.2cm]
$^b$Physik Department, TU M\"unchen, James-Franck-Stra{\ss}e, D-85748 Garching, Germany
}
\end{center}

\vskip 0.5cm

\begin{abstract}
  \noindent
  {We reemphasize the strong dependence of the branching ratios
$\mathcal{B}(K^+\to\pi^+\nu\bar\nu)$  and  $\mathcal{B}(K_L\to\pi^0\nu\bar\nu)$ 
on $|V_{cb}|$ that is stronger than in rare $B$ decays, in particular for $K_L\to\pi^0\nu\bar\nu$.}   Thereby
   the persistent tension between inclusive and exclusive determinations
  of $|V_{cb}|$ weakens the power of these theoretically
  clean decays   in the search for new physics (NP).
  We demonstrate how this uncertainty can be practically removed by
  considering within the SM suitable ratios of the two branching ratios  between each other and with other observables like the branching ratios for 
$K_S\to\mu^+\mu^-$, $B_{s,d}\to\mu^+\mu^-$ and $B\to K(K^*)\nu\bar\nu$.
  We use as basic CKM parameters $V_{us}$, $|V_{cb}|$ and the angles $\beta$ and $\gamma$ in the unitarity triangle (UT) with the latter two  determined through the measurements of tree-level $B$ decays. This avoids the use of the problematic $|V_{ub}|$. A ratio involving  $\mathcal{B}(K^+\to\pi^+\nu\bar\nu)$ and
  $\overline{\mathcal{B}}(B_{s}\to\mu^+\mu^-)$ while being $|V_{cb}|$-independent exhibits sizable dependence on   the angle $\gamma$.  It should be of interest for   several experimental groups  in the coming years.
  We point out that the $|V_{cb}|$-independent ratio of
  $\mathcal{B}(B^+\to K^+\nu\bar\nu)$ and
  $ \overline{\mathcal{B}}(B_{s}\to\mu^+\mu^-)$  from Belle II and LHCb signals a $1.8\sigma$ tension with   its SM value. {As a complementary test of the Standard Model we propose to extract $|V_{cb}|$ from different observables} as a function of $\beta$ and $\gamma$. We illustrate this with
    $\varepsilon_K$, $\Delta M_d$ and $\Delta M_s$  finding tensions between
    these three determinations of $|V_{cb}|$
  within the SM. We point out that from $\Delta M_s$ and $S_{\psi K_S}$ alone one finds  $|V_{cb}|=41.8(6)\times 10^{-3}$ and  $|V_{ub}|=3.65(12)\times 10^{-3}$. We stress the importance of {a precise measurement} of $\gamma$.
 {Assuming no NP in  $|\varepsilon_K|$ and $S_{\psi K_S}$ } we determine independently of $\vcb$ {and $\gamma$}: $\mathcal{B}(\kpn)_\text{SM}= (8.60\pm0.42)\times 10^{-11}$ and 
 $\mathcal{B}(\klpn)_\text{SM}=(2.94\pm 0.15)\times 10^{-11}$ {with only
   CKM uncertainty coming from $\beta$, that is already precisely known.} These are the most precise determinations to date. {Assuming no NP in $\Delta M_{s,d}$ allows to
   obtain analogous results for all $B$ decay  branching ratios considered in our paper without any CKM uncertainties}.

\end{abstract}

\thispagestyle{empty}
\newpage
\tableofcontents
\newpage
\setcounter{page}{1}

%--------+---------+---------+---------+---------+---------+---------+---------+
%
%
%
%--------+---------+---------+---------+---------+---------+---------+---------+
\section{Introduction}
The rare decays $\kpn$ and $\klpn$ played already for three decades an important  role in the tests of the Standard Model (SM) and of its various extensions
\cite{Buras:2004uu,Buras:2020xsm}. This is due to their theoretical cleanness
and GIM suppression of their branching ratios within the SM implying strong sensitivity to new physics (NP).

On the experimental side the most recent result for $\kpn$ from NA62 \cite{NA62:2021zjw} and the  $90\%$ C.L. upper bound on $\klpn$  from KOTO  \cite{Ahn:2018mvc} read respectively 
\be\label{EXP19}
\mathcal{B}(\kpn)_\text{exp}=(10.6^{+4.0}_{-3.5}\pm 0.9)\times 10^{-11}\,,\qquad
\mathcal{B}(\klpn)_\text{exp}\le 3.0\times 10^{-9}\,
\ee
and are to be compared with the SM predictions of {2015 \cite{Buras:2015qea}\footnote{A 2016 analysis in \cite{Bobeth:2016llm} found very similar
  results.} that are frequently quoted in the literature
\be\label{KSM}
\mathcal{B}(\kpn)_\text{SM}= (8.4\pm1.0)\times 10^{-11}\,,\qquad
\mathcal{B}(\klpn)_\text{SM}=(3.4\pm 0.6)\times 10^{-11}\,.
\ee
}
On the other hand the most recent updated predictions for both branching ratios
from \cite{Brod:2021hsj} read
\be\label{Brod}
\mathcal{B}(\kpn)_\text{SM}= (7.7\pm0.6)\times 10^{-11}\,,\qquad
\mathcal{B}(\klpn)_\text{SM}=(2.6\pm0.3)\times 10^{-11}\,.
\ee
They are  on the one hand significantly lower than the values in (\ref{KSM})
and on the other hand are much more accurate.

However, the inspection of the plots in \cite{Buras:2015qea} and {their updated versions in  Fig.~\ref{fig:CKMdependence}} of the present paper demonstrate very clearly\footnote{{As we will trade the dependence on $\vub$ for the one on
  $\beta$ we do not show the $\vub$ dependence of the branching ratios in the present paper. It can be
  found in \cite{Buras:2015qea}.}}
\begin{itemize}
\item
  strong dependence of $\mathcal{B}(\kpn)_\text{SM}$ on $\vcb$ and on the angle
  $\gamma$ in the unitarity triangle (UT), although only a weak dependence on {the angle $\beta$,}
  \item
  strong dependence of $\mathcal{B}(\klpn)_\text{SM}$ on $\vcb$ and {the angle $\beta$} in the UT but also significant dependence on $\gamma$.
\end{itemize}

{To obtain the result in (\ref{Brod})
  the authors of \cite{Brod:2021hsj} used the values of the CKM parameters
  from the CKMfitter adopted by PDG \cite{Zyla:2020zbs}.}
In particular the value
of $\vcb$ corresponding to (\ref{Brod}), $\vcb = (40.5 \pm 0.8) \cdt{-3}$,
is in the ballpark of {\em exclusive}
determinations of $\vcb$, as for example from $B\to D\ell\bar\nu$ \cite{Bordone:2019guc}. Had the authors of  \cite{Brod:2021hsj}
used the {\em inclusive} determination
of $\vcb$, that is the value $\vcb_{B\to X_c} = (42.16 \pm 0.50)\cdt{-3}$~\cite{Bordone:2021oof}, {close to the values obtained by Utfitter in their global analysis,} they would find significantly higher
branching ratio. {This is evident from}  Fig.~\ref{fig:CKMdependence}, where the values of branching ratios for $\kpn$ and $\klpn$ have been plotted as functions of $\vcb$
for different values of $\gamma$ and $\beta$. On the other hand the
  most recent exclusive value of $\vcb$ from FLAG reads { $\vcb=(39.48\pm0.68)\cdt{-3}$   \cite{Aoki:2021kgd}}   which would imply even lower values for branching ratios than given in (\ref{Brod}).

{This} {uncertainty in $\vcb$} is annoying in view of the very small theoretical uncertainties in these two decays, with QCD corrections
known at NLO \cite{Buchalla:1993bv,Buchalla:1993wq,Misiak:1999yg,Buchalla:1998ba}
and NNLO level \cite{Buras:2005gr,Buras:2006gb,Gorbahn:2004my} and
electroweak corrections at the NLO level \cite{Brod:2008ss,Brod:2010hi}.
Moreover isospin breaking effects and 
non-perturbative effects have been considered in \cite{Isidori:2005xm,Mescia:2007kn}
and further improvements are expected from lattice gauge theories \cite{Christ:2019dxu}.

It should also be emphasized that as the dominant CKM factor in rare Kaon
decays $V_{ts}^*V_{td}$ grows with $\vcb$ like $\vcb^2$, the branching ratio for $\klpn$
grows with $\vcb$ like $\vcb^4$ to be compared with $\vcb^2$ in the
case of $B_s\to\mu^+\mu^-$, where $V_{td}$ is replaced by $V_{tb}\approx 1$.
For $\kpn$ this dependence is weaker due to the presence of significant charm
contribution but still stronger than for $B_s\to\mu^+\mu^-$. Also
the short distance contribution to the $\ksm$ branching ratio grows like
$\vcb^4$ and the $\vcb$ dependence in the parameter $\varepsilon_K$, similar
to $\kpn$, although weaker than $\vcb^4$ due to the presence of charm
contribution,  is also stronger than in  $B_s\to\mu^+\mu^-$.

These strong parametric dependences on $\vcb$ in $\ksm$ and $\varepsilon_K$, combined with the uncertainty in $\vcb$,  are
also unfortunate  because of the recent progress promoting both to
precision observables. Indeed
\begin{itemize}
  \item
    it has been demonstrated in \cite{Dery:2021mct} that the short distance contribution to the $\ksm$ branching ratio can be extracted from data offering us still another precision observable subject only {to} the parametric uncertainties stressed above.
  \item
    the significant QCD uncertainty from pure charm contribution to $\varepsilon_K$ has been practically removed in \cite{Brod:2019rzc} through a clever but simple trick by using CKM unitarity differently than done until now in the literature. Moreover, the two-loop electroweak effects on the top contribution have been found to decrease $\varepsilon_K$ by less than $1\%$ \cite{Brod:2021qvc}. These reductions of theoretical uncertainties in $\varepsilon_K$
    will play a significant role in our analysis. {Therefore this  new
      improvement in \cite{Brod:2019rzc} should be incorporated in
    any global analysis like the ones used in the PDG.}
    \end{itemize}

{So far} we discussed only the $\vcb$ dependence but the $\vub$ parameter
is also relevant. In view of the tensions between exclusive and inclusive determinations of $\vub$ that are also sizable \cite{Ricciardi:2021shl}, we will, following
 \cite{Blanke:2018cya},
use as four basic CKM parameters
\be\label{4CKM}
\boxed{\lambda=\vus,\qquad \vcb, \qquad \beta, \qquad \gamma}
\ee
with $\beta$ and $\gamma$ being two angles in the UT, shown in  Fig.~\ref{UUTa}.
{Their} determination from mixing induced CP-asymmetries in tree-level $B$ decays {and using other tree-level strategies} is presently theoretically
cleaner than the determination of $\vub$. As demonstrated in \cite{Buras:2002yj}
the determination of the apex of the  UT, given as seen in Fig.~\ref{UUTa}
by $(\bar\rho,\bar\eta)$, by means of the measurements of $\beta$ and $\gamma$ in tree-level $B$ decays is very efficient. A recent review of such determinations of $\beta$ and $\gamma$
can be found in Chapter 8 of \cite{Buras:2020xsm}. {See also \cite{Descotes-Genon:2017thz,Cerri:2018ypt}.}

Recently, following the proposal in \cite{Buras:2003td}, it has been demonstrated in \cite{Bobeth:2021cxm} that considering the ratio
of the $B_s\to\mu^+\mu^-$ branching ratio to $\Delta M_s$ the parametric uncertainty in $B_s\to\mu^+\mu^-$ due to $\vcb$ could be totally eliminated allowing
confidently to determine possible tension for this ratio between its SM estimate and the data.

We would like to emphasize that in the case of lepton flavour violation and electric dipole moments, where the SM estimates are by orders of magnitude below
the present experimental upper bounds,  such parametric uncertainties as the one due to $\vcb$ in
rare decays of mesons are presently irrelevant. {But in the case
of $B_s\to\mu^+\mu^-$ and $\kpn$ the room left for NP is respectively 
below $30\%$ and $100\%$ of the SM value} and any reduction of parametric uncertainties
is important. The case of  $\klpn$ and $\ksm$, where the experimental upper bounds are still by at least two orders above the SM expectations, is different.
But it is expected that in  this decade the branching ratios for these decays will be measured and it is useful to be prepared for such measurements {already now.} The last statement applies also to $B\to K(K^*)\nu\bar\nu$ and  $B_d\to\mu^+\mu^-$ which will {play important roles in} flavour physics in the coming years.

The main goal of the present paper is the generalization of the strategy in
\cite{Buras:2003td,Bobeth:2021cxm} to the theoretically cleanest rare Kaon and $B$-meson decays including also the parameter
$\varepsilon_K$ which through the progress made in \cite{Brod:2019rzc} has been  promoted within the SM to the class of precision observables. Also
the mass differences $\Delta M_{s,d}$ will play a role in these strategies.
In doing
this we benefited from previous analyses, like the ones in \cite{Buchalla:1994tr,Buras:1994rj,Buchalla:1996fp,Buras:2002yj,Buras:2015qea,Blanke:2016bhf,Blanke:2018cya}. However, our paper should not be considered as an update of these strategies which
would be useful in itself.  In particular, in contrast to \cite{Blanke:2016bhf,Blanke:2018cya}, where the detailed dependence of various observables on $\vcb$ has
been investigated, our goal here is to eliminate this dependence by taking
suitable ratios of various observables and in the spirit of \cite{Buras:2003td,Bobeth:2021cxm} to develop strategies for {finding footprints} of NP in several observables independently of the value
of $\vcb$.

{Once the $\vcb$ dependence is eliminated, the $\vcb$-independent correlations between various observables depend on only three remaining parameters
  in (\ref{4CKM}). But the dependence on $\vus$ is negligible, the angle $\beta$ is already known from $S_{\psi K_S}$ asymmetry with respectable precision and
  there is a significant progress by the LHCb collaboration on the determination of $\gamma$ from tree-level strategies \cite{LHCb:2021dcr}:
  \be\label{betagamma}
  \boxed{\beta=(22.2\pm 0.7)^\circ, \qquad  \gamma = (65.4^{+3.8}_{-4.2})^\circ \,.}
  \ee
  Moreover, in the coming years the determination of $\gamma$ by the LHCb and Belle II
  should be significantly improved so that precision tests of the SM using our
  strategies will be possible.}

{Now comes an important issue that we would like to emphasize.}
One could ask the question whether such clean tests of the SM could be accomplished by performing a global analysis of  several processes as done in the standard analyses of the UT or the recent global fits testing the
violation of lepton flavour universality. Of course such a global analysis may
reveal any potential tension by lowering the
goodness of the SM fit. However, a clear-cut insight into the origin of
tensions 
is not so easily obtainable.
Indeed
such analyses involve usually CKM uncertainties, in particular the one from
$\vcb$, and also hadronic uncertainties present in other processes that
are larger than the ones in $\kpn$, $\klpn$, $K_S\to\mu^+\mu^-$, $B_{s,d}\to\mu^+\mu^-$,  $B\to K(K^*)\nu\bar\nu$, $\Delta M_{d,s}$ and $\varepsilon_K$. Moreover,
NP could enter many observables used in such global fits and the transparent 
identification of the impact of NP on a given observable is a challenge.
On the contrary, in the proposed strategies that involve ratios of observables, these uncertainties, in particular the one from $\vcb$,  cancel out
except for the bag factors {and weak decay constants} in $\varepsilon_K$ and $\Delta M_{d,s}$, which are already precisely known from LQCD
and importantly {their values do not depend on NP parameters}\footnote{{This applies also to hadronic matrix elements of new operators absent in the SM.}}. In this manner concentrating just on
the listed decays 
allows us to test the SM 
independently of the value of $\vcb$. These ratios could turn out to be
smoking guns of NP.

{However, the ratios of branching ratios  are not as interesting
  as branching ratios themselves. Fortunately our strategy allows to determine
  the latter in a $\vcb$-independent manner by using solely
  \be\label{4B}
  \boxed{|\varepsilon_K|, \qquad S_{\psi K_S},\qquad \Delta M_s,\qquad \Delta M_d\,.}
  \ee
  This strategy differs from usual strategies in that not tree-level decays
  but loop suppressed transitions are used to determine CKM parameters.
  But within the SM this strategy is legitimate and as the experimental data
  and theory for these observables, including both perturbative and non-perturbative QCD effects,
  have smaller uncertainties than tree-level decays one arrives at rather
  accurate predictions for rare decay branching ratios which is presently
  impossible   otherwise.

  {In this context we would like to comment on a recent analysis in
     \cite{Altmannshofer:2021uub} on  a  determination of $\vcb$
and $\vub$ from loop processes alone, rare decays and quark mixing, by assuming
no NP contributions to these observables. 
While this analysis is in fact  the generalization of one of the strategies
suggested in \cite{Buras:2015qea} to include additional processes and to
perform a global fit, our present strategy in using the observables in
(\ref{4B}) differs from the ones in \cite{Buras:2015qea} and 
\cite{Altmannshofer:2021uub} in the following manner.

We do not assume that NP is absent simultaneously in all four observables in
(\ref{4B}) because of some tensions between determinations of $\vcb$
through these observables which we will identify in Section~\ref{sec:3}.
Therefore, to obtain SM predictions for  rare Kaon decays we only assume
the absence of NP in $\varepsilon_K$ and $S_{\psi K_S}$. To obtain predictions
for $B_s\to \mu^+\mu^-$ and  $B_d\to \mu^+\mu^-$ we assume, following \cite{Buras:2003td}, the absence of NP in $\Delta M_s$ and $\Delta M_d$, respectively but
not simultaneously. In our view this strategy for finding SM predictions is presently more powerful than any global fit which would include decays like $B\to K\mu^+\mu^-$, $B\to K^*\mu^+\mu^-$ and $B_s\to\phi\mu^+\mu^-$ that exhibit significant contributions from NP.}

In fact, our strategy allows us to obtain one of 
  the most important results of our paper:} the
  most precise determination of the $\kpn$ and $\klpn$  branching ratios within the
  SM to date. From $\varepsilon_K$ and
  $S_{\psi K_S}$ alone with $60^\circ\le\gamma\le 75^\circ$ we find
\be\label{BV}
\boxed{\mathcal{B}(\kpn)_\text{SM}= {(8.60\pm 0.42)}\times 10^{-11}\,,\quad
\mathcal{B}(\klpn)_\text{SM}={(2.94\pm 0.15)}\times 10^{-11}\,,}
\ee
which supersede the {usually  quoted values in (\ref{KSM}). In fact
  in the first case the error is reduced by a factor of $2.4$ and in the second case by a factor of $4$.} {The agreement of the central value for
  $\kpn$ with the one in (\ref{KSM}) is accidental. The latter result
  was obtained by using some average values of $\vcb$ and $\vub$ from tree-level determinations and the 2015 value of $\gamma$ that was significantly higher
  than the one in (\ref{betagamma}).}

Most importantly these results are independent of the value of $\vcb$ {and the error includes the full variation of $\gamma$ in the range even larger than the usual CKM global fits}. The
crucial idea behind it is the elimination of the $\vcb$ dependence of
both branching ratios with the help of  $\varepsilon_K$ with an additional
bonus, very strong suppression of the $\gamma$ dependence of both branching ratios {so that the only relevant CKM uncertainty included in the error comes from $\beta$ that is already precisely known from the measurements of $S_{\psi K_S}$}. Indeed these results are more accurate than the ones in (\ref{KSM}) and
(\ref{Brod}) and are not subject to any uncertainties related to $\vcb$
and $\vub$. {As the $\vcb$ dependence of the $\klpn$  branching ratio is
  stronger than in the $\kpn$ case, the reduction of the error is larger.} Moreover the future measurement of $\gamma$ will only have a minor
impact on them. Further decrease of the errors can only be achieved
by a more precise determination of $\beta$ and the reduction of the remaining
non-perturbative uncertainties in $\kpn$ ($P_c$) and in $\varepsilon_K$ ($\kappa_\epsilon$).

{Proceeding in this manner and using the results of \cite{Bobeth:2021cxm}
together with very precise experimental values for $\Delta M_{s,d}$
we find
\be\label{LHCbTH}
\boxed{\overline{\mathcal{B}}(B_{s}\to\mu^+\mu^-)_{\rm SM} = (3.62^{+ 0.15}_{-0.10})\times 10^{-9},\qquad \mathcal{B}(B_{d}\to\mu^+\mu^-)_{\rm SM} = (0.99^{+ 0.05}_{-0.03})\ \times 10^{-10},}
\ee
}
\noindent
{which are in the ballpark of the SM values quoted in the literature \cite{Buchalla:1993bv,Buchalla:1998ba,Buras:2012ru,Bobeth:2013uxa,Beneke:2017vpq,Aebischer:2019mlg} but have the advantage of being independent of  the value of $\vcb$
and in fact of any CKM parameter. Similar to (\ref{BV}) the results in
(\ref{LHCbTH}) are most accurate to date.}

{As already pointed out in \cite{Bobeth:2021cxm} the ratio of
  $\overline{\mathcal{B}}(B_{s}\to\mu^+\mu^-)$ to $\Delta M_s$ is in $2.2\sigma$ tension
with the data. Assuming $\Delta M_s$ is SM-like and using for it the experimental data
exhibits in (\ref{LHCbTH}) this tension explicitly when compared
with the experimental data in (\ref{eq:WAV-Bs}).}

The outline of our paper is  as follows.
In Section~\ref{sec:2} we recall the formulae for the sides and the 
apex of the UT given in terms of the set (\ref{4CKM}). Subsequently we present
a number of very accurate formulae for rare $K$ and $B$ decays considered by us in terms of this
set of parameters. They in turn  allow us to derive   in a straightforward manner a number of accurate relations between various observables
 that are independent of $\vcb$ and often exhibit
very weak dependence on the remaining parameters. These relations are
valid only in the SM and their violation would signal NP at work.

{In Section~\ref{sec:3}, as a complementary test of {the SM},
    that in contrast to the usual UT analyses exhibits
    the $\vcb$ dependence, we propose to extract $\vcb$ from different processes as   a  function of $\beta$ and $\gamma$. This in turn allows the determination of $\vub$ as a function of these two UT angles. We illustrate this with
    $\varepsilon_K$, $\Delta M_d$ and $\Delta M_s$. This in turn using $\vcb$ from $\varepsilon_K$ allows
    to calculate $\kpn$ and $\klpn$  branching ratios  as functions of $\gamma$ and  $\beta$  without any explicit $\vcb$ and $\vub$ dependences. {As the $\gamma$ dependence is very weak, imposing the constraint on $\beta$ from $S_{\psi K_S}$
      we obtain, as seen in (\ref{BV}), the most precise estimate of both branching ratios to date.}

    For $B_{s,d}$ decays analogous use of $\Delta M_{s,d}$ eliminates the CKM dependence  from branching ratios \cite{Buras:2003td,Bobeth:2021cxm} leading
    to the result in (\ref{LHCbTH}). {Inserting the results in (\ref{BV})
      and (\ref{LHCbTH}) into the $\vcb$-independent ratios
      involving the remaining
      three decays, $\ksm$, $B^+\to K^+\nu\bar\nu$ and $B^0\to K^{0*}\nu\bar\nu$
      allows in turn to obtain  the most precise estimate of their
      branching ratios as well. The results for all branching ratios are
      summarized in Table~\ref{tab:SMBRs}.

 We also point out} that suitable ratios of $\varepsilon_K$ and $\Delta M_{d,s}$  exhibit visible
 tensions between these three observables independently of $\vcb$ and $\gamma$ when
 the constraint on the angle $\beta$ in (\ref{betagamma}) is taken into account.}

In Section~\ref{sec:4B} we present first {the Table~\ref{tab:critical} which
  summarizes various powers entering the parametric power low expressions for the observables. They could be named   {\em critical exponents of flavour physics}. Subsequently we present} a guide to $\vcb$-independent relations
found in the text that indicates with the help of the Table~\ref{tab:guide}
which of the relations found by us has weak, strong or none dependence
on $\beta$ and $\gamma$. This table allows to find in no time the
analytic expressions for each relation in the text and the corresponding 
plot as a function of $\gamma$ for different values of $\beta$.
The study of the impact of NP on our analysis in specific models is left for the future. We conclude in Section~\ref{sec:4}.
  In the Appendix~\ref{App} we list
  the  expressions for the $m_t$ and $m_c$ dependent functions which enter our analysis.

  \section{Rare Kaon and $B$ decays: a $\vcb$-independent Study}\label{sec:2}
  \subsection{{A} Useful Parametrization of the UT}
  In finding the $\vcb$-independent correlations between various observables
  it is useful to use the so-called improved Wolfenstein parametrization \cite{Buras:1994ec}   of the CKM matrix that is much more precise than the original
  Wolfenstein parametrization \cite{Wolfenstein:1983yz}. {While being
    not exact, it allows for a much better insight into the $\vcb$-dependence
    of various branching ratios than it is possible using the standard
    parametrization of the CKM matrix.}

\begin{figure}
\centering
\includegraphics[width = 0.55\textwidth]{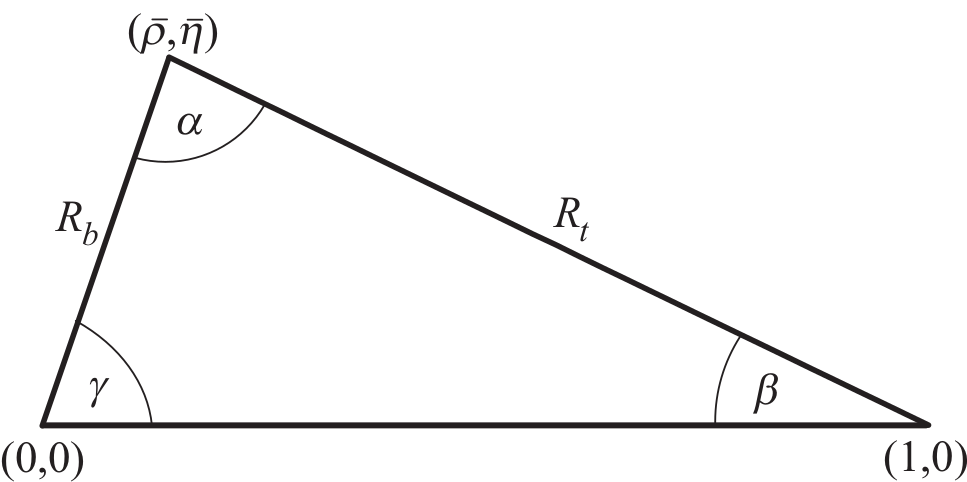}
 \caption{\it The Unitarity Triangle. }\label{UUTa}
\end{figure}

Using it we recall first the standard expressions for the two sides of the rescaled
UT shown in Fig.~\ref{UUTa} in terms of the elements of the CKM matrix with $\lambda=\vus$. These two
sides,  denoted by $R_t$ and $R_b$,
are given by
\begin{equation}\label{2.95}
R_t \equiv \frac{| V_{td}^{}V^*_{tb}|}{| V_{cd}^{}V^*_{cb}|} =
 \sqrt{(1-\bar\rho)^2 +\bar\eta^2}
=\frac{1}{\lambda} \left| \frac{V_{td}}{V_{cb}} \right|,
\end{equation}
\begin{equation}\label{2.94}
R_b \equiv \frac{| V_{ud}^{}V^*_{ub}|}{| V_{cd}^{}V^*_{cb}|}
= \sqrt{\bar\rho^2 +\bar\eta^2}
= (1-\frac{\lambda^2}{2})\frac{1}{\lambda}
\left| \frac{V_{ub}}{V_{cb}} \right|.
\end{equation}

{$R_t$ and $R_b$ can be solely expressed 
in terms of the angles
$\beta$ and $\gamma$, as follows
\cite{Buras:2002yj}
\be\label{RtRb}
R_t=\frac{\sin\gamma}{\sin(\beta+\gamma)}\approx \sin\gamma,
\qquad R_b=\frac{\sin\beta}{\sin(\beta+\gamma)}\approx \sin\beta\,.
\ee
We observe that $R_t$ depends dominantly on $\gamma$, while $R_b$ on $\beta$.
These approximations follow from the experimental fact that
$\beta+\gamma\approx 90^\circ$  and it is an excellent approximation to set $\sin(\beta+\gamma)=1$ in the formulae below although we will not do it in the
numerical evaluations.

On the other hand: 
\be\label{S4}
\bar\rho = 1-R_t\cos\beta,
\qquad \bar\eta=R_t\sin\beta.
\ee

Consequently, $\vtd$,  $\vub$ and $\vts$ } can be entirely expressed in terms of the parameters in (\ref{4CKM}) 
\be\label{vtdvub}
\boxed{\vtd=\lambda  \vcb\sin\gamma,\qquad
\vub={\lambda\sqrt{\sigma}}  \vcb\sin\beta,\qquad \sigma = \left( \frac{1}{1- \frac{\lambda^2}{2}} \right)^2\,,}
\ee

\be\label{vts}
\boxed{\vts=G(\beta,\gamma)\vcb, 
\qquad G(\beta,\gamma)=
1 +\frac{\lambda^2}{2}(1-2 \sin\gamma\cos\beta)\,,}
\ee
where the approximations in (\ref{RtRb}) have been used and in the expression for $\vts$ terms of $\ord(\lambda^4)$ have been neglected.

In turn, to an excellent accuracy of $0.2\%$ we also find 
 for the imaginary part of $\lambda_t=V_{td}V^*_{ts}$
\be\label{imlt}
 \boxed{{\IM}\lambda_t=\vub \vcb\sin\gamma={\lambda\sqrt{\sigma}} \sin\beta \sin\gamma\vcb^2\,.}
 \ee

 {In order to increase the transparency of numerous formulae in our paper
 we will use the following {\em reference} values for the variables
 in (\ref{4CKM})
\be\label{REFCKM}
\boxed{\lambda=0.225,\qquad \vcb=41.0\times 10^{-3}, \qquad \beta=22.2^\circ, \qquad  \gamma = 67^\circ \,.}
\ee
}

\begin{table}[!tb]
\center{\begin{tabular}{|l|l|}
\hline
$m_{B_s} = 5366.8(2)\mev$\hfill\cite{Zyla:2020zbs}	&  $m_{B_d}=5279.58(17)\mev$\hfill\cite{Zyla:2020zbs}\\
$\Delta M_s = 17.749(20) \,\text{ps}^{-1}$\hfill \cite{Zyla:2020zbs}	&  $\Delta M_d = 0.5065(19) \,\text{ps}^{-1}$\hfill \cite{Zyla:2020zbs}\\
{$\Delta M_K = 0.005292(9) \,\text{ps}^{-1}$}\hfill \cite{Zyla:2020zbs}	&  {$m_{K^0}=497.61(1)\mev$}\hfill \cite{Zyla:2020zbs}\\
$S_{\psi K_S}= 0.699(17)$\hfill\cite{Zyla:2020zbs}
		&  {$F_K=155.7(3)\mev$\hfill  \cite{Aoki:2019cca}}\\
	$|V_{us}|=0.2253(8)$\hfill\cite{Zyla:2020zbs} &
 $|\eps_K|= 2.228(11)\cdot 10^{-3}$\hfill\cite{Zyla:2020zbs}\\
$F_{B_s}$ = $230.3(1.3)\mev$ \hfill \cite{Aoki:2019cca} & $F_{B_d}$ = $190.0(1.3)\mev$ \hfill \cite{Aoki:2019cca}  \\
 $B_s(4.18 \GeV)=0.849(23)$\hfill \cite{DiLuzio:2019jyq}&
$B_d(4.18 \GeV) =0.835(28)$\hfill \cite{DiLuzio:2019jyq}
\\
  $\hat B_s=1.291(35)$\hfill\cite{Bobeth:2021cxm}            &
 $\hat B_d=1.269(43)$ \hfill\cite{Bobeth:2021cxm}           
\\
{$m_t(m_t)=162.83(67)\GeV$\hfill\cite{Brod:2021hsj} }  & {$m_c(m_c)=1.279(13)\GeV$} \\
{$S_{tt}(x_t)=2.303$} & {$S_{ut}(x_c,x_t)=-1.983\times 10^{-3}$} \\
    $\eta_{tt}=0.55(2)$\hfill\cite{Brod:2019rzc} & $\eta_{ut}= 0.402(5)$\hfill\cite{Brod:2019rzc}\\
$\kappa_\varepsilon = 0.94(2)$\hfill \cite{Buras:2010pza}	&
$\eta_B=0.55(1)$\hfill\cite{Buras:1990fn,Urban:1997gw}\\
$\tau_{B_s}= {1.515(4)}\,\text{ps}$\hfill\cite{Zyla:2020zbs}  & $\tau_{B_d}= {1.519(4)}\,\text{ps}$\hfill\cite{Zyla:2020zbs}   
\\	       
\hline
\end{tabular}  }
\caption {\textit{Values of the experimental and theoretical
    quantities used as input parameters. For future 
updates see PDG \cite{Zyla:2020zbs}  and HFLAV  \cite{Aoki:2019cca}. 
}}
\label{tab:input}
\end{table}

{We collect other  parameters used by us in Table~\ref{tab:input}.}

\subsection{Rare Kaon Decays}
\boldmath
\subsubsection{$\kpn$ and $\klpn$}
\unboldmath
For the branching ratios for $\kpn$ and $\klpn$ decays, the formulae with the exact dependence on the CKM parameters are given in~\cite{Buras:2020xsm}. However, for the search of the $\vcb$-independent relations
it is useful to use the improved Wolfenstein parametrization and in particular
the formulae in (\ref{S4}). Then the exact
formulae for the branching ratios in questions are approximated by expressions
that are particularly useful for our analysis {and are shown in the following}.

In the case of $\kpn$
we recall the following formula \cite{Buchalla:1998ba,D'Ambrosio:2001zh} that summarizes the dependence of $\mathcal{B}(K^+ \rightarrow \pi^+ \nu \bar{\nu})$ on $R_t$, $\beta$ and $V_{cb}$:
\begin{align}
\label{AIACD}
\mathcal{B}(K^+ \rightarrow \pi^+ \nu \bar{\nu}) =\,
&(1+\Delta_{\text{EM}})\frac{\kappa_+}{\lambda^8}~\vcb^4 X(x_t)^2
\Bigg[\sigma R^2_t\sin^2\beta\notag\\
&+
 \frac{1}{\sigma} \left(R_t\cos\beta +
\frac{\lambda^4P_c(X)}{\vcb^2X(x_t)}\right)^2\Bigg],
\end{align}
where \cite{Buras:2005gr,Buras:2006gb,Isidori:2005xm,Mescia:2007kn,Brod:2008ss} 
\be
\kappa_+={ (5.173\pm 0.025 )\times 10^{-11}\left[\frac{\lambda}{0.225}\right]^8},\qquad
P_c(X)=(0.405\pm0.024){  \left[\frac{0.225}{\lambda}\right]^4  }     
\ee
   {and $\Delta_{\text{EM}}=-0.003$}. {The formula for $X(x_t)$ is given in the Appendix~\ref{App}.}

{The expression in (\ref{AIACD})} can be considered as the fundamental formula for a correlation between 
$\mathcal{B}(\kpn)$, $\beta$ and any observable used to determine $R_t$ or equivalently $\gamma$ as seen in (\ref{RtRb}).
It is valid also in all models with CMFV \cite{Buras:2000dm} where $X(x_t)$ is replaced by a real function $X(x_t,v)$ with $v$ collecting new physics parameters. When this formula was proposed twenty years ago, it contained significant uncertainties 
in $R_t$ determined through $\Delta M_d/\Delta M_s$, in  $P_c(X)$ known only at 
NLO at that time,  in $\kappa_+$ and in $\vcb$. The first three uncertainties have 
been significantly reduced since then leaving $\vcb$ as the main uncertainty. The above equation provides an approximation of the exact expression in~\cite{Buras:2020xsm} up to $1\%$.

In the case of $\klpn$, using the exact expression for the branching ratio together with ${\rm Im}\lambda_t$ in (\ref{imlt}),
we have
\begin{equation}\label{bklpn}
\mathcal{B}(\klpn)=\kappa_L \vcb^4
\left[\frac{{\lambda\sqrt{\sigma}}\sin\beta\sin\gamma}{\lambda^5}X(x_t)\right]^2,
\end{equation}
\noindent
where \cite{Mescia:2007kn}
\begin{equation}\label{kapl}
\kappa_L=
(2.231\pm 0.013)\times 10^{-10}\left[\frac{\lambda}{0.225}\right]^8.
\end{equation}
As we will see below, the fact that the function $X(x_t)$ enters universally  $\kpn$ and
  $\klpn$ branching ratios implies a practically $m_t$-independent 
  relation between {them}.

Due to the absence of $P_c(X)$ in (\ref{bklpn}), 
$\mathcal{B}(K_L\to\pi^0\nu\bar\nu)$ 
has essentially no theoretical uncertainties. It is only affected by
parametric uncertainties coming from $\vcb$, $\beta$ and to a lesser extent from
$\gamma$ and $m_t$.

 For our purposes we follow  \cite{Buras:2015qea} and cast the formulae
 (\ref{AIACD}) and (\ref{bklpn}) into a {semi-numerical} form that expresses
 the dominant parametric uncertainties.
 Relative to \cite{Buras:2015qea}, in the case of 
$\kpn$
 {we just change the central values of $\vcb$ and $\gamma$ into the {reference ones in (\ref{REFCKM})} 
   and we evaluate the central value for the branching ratio using (\ref{AIACD}).} {Furthermore, we express the $\gamma$-dependence with a sine function.} In the case of
$\klpn$, which was given in  \cite{Buras:2015qea} in terms of $\vub$, $\vcb$ and
$\gamma$, we {also} trade the dependence on $\vub$ for the one on $\beta$ using
(\ref{vtdvub}).
We find
\begin{align}
   { \mathcal{B}(\kpn) = {(7.92 \pm 0.28)} \times 10^{-11} \,
    \bigg[\frac{\left|V_{cb}\right|}{41.0\times 10^{-3}}\bigg]^{2.8}
    \bigg[\frac{\sin\gamma}{\sin 67^\circ}\bigg]^{1.39}},\label{kplusApprox}
\end{align}
\begin{align}
    \mathcal{B}(\klpn) ={(2.61 \pm 0.04)} \times 10^{-11} \,
    &{\bigg[\frac{\left|V_{cb}\right|}{41.0\times 10^{-3}}\bigg]^4
\bigg[\frac{\sin\gamma}{\sin(67^\circ)}\bigg]^{2}
    \bigg[\frac{\sin\beta}{\sin({22.2^\circ})}\bigg]^2},\label{k0Approx}
\end{align}
where we do not show explicitly the parametric dependence on $\lambda$ and set $\lambda=0.225$.

One can check using exact expressions that the $\lambda$ dependence of $ \mathcal{B}(\kpn)$ is very weak, due to partial cancellations among different
contributions. Even setting  $\lambda=0$ this branching ratio changes only by 4\%.
The parametric relation for $\mathcal{B}(\klpn)$ is exact, while for $\mathcal{B}(\kpn)$ it gives an excellent approximation:
{with respect to (\ref{AIACD}), for the large ranges {$38 \leq |V_{cb}|\times 10^{3} \leq 43$ and $60^\circ \leq \gamma \leq 75^\circ$ it is accurate to $1.5\%$ scanning one parameter at a time and to $2.5\%$ letting both of them vary simultaneously in the corresponding intervals.} {The non-integer exponents, here and in similar equations in the following, are indeed fitted to describe as {power-law} functions of parameters some more complicated exact expressions, with the best possible accuracy.} Note that (\ref{kplusApprox}) provides an even better approximation of the exact branching ratio~\cite{Buras:2020xsm}, up to $1\%$ scanning one parameter at a time and to $1.5\%$ letting both of them vary simultaneously in the corresponding intervals.}
{In the case of $\mathcal{B}(\kpn)$ the dependence on $\beta$ is very weak as
one can verify even analytically by inspecting the formula (\ref{AIACD}).
Therefore 
we have absorbed it into the non-parametric error.} 

\begin{figure}[t!]
\centering%
\includegraphics[width=0.48\textwidth]{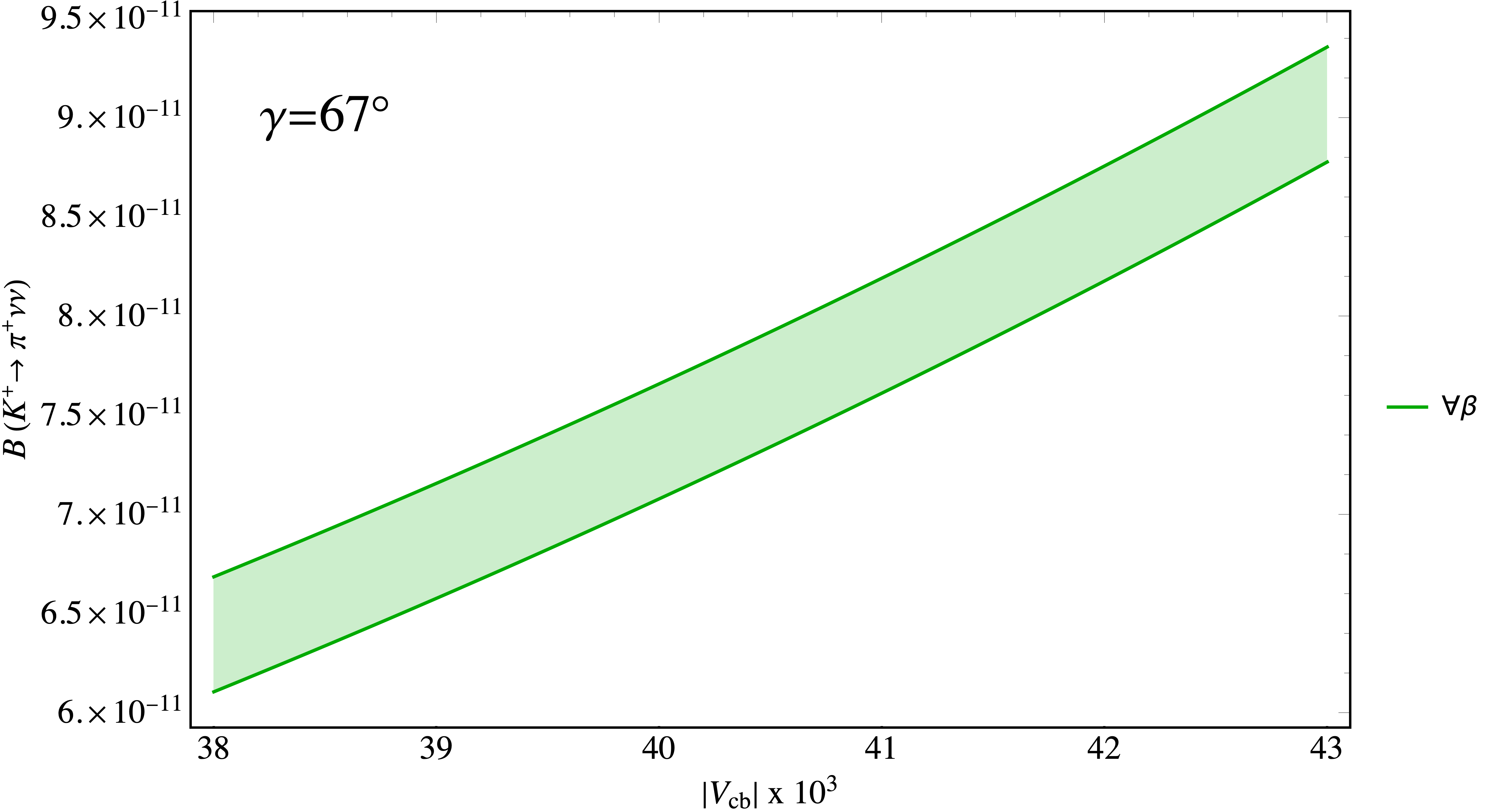}%
\hfill%
\includegraphics[width=0.48\textwidth]{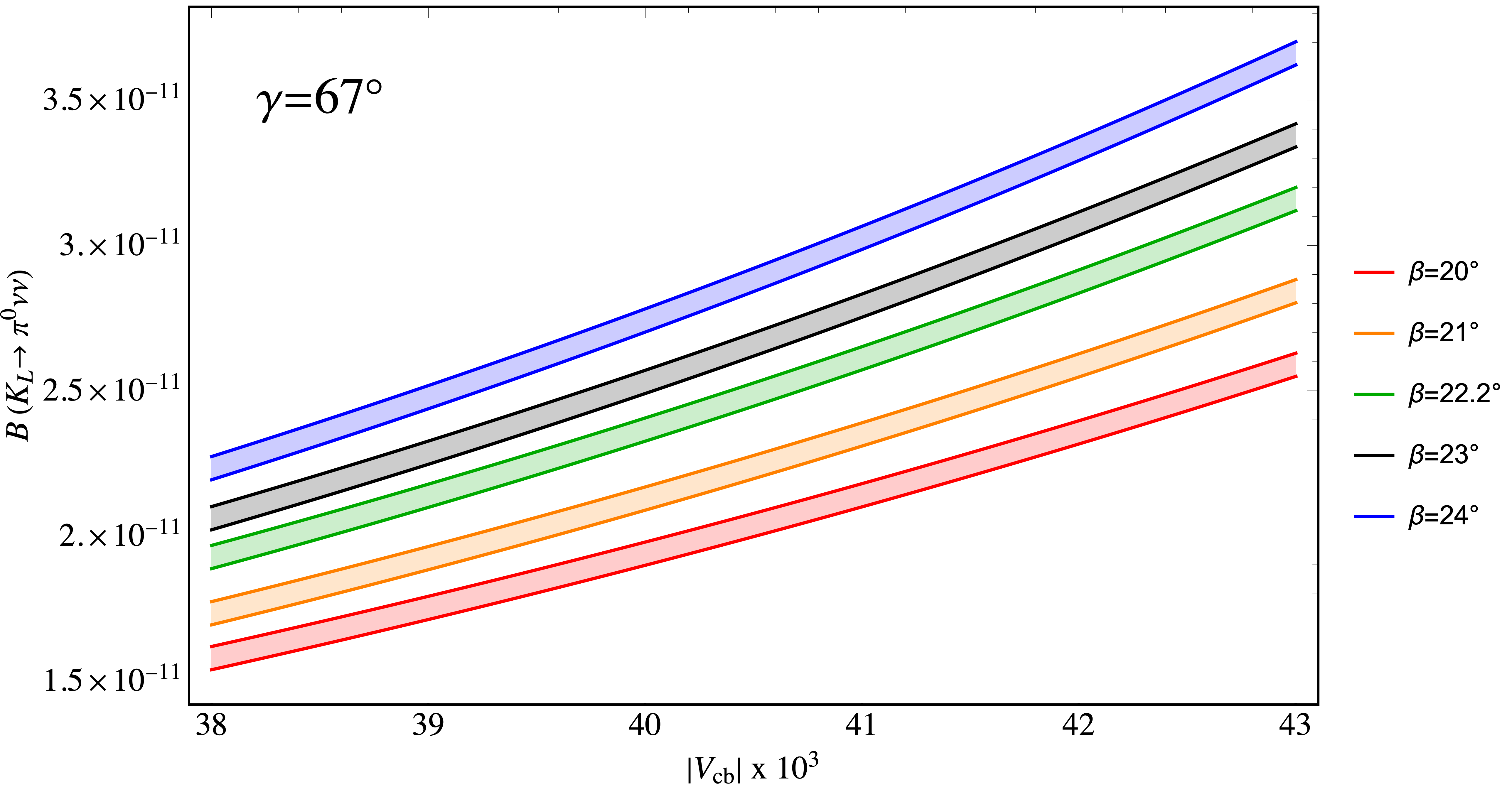}\\
\includegraphics[width=0.48\textwidth]{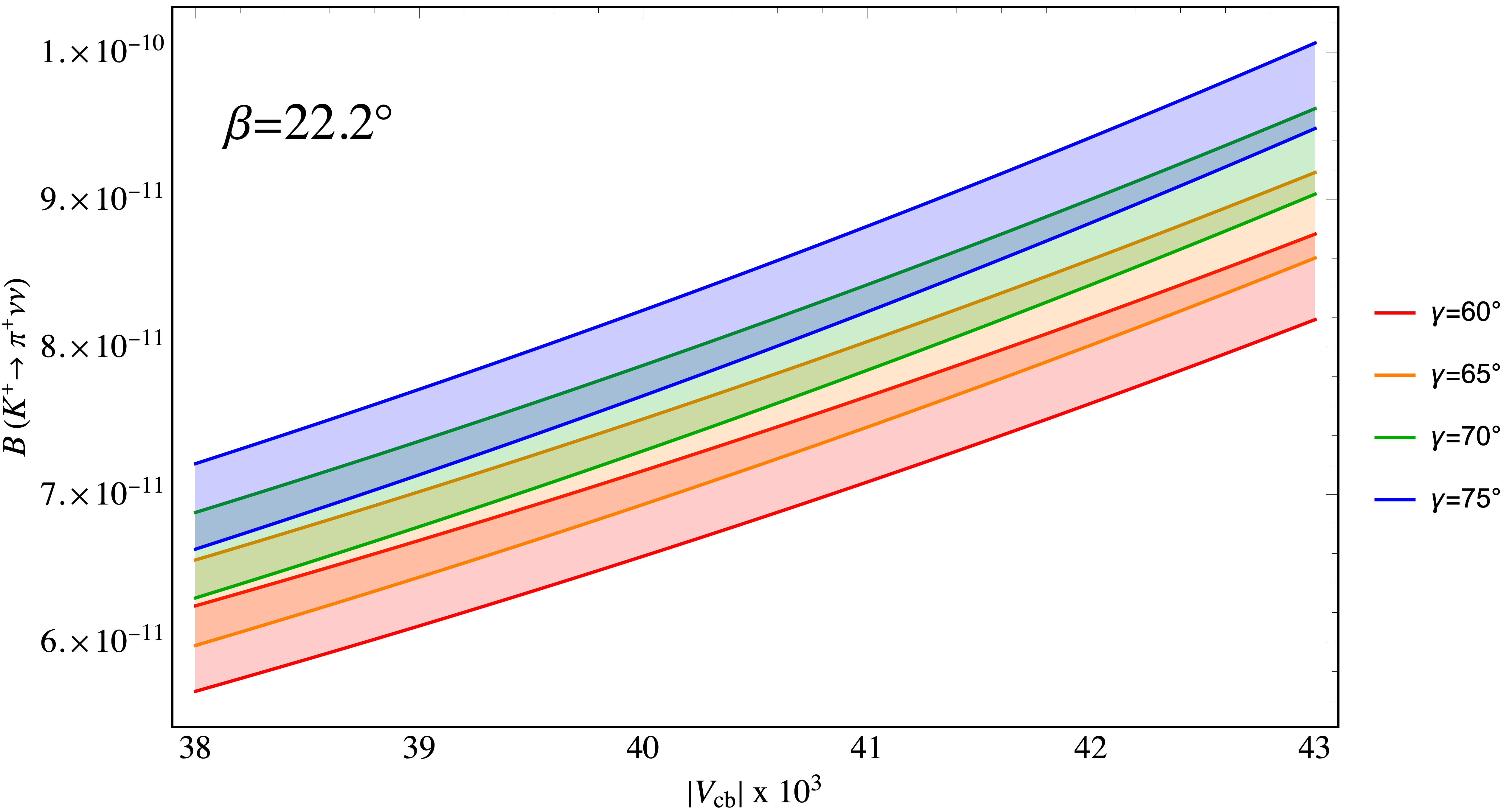}%
\hfill%
\includegraphics[width=0.48\textwidth]{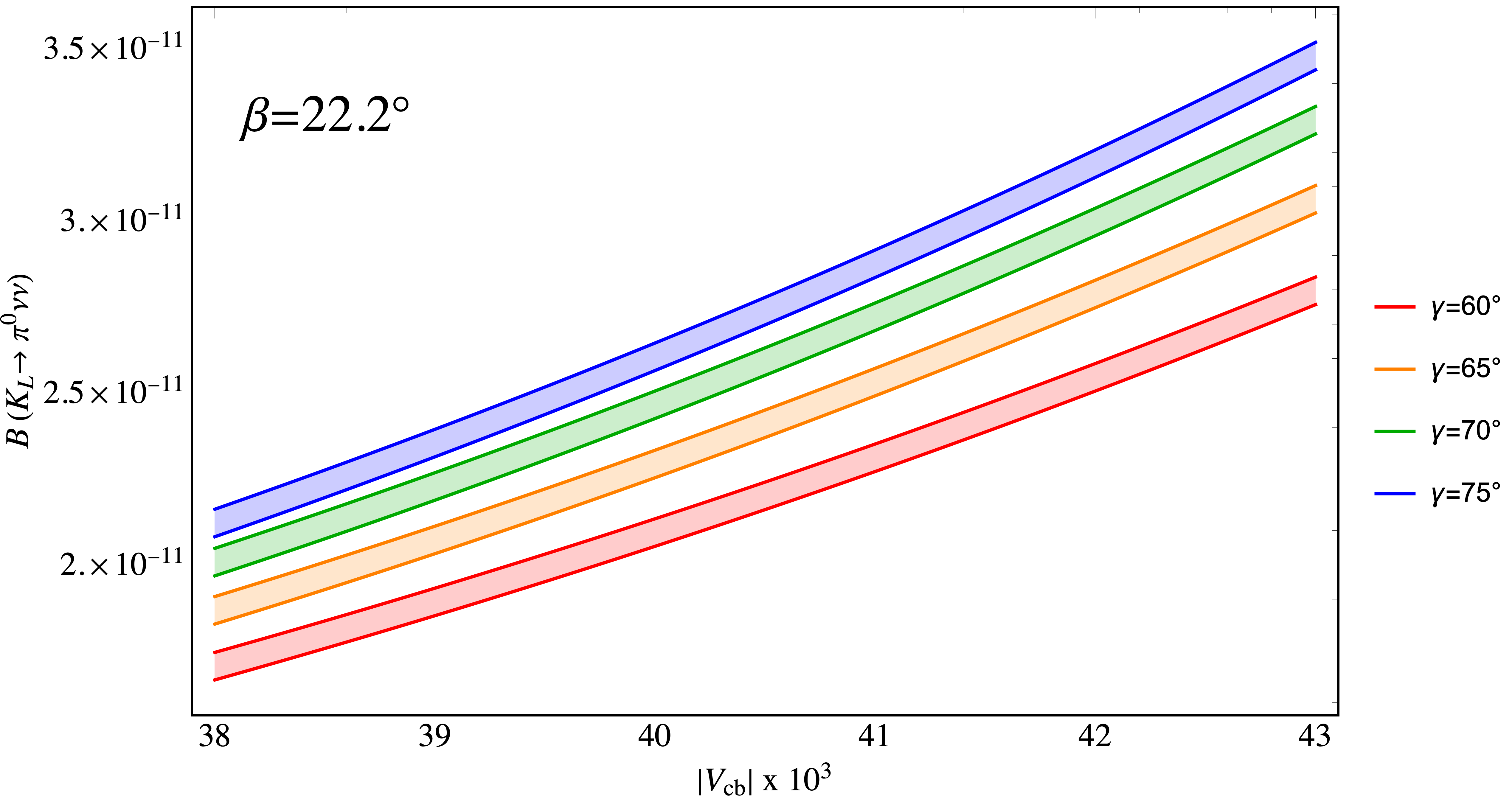}%
\caption{\it {The dependence of the branching ratios $\mathcal{B}(\kpn)$ (left panels) and $\mathcal{B}(\klpn)$ (right panels) on  $|V_{cb}|$ for different values of   $\beta=20.0^\circ,21.0^\circ, 22.0^\circ, 23.0^\circ, 24.0^\circ$  at fixed $\gamma=67^\circ$ and for different values of $\gamma=60.0^\circ, 65.0^\circ, 70.0^\circ, 75^\circ$  at fixed $\beta=22.2^\circ$ . {The width of the bands represents the uncertainties whose origin is not related to the $\gamma$, $\beta$ and $\vcb$ parameters.}} 
\label{fig:CKMdependence}}
\end{figure}

The exact dependence of both branching ratios on $\vcb$ for different $\gamma$
and $\beta$ is shown in Fig.~\ref{fig:CKMdependence}.  We observe the
pattern summarized at the beginning of our paper. Furthermore, one can notice that the branching ratio for $\kpn$  is almost exactly independent {of} the $\beta$ angle, as commented above. {One can also see that the largest uncertainties on these branching ratios {are}  due to the $\vcb$ parameter: the {elimination} of this source of error is the main focus of our work.}

The dependence of the branching ratios for $\kpn$ and $\klpn$ on $\gamma$, $\beta$ and $\vcb$  has also been studied in \cite{Blanke:2018cya}.  There are other useful results in
 that paper, in  \cite{Buras:2015qea},  \cite{Blanke:2016bhf} and \cite{Buras:2015yca}. In the latter paper several simplified models have been presented.

{The theoretically clean character of $\klpn$ and its very strong dependence
on $\vcb$ could in principle allow a precise measurement of $\vcb$ \cite{Buras:1994rj}  by inverting (\ref{bklpn}) to obtain
\begin{equation}\label{bklpn1}
  \vcb^2= \sqrt{\frac{\mathcal{B}(\klpn)}{\kappa_L}}
  \frac{\lambda^5}{{\lambda\sqrt{\sigma}}\sin\beta\sin\gamma X(x_t)}.
\end{equation}
{Note that a $10\%$ measurement of the branching ratio allows to determine
  $\vcb$ with the precision of $2.5\%$.
This strategy cannot be executed at present and it is likely
polluted by NP contributions. But 
 inserting (\ref{bklpn1}) into (\ref{AIACD}) allows to derive the expression for
$\kpn$ branching ratio in terms of the $\klpn$ one.}

To this end it is useful to define  the ``reduced'' branching ratios \cite{Buchalla:1994tr}
\begin{equation}\label{b1b2}
B_1=\frac{\mathcal{B}(\kpn)}{\kappa_+(1+\Delta_{\text{EM}})},\qquad
B_2=\frac{\mathcal{B}(\klpn)}{\kappa_L}.
\end{equation}
We find then
\be\label{B1vsB2}
B_1=B_2\left[1+\frac{1}{\sigma^2}(\cot\beta+\frac{\sqrt{\sigma} P_c(X)}{\sqrt{B_2}})^2\right],
\ee
{with $\sigma$ defined in (\ref{vtdvub})}.

It should be emphasized that this relation is independent of $\vcb$, $\gamma$
and $m_t$. However, as  (\ref{AIACD}) is not exact also this relation is an
approximation. Albeit, an excellent one, with only $1\%$ error. Therefore it can  be used in principle to determine the angle $\beta$, the sole parameter in this formula \cite{Buchalla:1994tr}. Here we just present it as an elegant formula for the {well known relation
between  $\kpn$ branching ratio and the $\klpn$ one within the SM and models with CMFV}, stressing its
very weak dependence on  $\vcb$, $\gamma$
and $m_t$.

{Alternatively using (\ref{kplusApprox}) and (\ref{k0Approx}) one can eliminate $\vcb$ to
find  
{\begin{align}
\mathcal{B}(\kpn) &= {(7.92\pm 0.30)}\times 10^{-11}   \left[\frac{\sin 22.2^\circ}{\sin \beta}\right]^{1.4} \left[\frac{\mathcal{B}(\klpn)}{{2.61}\times 10^{-11}}\right]^{0.7}\label{master0}.
\end{align}}
{This formula is also independent {of} $\gamma$ and reproduces (\ref{B1vsB2}) with an accuracy in the ballpark of 4\%; it is provided {here} in order to show the correlation between the two observables in a more transparent {manner}, while for the numerical analysis their exact expressions are used. The uncertainty shown here has been computed by propagating the non-parametric errors of the two involved branching ratios. The same procedure will be used for all the equations presenting correlations or ratios between observables.} The above relation motivates us to define the approximately $\vcb$-independent ratio
\be
\boxed{R_0=\frac{\mathcal{B}(\kpn)}{\mathcal{B}(\klpn)^{0.7}}\, ,}
\label{eq:R0}
\ee
whose $\beta$ dependence is shown in Fig.~\ref{fig:R0}. The coloured bands represent the variation of $R_0$, with fixed $\gamma$, when $\vcb$ takes values in $38 < \vcb \times 10^3< 43$, which is smaller than 0.5\%. Different bands correspond to various $\gamma$ values: when $60^\circ<\gamma<75^\circ$ the variation of $R_0$ is within the per-cent level. The numerical analysis, thus, shows that $R_0$ is indeed $\vcb$ and $\gamma$ independent to an excellent accuracy. On the other hand, the uncertainty related to the errors on parameters different from $\beta$, $\gamma$ and $\vcb$, represented in gray in the figure, is much larger, in the ballpark of 5\%.} {Restricting the value of $\beta$ to the one in (\ref{betagamma}) and including all other uncertainties we find
  \be
  (R_0)_{\rm SM}={(2.03\pm 0.11)}\times 10^{-3}\,.
  \ee}
\begin{figure}[t]
\centering%
\includegraphics[width=0.65\textwidth]{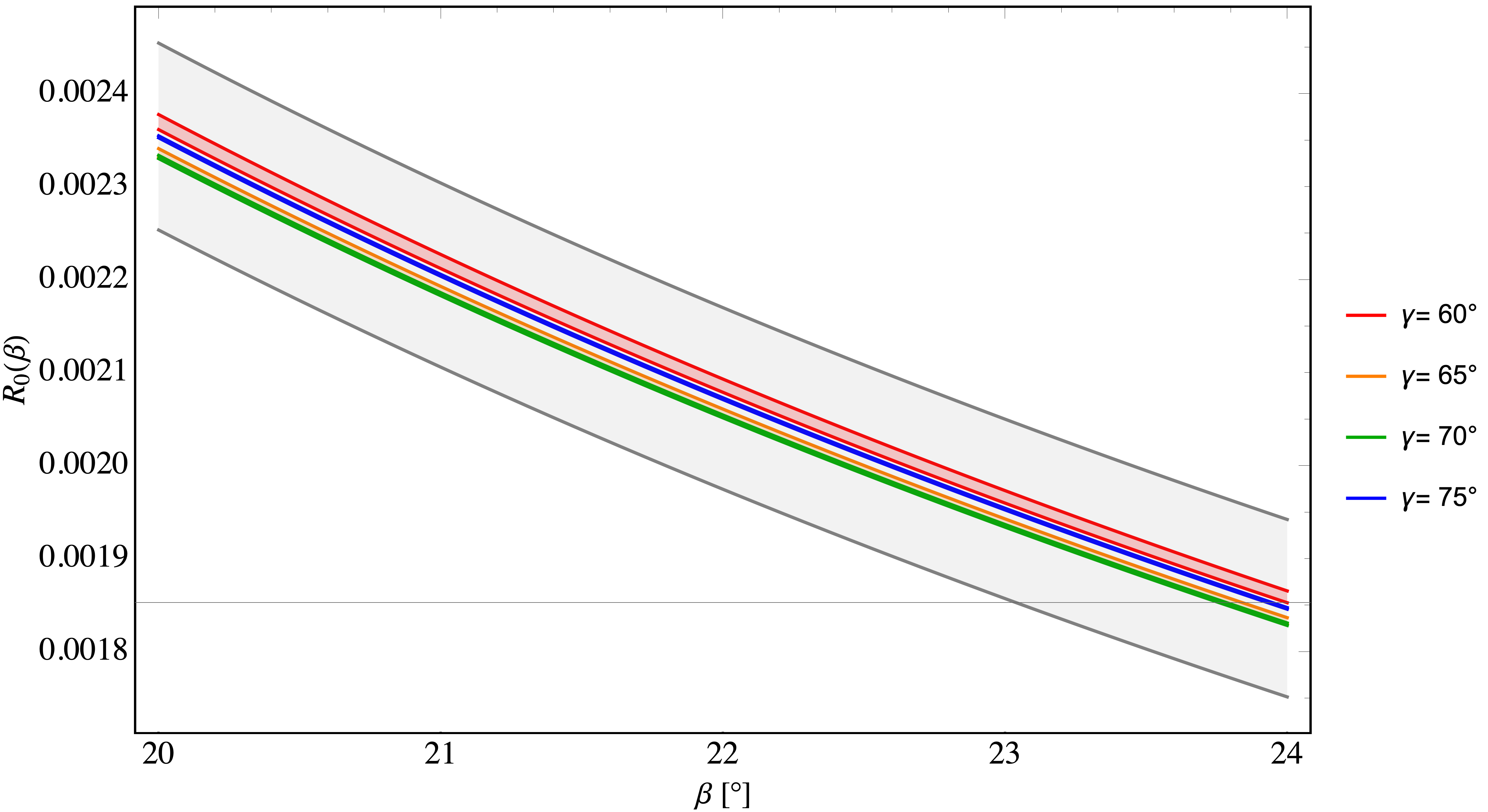}%
\caption{\it The ratio $R_0$ given in (\ref{eq:R0}) as a function of $\beta$ for different  values of $\gamma$ within the SM.
  The coloured bands correspond to $38 \leq |V_{cb}|\times 10^{3} \leq 43$. {The gray band represents the non-parametric uncertainty.} \label{fig:R0}}
\end{figure}
\begin{figure}[t]
\centering%
\includegraphics[width=0.85\textwidth]{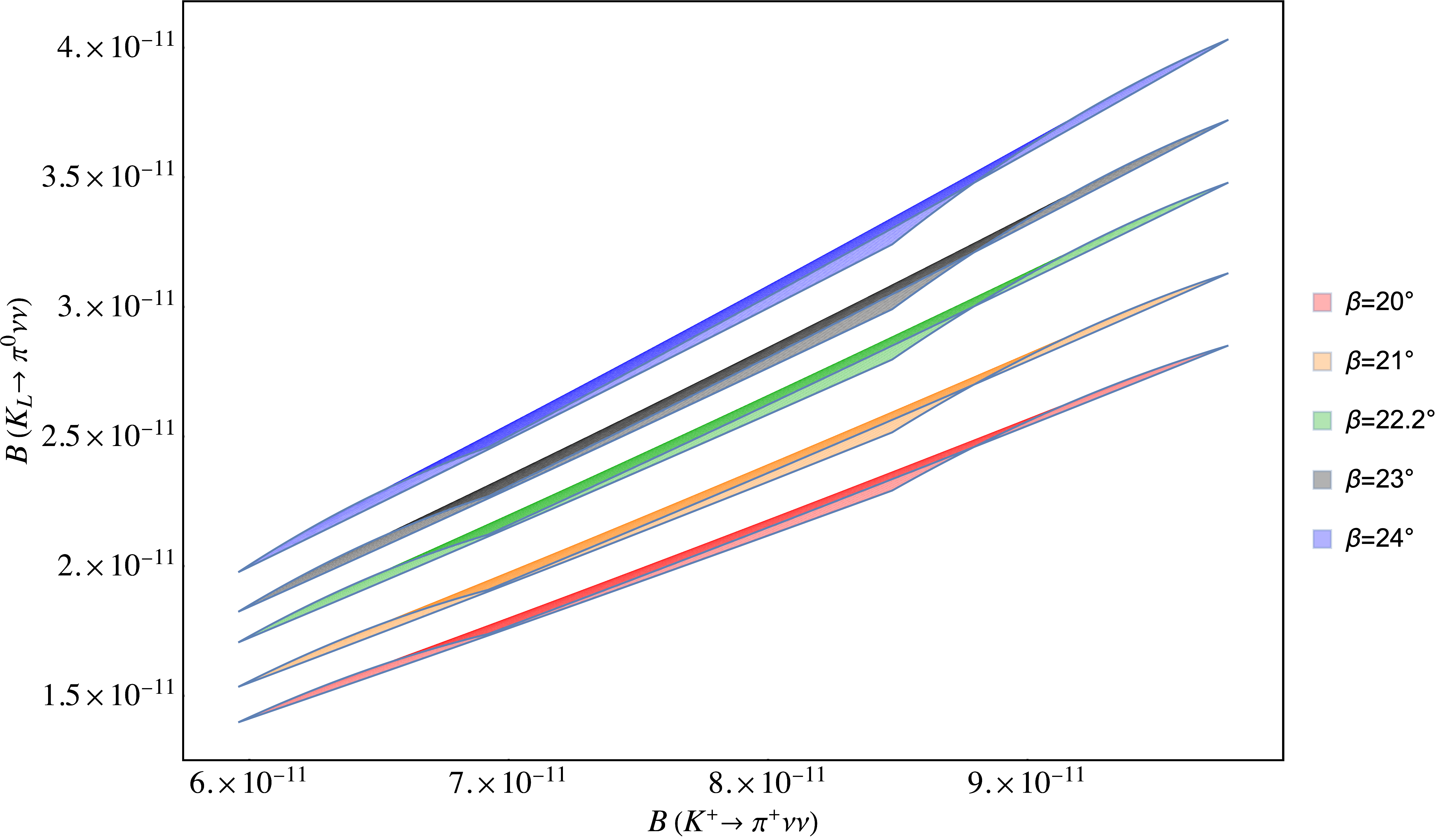}%
\caption{\it The correlation between branching ratios for $\klpn$ and $\kpn$
  as given in (\ref{B1vsB2}) for different  values of $\beta$ within the SM.
  The ranges of branching ratios correspond to $38 \leq |V_{cb}|\times 10^{3} \leq 43$ {and $60^\circ\leq\gamma \leq 75^\circ$}. \label{fig:3}}
\end{figure}
In Fig.~\ref{fig:3} we show the usual plot, {representing (\ref{B1vsB2})},  that correlates the branching ratios for
$\klpn$ and $\kpn$ in the SM for fixed values of $\beta$ \cite{Buchalla:1994tr}. The SM values depend
on $\vcb$ but the {positions of the straight lines depend  basically only  on $\beta$. Moreover, they have practically a universal slope.}
Also the dependence on $\gamma$ {almost perfectly} cancels out. {However, the cancellation is exact using (\ref{AIACD}) for $\mathcal{B}(\kpn)$, while one can see in the plot that using the exact expression there is a residual weak $\gamma$-dependence.} {The uncertainties not related to $\vcb$, $\gamma$ and $\beta$ are not shown in this figure.}

{A given line in  Fig.~\ref{fig:3} reminds us at first sight the
  correlation between $\kpn$ and $\klpn$ branching ratios in models with MFV \cite{Buras:2001af}
  for $X(x_t)>0$ and in the plots showing this correlations in different
  models, like in \cite{Buras:2015yca,Blanke:2009pq}, the SM value is represented by a point.
  But one should realize that in that papers the lines are obtained by
  varying $X$ while keeping $\vcb$ and $\beta$ fixed. On the other hand in Fig.~\ref{fig:3} while
  $X$ is kept at its SM value both $\vcb$ and $\beta$ are varied. In other
  words the SM point in the plots in \cite{Buras:2015yca,Blanke:2009pq} and similar plots   found in the literature is rather uncertain and  Fig.~\ref{fig:3}
  signals this uncertainty.} {Inspecting formulae (\ref{AIACD}) and (\ref{bklpn}) one finds that for fixed $\beta$ the position on a given straight line Fig.~\ref{fig:3} is
  determined by the combination $\vcb^2 X(x_t)$.}
 
\boldmath
\subsubsection{$K_{S}\to \mu^+\mu^-$}\label{sec:KSmm}
\unboldmath
This  decay  provides another sensitive probe of imaginary parts of
short-distance couplings. Its branching ratio receives long-distance (LD) and
short-distance (SD) contributions, which are added incoherently in the total
rate \cite{Ecker:1991ru, Isidori:2003ts}. This is in contrast to the decay
$\klm$, where LD and SD amplitudes interfere with each other; moreover $\BR(\klm)$ is
sensitive to real parts of couplings. The SD part of $\BR(\ksm)$ is given as
$(\lambda_t=V^*_{ts}V_{td})$
\begin{equation}
  \label{eq:ksm-br-SD}
 \BR(\ksm)_{\rm SD} 
  = \tau_{K_S} \frac{G_F^2 \alpha^2}{8 \pi^3\sin^4\theta_W} m_K F_K^2 \sqrt{1-4\frac{m^2_\mu}{m_K^2}} m_\mu^2\,
    \mbox{Im}^2 \left[\lambda_t  Y(x_t) \right] ,
\end{equation}
which, applying (\ref{imlt}), can be expressed as
{
\be
\label{eq:ksm-br-SDAppr}
 \BR(\ksm)_{\rm SD} 
=1.04\times 10^{-5}\vcb^4
\left[{\lambda\sqrt{\sigma}}\,{\sin\gamma\sin\beta} \,Y(x_t)\right]^2 \, ,
\ee
} where $Y(x_t)$ is given in the Appendix~\ref{App}. 

In 2019  the LHCb collaboration improved the upper bound on $\ksm$
by one order of magnitude \cite{Aaij:2017tia}
\begin{align}
  \label{ksmbound}
  \BR(\ksm)_{\rm LHCb} & < 0.8\, (1.0) \times 10^{-9} &
  \mbox{at} \; 90\%\, (95\%) \; \mbox{C.L.}
\end{align}
to be compared with the SM prediction \cite{Isidori:2003ts, DAmbrosio:2017klp}
\begin{align}\label{ISDA}
  \BR(\ksm)_{\rm SM} &
  = (4.99_{\rm LD} + 0.19_{\rm SD}) \times 10^{-12} 
  = (5.2 \pm 1.5) \times 10^{-12}.
\end{align}

Recently it has been demonstrated in \cite{Dery:2021mct} that the short distance contribution
in (\ref{eq:ksm-br-SD}) can be extracted from data offering us still another
precision observable. Here we point out that in the SM the ratio (see (\ref{bklpn}) and (\ref{eq:ksm-br-SDAppr}))
\be\label{SR1}
\boxed{R_{\rm SL}=\frac{\BR(\ksm)_{\rm SD}}{\BR(\klpn)}=1.55\times 10^{-2}\,\left[\frac{\lambda}{0.225}\right]^2
\left[\frac{Y(x_t)}{X(x_t)}\right]^2\,}
\ee
is independent of any SM parameter except for $m_t$ and $\lambda$ which are both precisely known. 

%Beyond the SM, we define
%\be
%X_{\rm eff}\equiv \lambda_t  X(x_t,v), \qquad Y_{\rm eff}\equiv \lambda_t  Y(x_t,v),
%\ee
%where $X(x_t,v)$ and $Y(x_t,v)$ include both SM and BSM contributions and
%$v$ denotes the collection of new physics parameters. Then (\ref{SI1}) is 
%generalized in the case of vector currents to
%\be\label{Simple2}
%\frac{\BR(\ksm)_{\rm SD}}{\BR(\klpn)}=1.56\times 10^{-2}\,\left[\frac{\lambda}{0.225}\right]^2BV
%\left[\frac{{\IM}(Y_{\rm eff})}{{\IM}(X_{\rm eff})}\right]^2\,.
%\ee

\boldmath
\subsection{Correlations with $B_{s,d}\to\mu^+\mu^-$}
\unboldmath
Using the exact formulae {in \cite{Buras:2020xsm}} that are based
on the calculations over three decades by several groups
\cite{Buchalla:1993bv,Buchalla:1998ba, Buras:2012ru, Bobeth:2013uxa, Bobeth:2013tba, Hermann:2013kca,Beneke:2017vpq, Beneke:2019slt}
the dependence of the branching ratio for $B_s\to\mu^+\mu^-$
on the input parameters involved can be transparently summarized as follows 
 \cite{Bobeth:2013uxa} 
\begin{equation}
  \overline{\mathcal{B}}(B_{s}\to\mu^+\mu^-)_{\rm SM} = ({3.47}\pm0.06)\times 10^{-9}
\left(\frac{F_{B_s}}{{230.3}\mev}\right)^2 \left|\frac{V_{tb}^*V_{ts}}{{0.0402}}\right|^2\bar R_s
\label{BRtheoRpar}
\ee
\noindent
where
\be
\label{Rs}
\bar R_s=
\left(\frac{\tau_{B_s}}{{1.515} {\rm ps}}\right)\left(\frac{{0.935}}{r(y_s)}\right)
 \left(\frac{m_t(m_t)}{162.83 \GeV}\right)^{3.02}\left(\frac{\alpha_s(M_Z)}{0.1184}\right)^{0.032} \,.
 \ee
 Here $r(y_s)$ summarizes $\Delta\Gamma_s$ effects with $r(y_s)={0.935}\pm0.007$ within the SM  \cite{DescotesGenon:2011pb,deBruyn:2012wj,deBruyn:2012wk}.

Similarly one finds \cite{Bobeth:2013uxa}
\begin{equation}
  {\mathcal{B}}(B_{d}\to\mu^+\mu^-)_{\rm SM} = ({0.968}\pm0.02)\times 10^{-10}
\left(\frac{F_{B_d}}{{190.0}\mev}\right)^2 \left|\frac{V_{tb}^*V_{td}}{0.00848}\right|^2\bar R_d \,.
\label{BRtheoRpard}
\ee
\noindent
As  to an excellent {accuracy}  $r(y_d)=1$, one has this time
\be
\label{Rd}
\bar R_d=
\left(\frac{\tau_{B_d}}{1.519 {\rm ps}}\right)
 \left(\frac{m_t(m_t)}{162.83 \GeV}\right)^{3.02}\left(\frac{\alpha_s(M_Z)}{0.1184}\right)^{0.032} \,.
\ee

{These two branching ratios are shown as functions of $\vcb$ in Fig.~\ref{fig:CKMdependenceB}, for different values of $\gamma$ with fixed $\beta$ and viceversa. One can notice that the $\beta$ dependence is almost negligible for both observables, while the $\gamma$ dependence is stronger, especially in the case of the $B_d$ decay. Furthermore, very importantly, it is evident that {the size
    of possible anomaly in $B_s\to\mu^+\mu^-$  depends strongly  on the value of $\vcb$ as emphasized recently in \cite{Bobeth:2021cxm}. For the values
    of $\vcb$ from inclusive determinations an anomaly at the level of
    $2\sigma$ can be concluded  \cite{Geng:2021nhg,Altmannshofer:2021qrr}, while
for $\vcb$ values smaller than $40.3 \times 10^{-3}$ the SM prediction agrees with the experimental measurement of $B_s\to \mu^+\mu^-$ decay at $1\sigma$ level. {Finally, for} the FLAG's $\vcb$ value in the ballpark
of $39 \times 10^{-3}$ perfect agreement of the SM with the data is obtained.
Only by constructing the ratio $R_s$ in (\ref{CMFV6}) an anomaly at the
level of $2.2 \sigma$ independently of $\vcb$ can be concluded \cite{Bobeth:2021cxm}\footnote{On the other hand as shown in \cite{Buras:2022wpw} this anomaly increases to $2.7 \sigma$ if only 2+1+1 hadronic matrix element in
    $\Delta M_{s}$ from HPQCD collaboration \cite{Dowdall:2019bea} is used and not the average of     2+1+1 and 2+1 LQCD data as done here.}. {See also comments after (\ref{LHCbTH}).}
  }

\begin{figure}[t!]
\centering%
\includegraphics[width=0.48\textwidth]{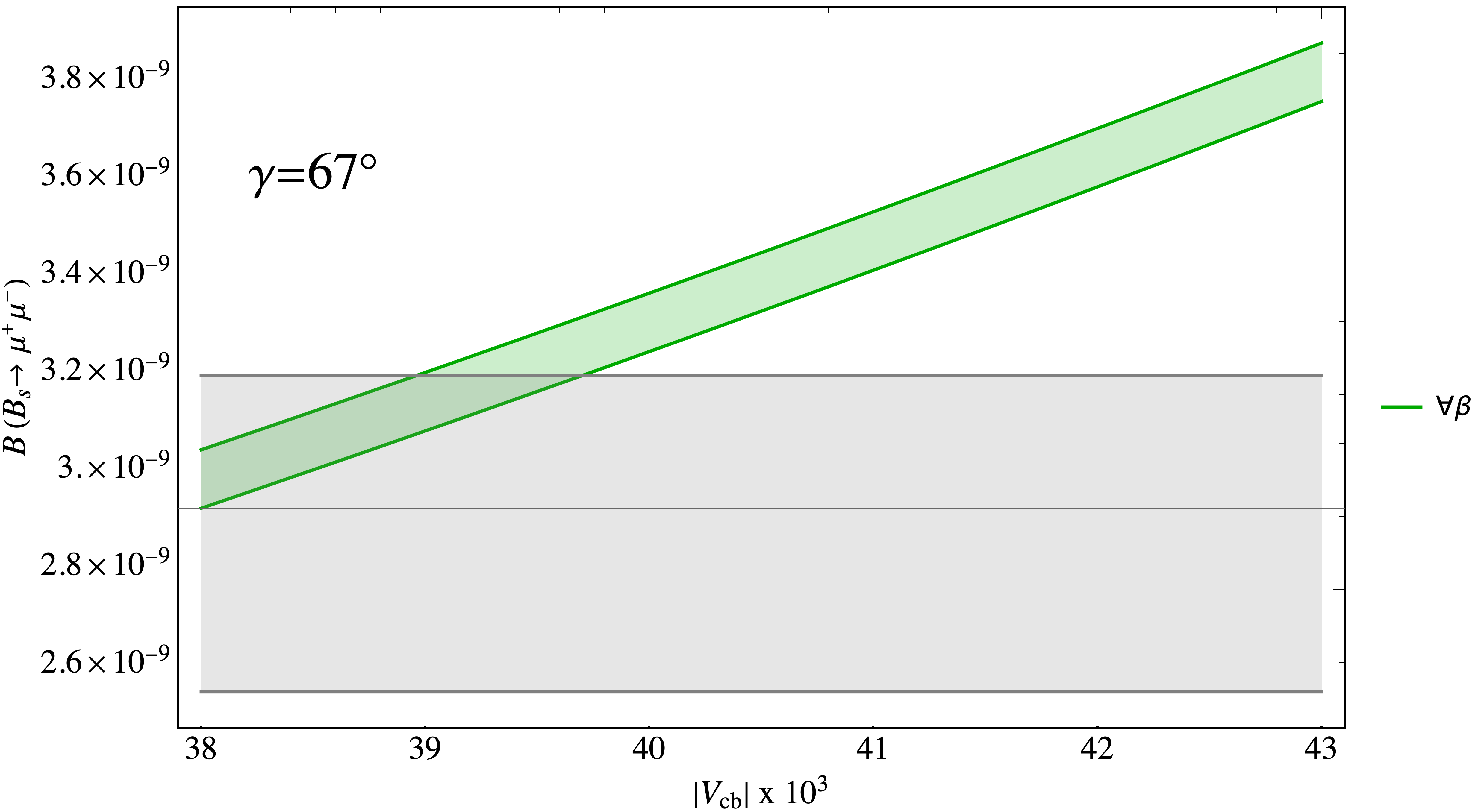}%
\hfill%
\includegraphics[width=0.48\textwidth]{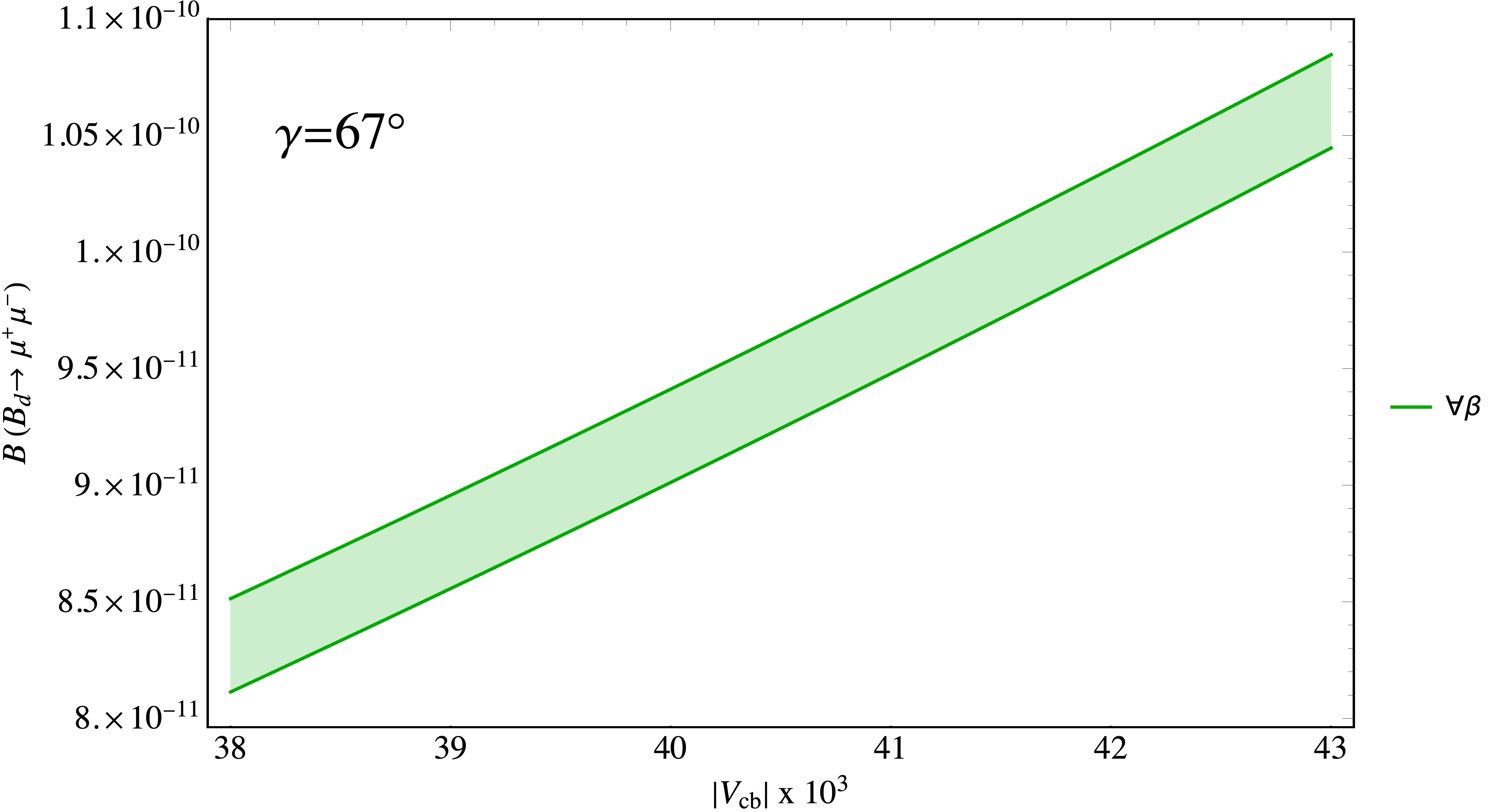}\\
\includegraphics[width=0.48\textwidth]{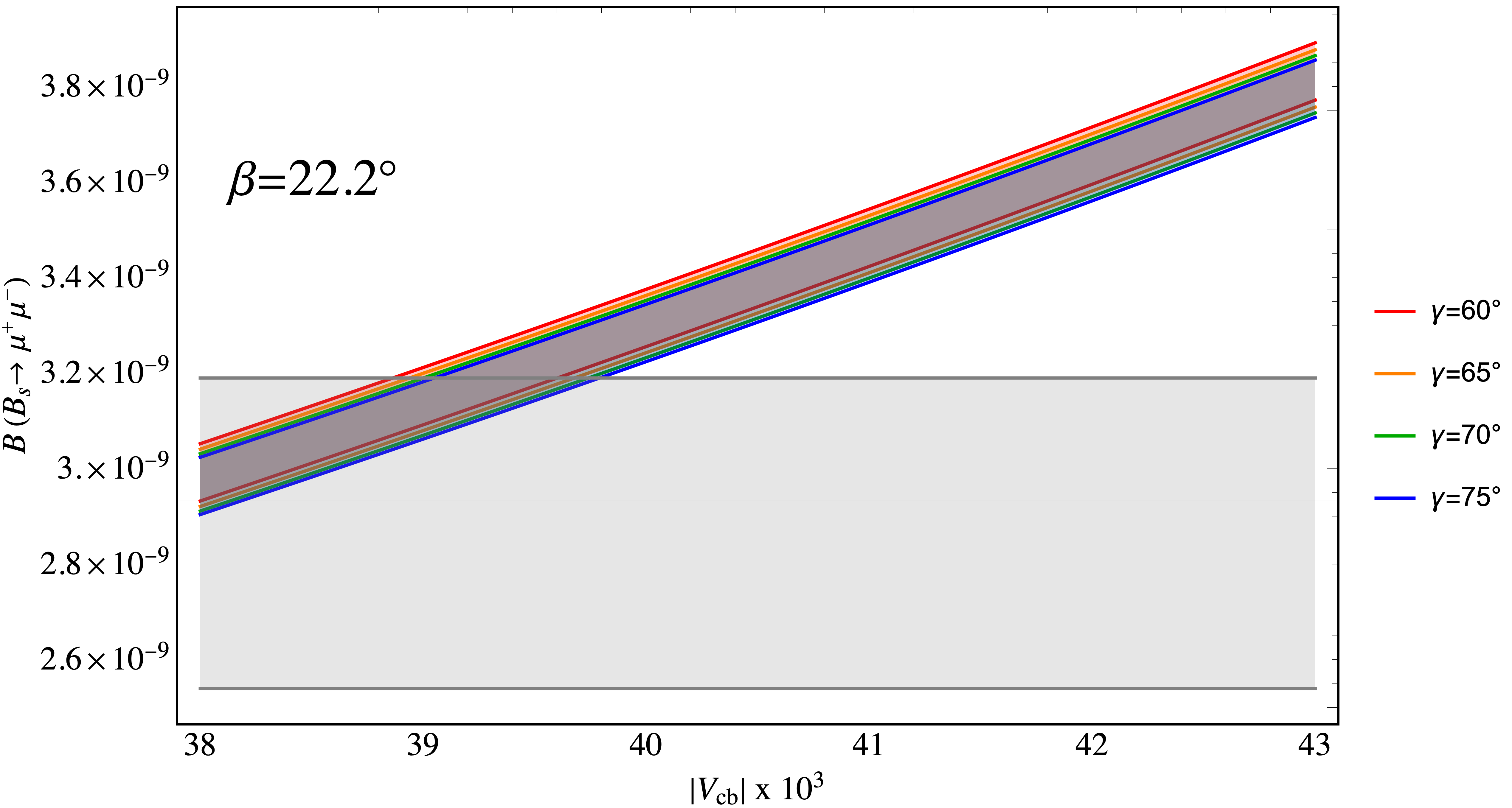}%
\hfill%
\includegraphics[width=0.48\textwidth]{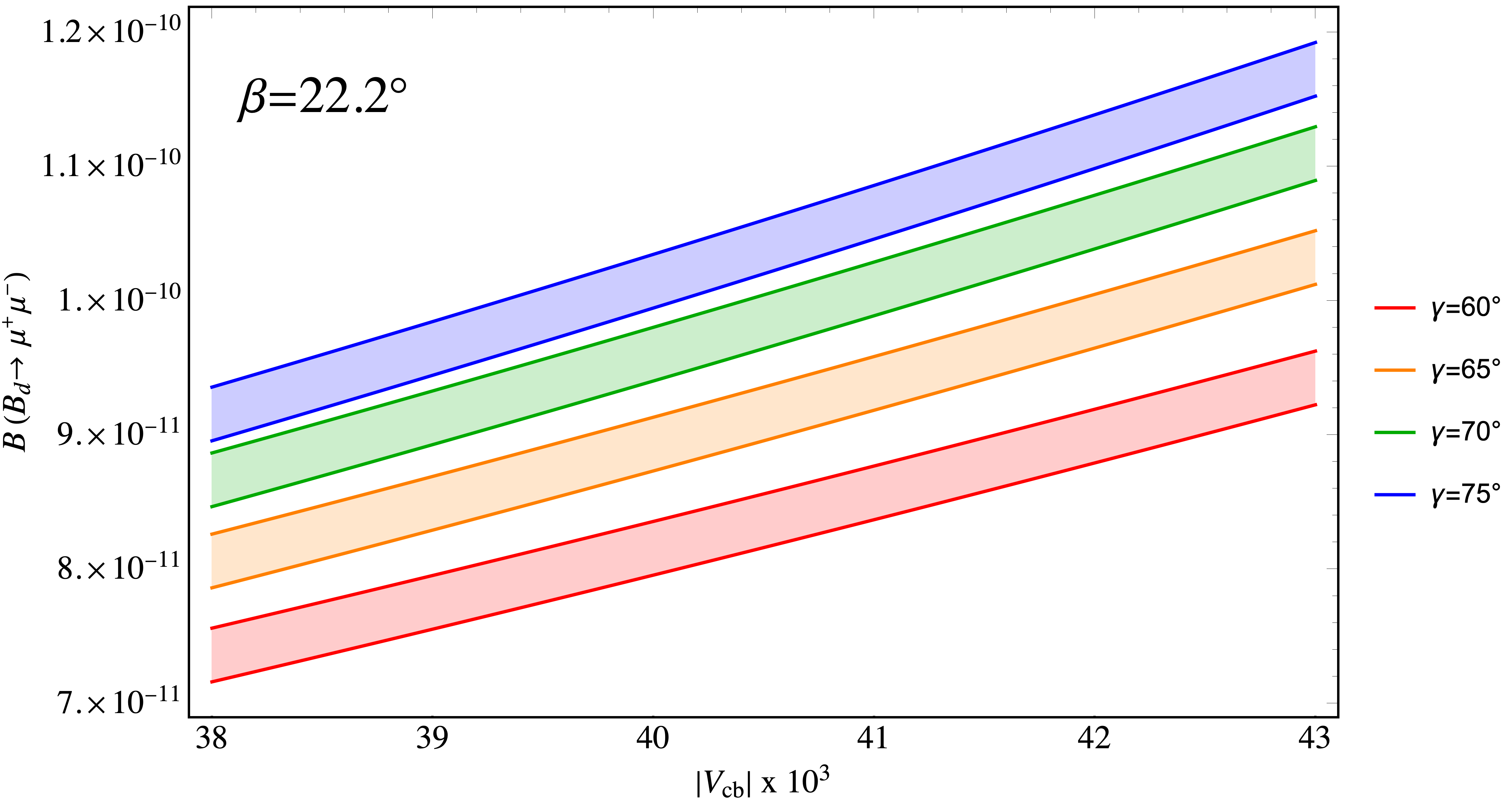}%
\caption{\it {The dependence of the branching ratios $ \overline{\mathcal{B}}(B_{s}\to\mu^+\mu^-)$ (left panels) and $ {\mathcal{B}}(B_{d}\to\mu^+\mu^-)$ (right panels) on  $|V_{cb}|$ for different values of   $\beta=20.0^\circ,21.0^\circ, 22.0^\circ, 23.0^\circ, 24.0^\circ$  at fixed $\gamma=67^\circ$ and for different values of $\gamma=60.0^\circ, 65.0^\circ, 70.0^\circ, 75^\circ$  at fixed $\beta=22.2^\circ$ . {The width of the bands represents the uncertainties whose origin is not related to the $\gamma$, $\beta$ and $\vcb$ parameters. The gray horizontal bands in the left panels shows the $1\sigma$ experimental range (see (\ref{eq:WAV-Bs})).}} 
\label{fig:CKMdependenceB}}
\end{figure}

Following the same strategy as for the analysis of the $\mathcal{B}(\kpn)$-$\mathcal{B}(\klpn)$ correlation, one can take advantage from the fact that $\overline{\mathcal{B}}(B_s\to\mu^+\mu^-)$ is an exact quadratic function of $\vcb$ (see (\ref{vts}) and (\ref{BRtheoRpar})), and find 
\be
\label{eq:vcb-B3}
\vcb^2= \frac{\overline{\mathcal{B}}(B_s\to\mu^+\mu^-)}{{2.14} \times 10^{-6}\bar R_s} \left(\frac{{230.3}\mev}{F_{B_s}} \frac{1}{G(\beta,\gamma)}\right)^2  \, ,
\ee
where $G(\beta,\gamma)$ is defined as in (\ref{vts}).

Inserting the above expression into (\ref{AIACD}), one can derive the relation between the branching ratios for $K^+\to \pi^+ \nu\bar{\nu}$ and $B_s\to\mu^+\mu^-$ decays. Defining
\be
B_3= \frac{\overline{\mathcal{B}}(B_s\to\mu^+\mu^-)}{{2.14} \times 10^{-6}{\bar R_s}}\, ,
\ee
one obtains {
\begin{align}
\label{eq:B1B3}
& B_1=\frac{X(x_t)^2}{\lambda^8} B_3^2 \left(\frac{{230.3}\mev}{F_{B_s}} \frac{1}{G(\beta,\gamma)}\right)^4  \nonumber\\
 \times\Big[& \sigma \sin^2\gamma \sin^2\beta+\frac{1}{\sigma}\left(\sin\gamma \cos\beta+\frac{\lambda^4 P_c(X)}{B_3 X(x_t)} \left(\frac{F_{B_s}}{{230.3}\mev} G(\beta,\gamma)\right)^2\right)^2 \Big]\, ,
\end{align}
with $\sigma$ and $G(\beta,\gamma)$ defined previously.}

Alternatively using (\ref{kplusApprox}) and (\ref{BRtheoRpar}) one can eliminate $\vcb$ to
find \cite{Buras:2015qea}\footnote{Relative to \cite{Buras:2015qea} we just adjusted the central value of $\gamma$ to previous formulae.} 
{\begin{align}
\mathcal{B}(\kpn) &= {(7.92\pm 0.34)}\times 10^{-11}   \left[\frac{\sin\gamma}{\sin 67^\circ}\right]^{1.39} \left[\frac{G(22.2^\circ,67^\circ)}{G(\beta,\gamma)}\right]^{2.8}\notag\\
&\qquad\qquad\qquad\quad\times\left[\frac{\overline{\mathcal{B}}(B_s\to\mu^+\mu^-)}{{3.47}\times 10^{-9}\bar R_s}\right]^{1.4}\left[\frac{{230.3}\mev}{F_{B_s}}\right]^{2.8}\label{master1}.
\end{align}}
%{The uncertainty shown here has been computed by propagating the non-parametric errors of the two involved branching ratios. The same procedure will be used for all the equations presenting correlations or ratios between observables.}
The above expression reproduces the correlation of (\ref{eq:B1B3}) with an accuracy {of less than $0.5\%$}, when $\gamma$ varies in the range $60^\circ \leq \gamma\leq 75^\circ$ and the branching ratio $\overline{\mathcal{B}}(B_s\to\mu^+\mu^-)$ takes values in its $1\sigma$ interval.
Here, the dependence on $\gamma$ is slightly different with respect to (\ref{kplusApprox}), due to the {$\gamma$ dependence of the $\vts$ element} entering in $\overline{\mathcal{B}}(B_s\to\mu^+\mu^-)$. 
%\ev{I would put the updated central value of $\mathcal{B}(\kpn)$ in front, like $8.30 \times 10^{-11}$. However, with this values and the same 0.81 $\gamma$ exponent, we have a deviation of 4.7\% for $\gamma=80^\circ$, which becomes 2.5\% for $\gamma=75^\circ$. One, fixing 8.30 in front, could adjust the $\gamma$ exponent to have a better approximation. For $\gamma=80^\circ$ I haven't found any satisfactory result, but keeping $\gamma \leq 75^\circ$, with 0.76 exponent we have an accuracy of $\sim 2.2\%$.}

%\ajb{Possibly varying $\gamma$ only up to 75 would improve accuracy.} \ev{Yes, a bit: 1.7\% instead of 2.1\%.}

Proceeding in the same manner with $B_d\to\mu^+\mu^-$ and defining
\be
B_4= \frac{\mathcal{B}(B_d\to\mu^+\mu^-)}{{1.34} \times 10^{-6}\bar R_d}\, ,
\ee
we find analogous equations to (\ref{eq:B1B3}) and (\ref{master1}), namely
\begin{align}\label{eq:B1B4}
& B_1=\frac{X(x_t)^2}{\lambda^8} B_4^2 \left(\frac{{190.0}\mev}{F_{B_d}} \frac{1}{\lambda}\right)^4 \frac{1}{\sin^2\gamma} \nonumber\\
 \times\Big[& \sigma \sin^2\beta+\frac{1}{\sigma}\left( \cos\beta+\frac{\lambda^4 P_c(X)}{B_4 X(x_t)} \left(\frac{F_{B_d}}{{190.0}\mev} \lambda\right)^2\sin\gamma\right)^2 \Big]\, 
\end{align}
and 
{\begin{align}
\mathcal{B}(\kpn) &= {(7.92\pm 0.36)}\times 10^{-11}     \left[\frac{\sin (67^\circ)}{\sin\gamma}\right]^{1.41}\notag\\
&\qquad\qquad\qquad\quad\times\left[\frac{{\mathcal{B}}(B_d\to\mu^+\mu^-)}{{0.968}\times 10^{-10}\bar R_d}\right]^{1.4}\left[\frac{{190.0}\mev}{F_{B_d}}\right]^{2.8}\label{master2}.
\end{align}}
 The last expression provides an approximation of (\ref{eq:B1B4}) accurate to {$0.5\%$}, when $\gamma$ varies in the range $60^\circ \leq \gamma\leq 75^\circ$ and the branching ratio ${\mathcal{B}}(B_d\to\mu^+\mu^-)$ takes values in its $1\sigma$ interval.

Note that  the  simple relations in (\ref{master1}) and (\ref{master2})
are independent of $\vcb$
and in fact represent exact expressions to an excellent accuracy. This motivates us to define the following two $\vcb$-independent ratios
\be\label{R12}
 \boxed{R_1(\beta,\gamma)=\frac{\mathcal{B}(\kpn)}{\left[{\overline{\mathcal{B}}}(B_s\to\mu^+\mu^-)\right]^{1.4}},\qquad
   R_2(\beta,\gamma)=\frac{\mathcal{B}(\kpn)}{\left[{\mathcal{B}}(B_d\to\mu^+\mu^-)\right]^{1.4}}.}
   \ee
 In particular the ratio $R_1$  should be of interest in 
the coming years due to  
the improved  measurement of $\kpn$ by NA62, of $B_s\to\mu^+\mu^-$ by LHCb, CMS and ATLAS and of $\gamma$ by LHCb {and Belle II}. Moreover the accuracy of the last factor in (\ref{master1}) has been improved
by LQCD since the 2015 analysis in \cite{Buras:2015qea}. 

In our numerical analysis, where the exact expressions for the branching ratios are used, the ratios above indeed turn out to be $\vcb$-independent to {an accuracy better than per-cent level}. {In Fig.~\ref{fig:3a} we show the ratios $R_1$ and $R_2$ defined in (\ref{R12}) as functions  of $\gamma$ for different values of $\beta$ within the SM.} {The coloured bands correspond to the scanning of $\vcb$ in the interval $[38,43]\times 10^{-3}$, for each fixed value of $\beta$. One can notice that the induced variations of the $R_1$ and $R_2$ ratios {with $\vcb$} are indeed of the per-mille level and one order of magnitude smaller than the {per-cent level} uncertainty related to parameters different from $\vcb$, $\gamma$ and $\beta$, which is represented by the gray band. We observe, furthermore, that $R_2$ does not depend on $\beta$, as expected from the fact that $\mathcal{B}(B_d\to\mu^+\mu^-)$ is a function of $V_{tb}$ and {$\vtd$} only and $\mathcal{B}(\kpn)$ is almost exactly $\beta$-independent. {Instead, the largest uncertainty for these two ratios is {associated with} the variation of the $\gamma$ angle.}}

{In Fig.~\ref{fig:3abis} we show the correlations between $\mathcal{B}(\kpn)$ and $\overline{\mathcal{B}}(B_s\to\mu^+\mu^-)^{1.4}$ and between $\mathcal{B}(\kpn)$ and ${\mathcal{B}}(B_d\to\mu^+\mu^-)^{1.4}$. One can notice that there is indeed a linear correlation between the two quantities, where different slopes of the lines correspond to different values of $\gamma$, with a $\beta$ dependence that is almost perfectly negligible. The shown ranges of values for the observables, namely the different points of the depicted segments, are given by the $\vcb$ variation in $38 < \vcb \times 10^3 < 43$. Other kinds of uncertainty are not shown here.} {In studying these plots and analogous plots below one should remember
  that one of the branching ratios is raised to an appropriate power that allows
  to remove the $\vcb$-dependence from the correlation between the two branching ratios in question. In the case at hand this power is $1.4$ and the gray
  area in the left plot in  Fig.~\ref{fig:3abis} corresponds
  to $1\sigma$ experimental range in
\begin{align}
  \label{eq:WAV-Bs}
  \oL{\cB}(B_s \to \mu^+\mu^-) & 
  = (2.85\; {}^{+0.34}_{-0.31}) \cdt{-9} \,,
\end{align}
obtained in  \cite{Hurth:2021nsi} on the basis  of LHCb, CMS and ATLAS data
\cite{LHCb:2021awg,CMS:2020rox,ATLAS:2020acx}. 
Similar averages have been provided in \cite{Geng:2021nhg} and \cite{Altmannshofer:2021qrr}.}

{The correlation in the left panel in Fig.~\ref{fig:3abis} is {one of the most interesting results of our paper} for coming years because the measurements of the branching ratio for $B_s\to\mu^+\mu^-$ should be improved at the LHC,
  $\gamma$ measured to high accuracy by LHCb and Belle II experiments
  and the branching ratio for $\kpn$ by NA62.
  While this correlation is independent of $\vcb$, an improved determination
  of $\vcb$ would allow for the absolute determination of both branching ratios in
  the SM resulting in a point on one of the lines chosen by the future improved
  measurement of $\gamma$.} {However, already now, as demonstrated in Section~\ref{sec:3}, we can find, imposing the agreement of the SM with the data on
  $\varepsilon_K$, $\Delta M_s$ and $S_{\psi K_S}$, the SM range represented
  by the rectangle in the left panel of Fig.~\ref{fig:3abis}. Its position is in fact independent of both $\vcb$ and $\gamma$ and exposes clearly an anomaly
  in $B_s\to\mu^+\mu^-$. Future improved measurements of both branching ratios
  by LHCb and NA62 will hopefully enhance this anomaly.}

\begin{figure}[t]
\centering%
\includegraphics[width=0.455\textwidth]{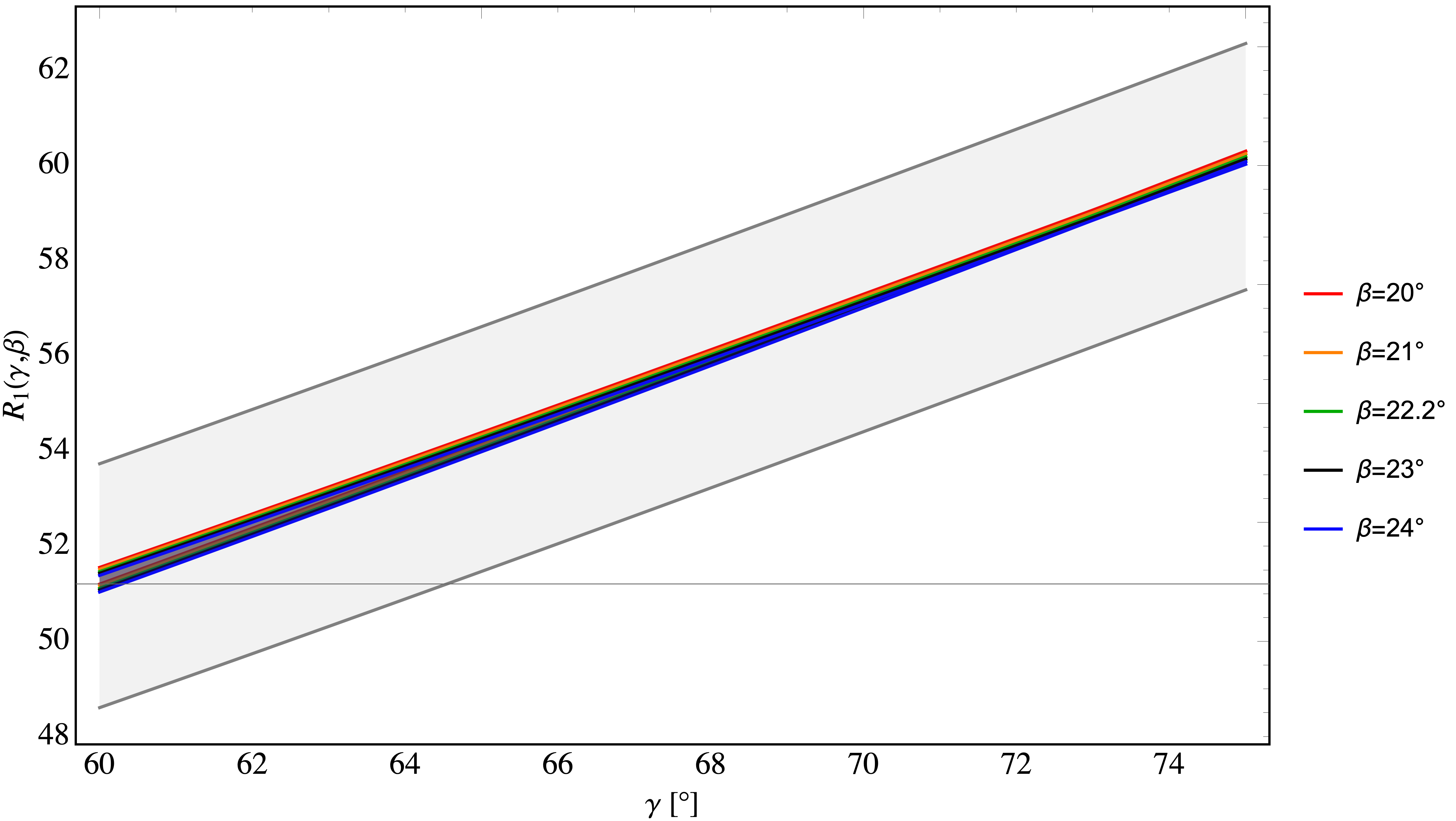}
\includegraphics[width=0.45\textwidth]{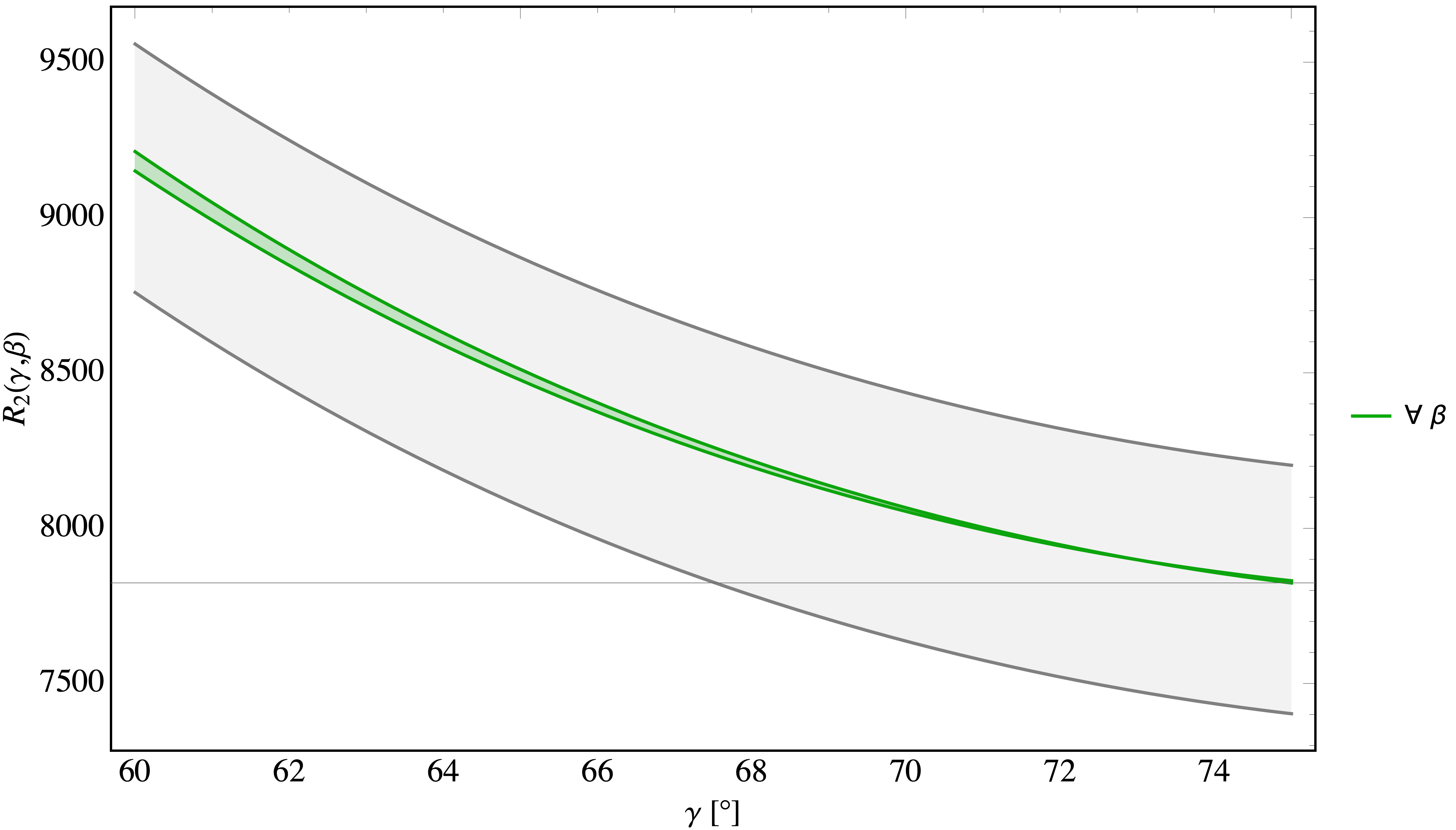}%
\caption{\it The ratios $R_1$ and $R_2$ defined in (\ref{R12}) as functions
  of $\gamma$ for different values of $\beta$ within the SM. {The coloured bands correspond to $38 < \vcb \times 10^{3}< 43$, while the gray ones give the non-parametric uncertainties.} \label{fig:3a}}
\end{figure}

\begin{figure}[t]
\centering%
\includegraphics[width=0.455\textwidth]{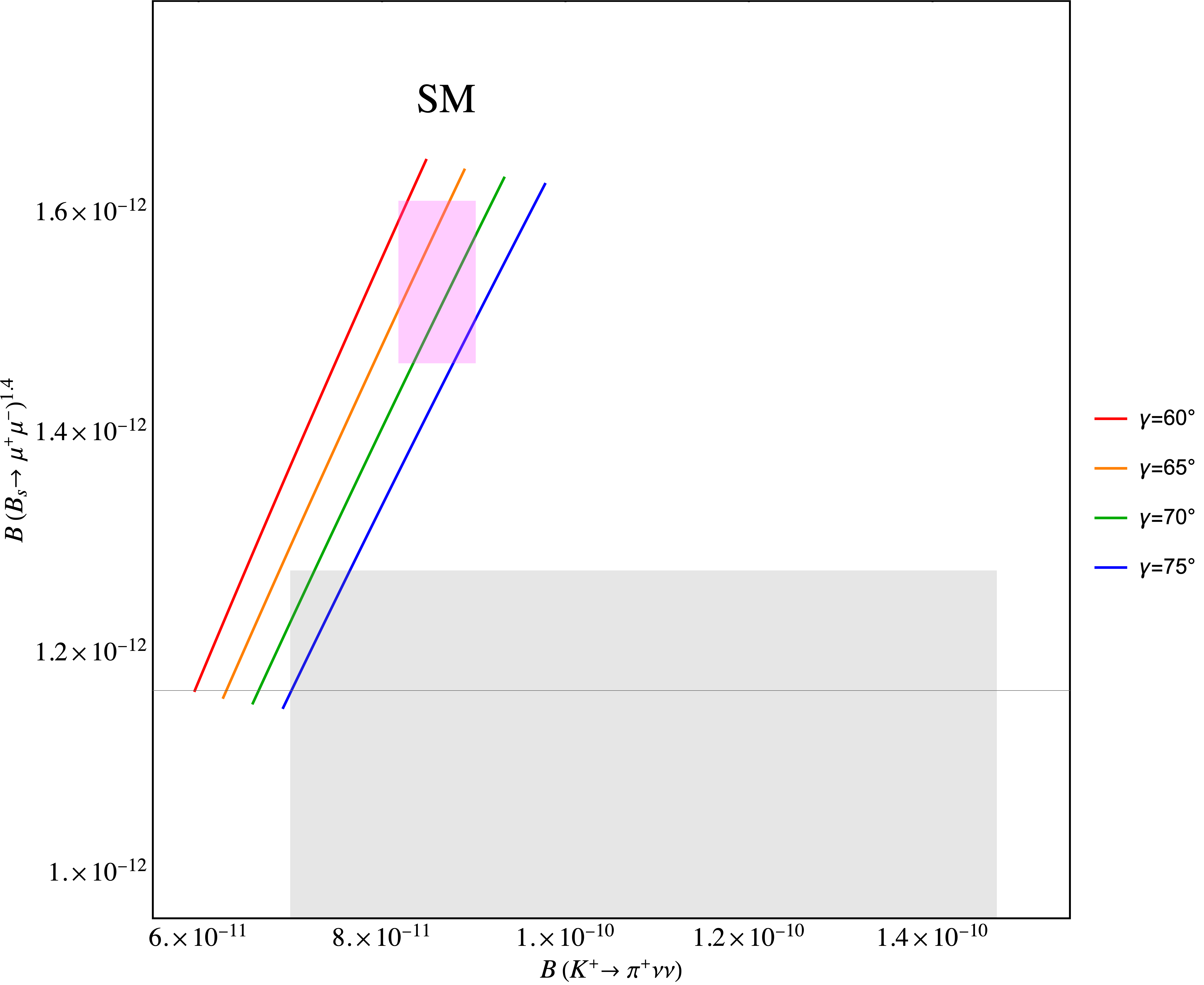}
\includegraphics[width=0.45\textwidth]{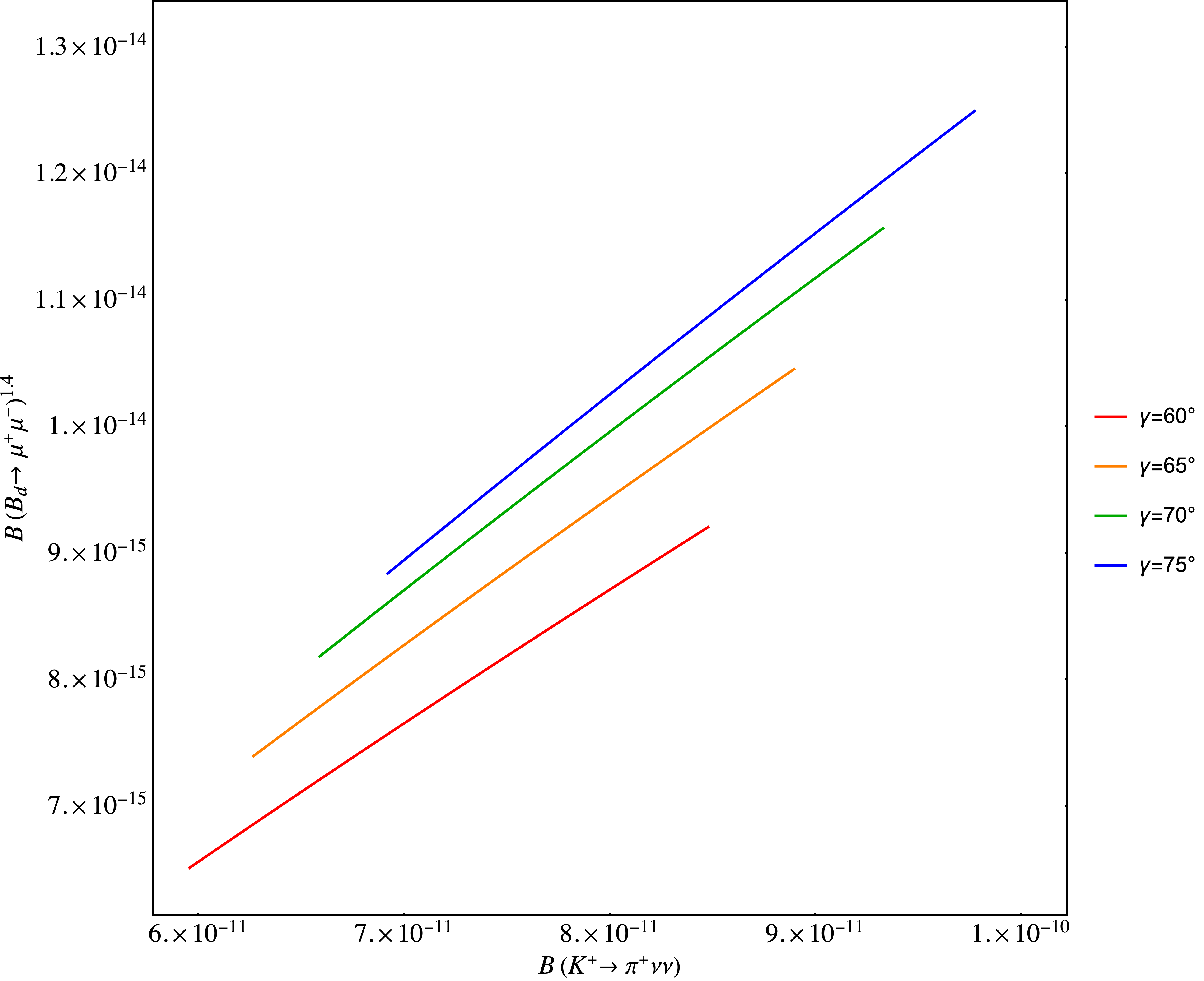}%
\caption{\it {The correlations of $\mathcal{B}(\kpn)$ with $\overline{\mathcal{B}}(B_s\to\mu^+\mu^-)^{1.4}$ (left panel) and with ${\mathcal{B}}(B_d\to\mu^+\mu^-)^{1.4}$ (right panel) as given in (\ref{eq:B1B3}) and (\ref{eq:B1B4}), for different  values of $\gamma$ within the SM.
    The ranges of branching ratios correspond to $38 \leq |V_{cb}|\times 10^{3} \leq 43$ {and $20^\circ\leq\beta \leq 24^\circ$.}} The gray area represents the
present experimental situation.} \label{fig:3abis}
\end{figure}

Now, it is likely that the branching ratios for $\kpn$ and $B_s\to\mu^+\mu^-$ will be measured accurately well ahead of the one for $B_d\to\mu^+\mu^-$
  and waiting for the latter measurement the triple correlation between the
  three branching ratios will be useful. {Improving on a similar 
    correlation in \cite{Buras:2015qea} we find}  
 { \begin{align}
\mathcal{B}(\kpn) &= {(7.92 \pm 0.25)}\times 10^{-11}  \left[\frac{\overline{\mathcal{B}}(B_s\to\mu^+\mu^-)}{{3.47}\times 10^{-9}\bar R_s}\right]^{0.74}\left[\frac{{230.3}\mev}{F_{B_s}}\right]^{1.4}\notag\\
&\qquad\quad\times\left[\frac{\mathcal{B}(B_d\to\mu^+\mu^-)}{{0.968}\times 10^{-10}\bar R_d}\right]^{0.74}\left[\frac{{190.0}\mev}{F_{B_d}} \right]^{1.4} \frac{H(\beta,\gamma)}{H(22.2^\circ,67^\circ)},\label{master3}
  \end{align}}
  where
  {\be
  \label{eq:H}
  H(\beta,\gamma)= \frac{1}{(G(\beta,\gamma))^{1.4}}  .
  \ee}
{In Fig.~\ref{fig:H} we show the $\gamma$ dependence of $H(\beta,\gamma)$
for $\beta=22.2^\circ$, which is {very} weak, implying a variation of less than {$1\%$} for $H(\beta,\gamma)$ when $\gamma$ takes values in $[60^\circ,75^\circ]$. The $\beta$ dependence is even weaker, less than $0.5\%$ for $\beta$ in $[20^\circ,24^\circ]$ and it is not shown.}

\begin{figure}[t]
\centering%
\includegraphics[width=0.45\textwidth]{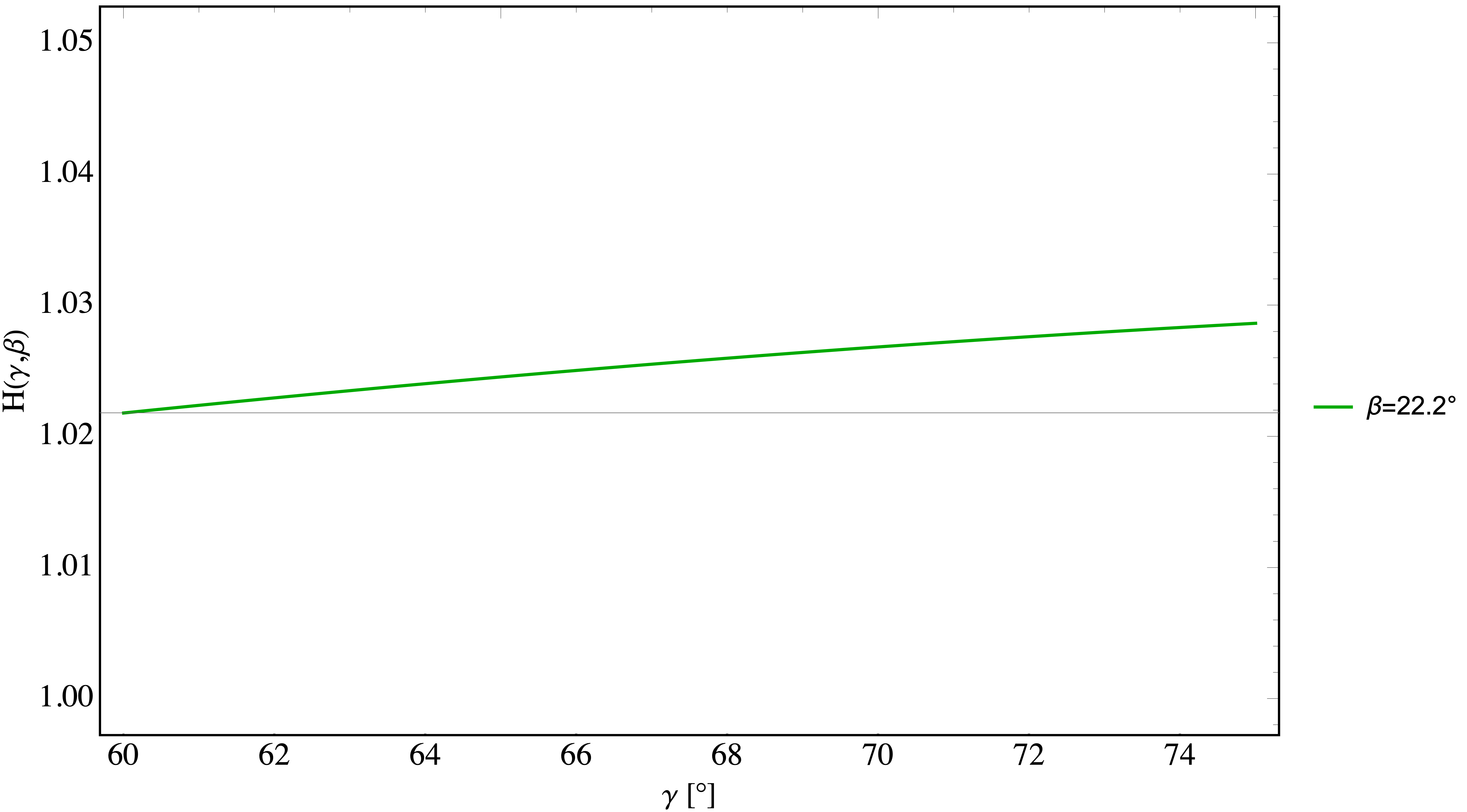}
%\includegraphics[width=0.45\textwidth]{PlotHb.png}%
%\caption{\it The function $H(\beta,\gamma)$ defined in (\ref{eq:H}) as functions
  %of $\gamma$ for $\beta=22.2^\circ$ (left) and as a function of $\beta$ with $\gamma=70^\circ$ (right).}
  \caption{\it The function $H(\beta,\gamma)$ defined in (\ref{eq:H}) as functions
  of $\gamma$ for $\beta=22.2^\circ$.}
  \label{fig:H}
\end{figure}
  
  Note that having in the future
  very accurate values for ${F_{B_s}}$, ${F_{B_d}}$ and
  the experimental branching ratios for $B_s\to\mu^+\mu^-$ and $\kpn$ will allow to predict the branching ratio for $B_d\to\mu^+\mu^-$ with high precision in the SM  {practically without any dependence on 
    CKM parameters due to the very weak dependence of $H(\beta,\gamma)$ on
    $\beta$ and $\gamma$.}

Similarly, one can define $\vcb$-independent ratios
 \be\label{R34}
  \boxed{R_3(\beta,\gamma)=\frac{\mathcal{B}(\klpn)}{\left[{\overline{\mathcal{B}}}(B_s\to\mu^+\mu^-)\right]^{2}},\qquad
   R_4(\beta,\gamma)=\frac{\mathcal{B}(\klpn)}{\left[{\mathcal{B}}(B_d\to\mu^+\mu^-)\right]^{2}}\,.}
   \ee
{\be\label{R3}
R_3(\beta,\gamma)=\frac{{({2.17}\pm 0.09)}\times 10^6}{\bar R_s^2}\left[{\frac{\sin\gamma\sin\beta}{\sin (67^\circ)\sin (22.2^\circ)}}\right]^2 \left[\frac{G(22.2^\circ,67^\circ)}{G(\beta,\gamma)}\right]^4\left(\frac{{230.3}\mev}{F_{B_s}}\right)^4\,,
\ee
\be\label{R4}
R_4(\beta,\gamma)=\frac{{({2.79\pm 0.13})}\times 10^{9}}{\bar R_d^2}
  \left[{\frac{\sin (67^\circ)}{\sin\gamma}\frac{\sin\beta}{\sin (22.2^\circ)}}\right]^2\left(\frac{{190.0}\mev}{F_{B_d}}\right)^4.
  \ee}

  \begin{figure}[t!]
\centering%
\includegraphics[width=0.48\textwidth]{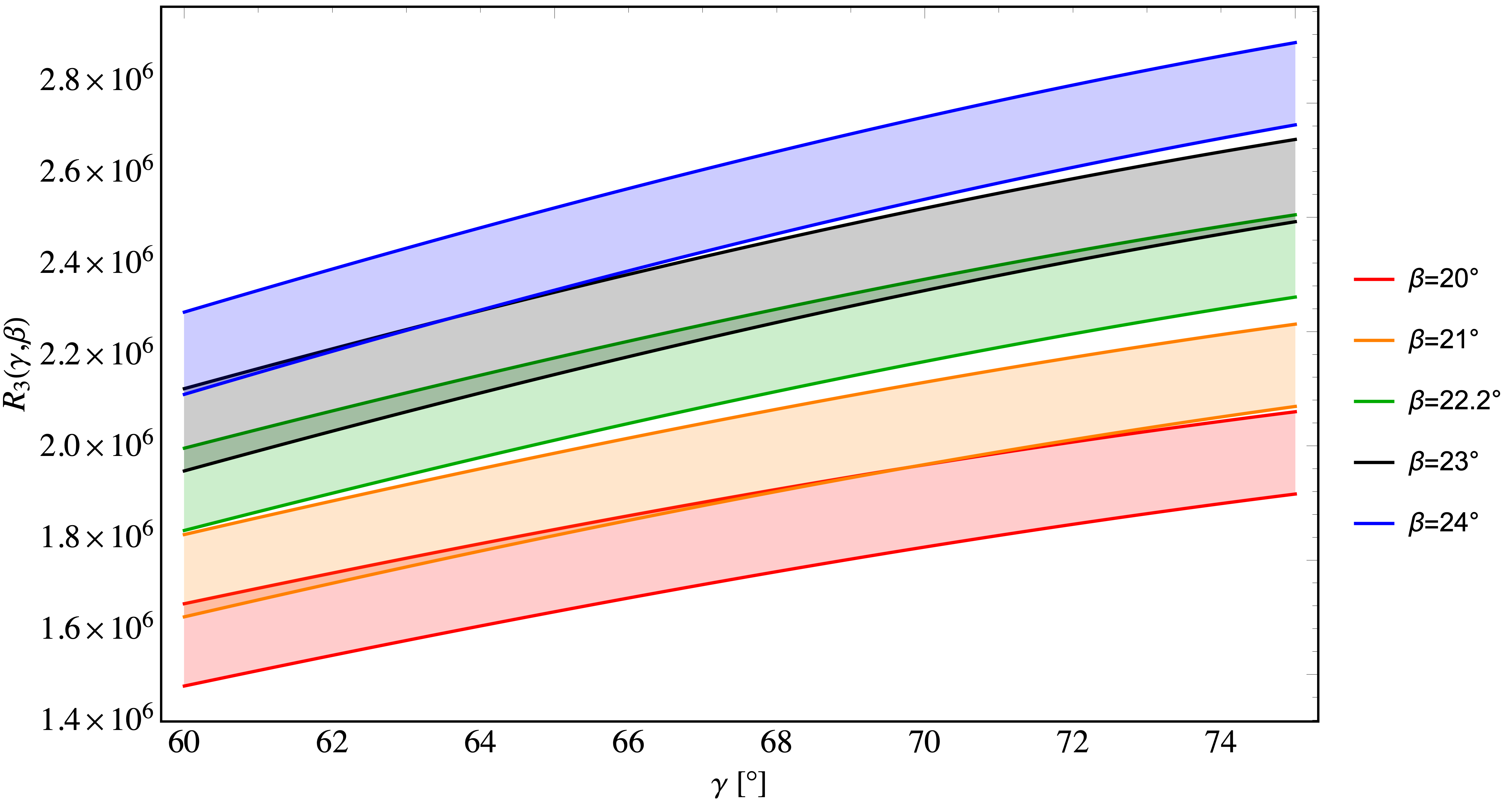}%
\hfill%
\includegraphics[width=0.48\textwidth]{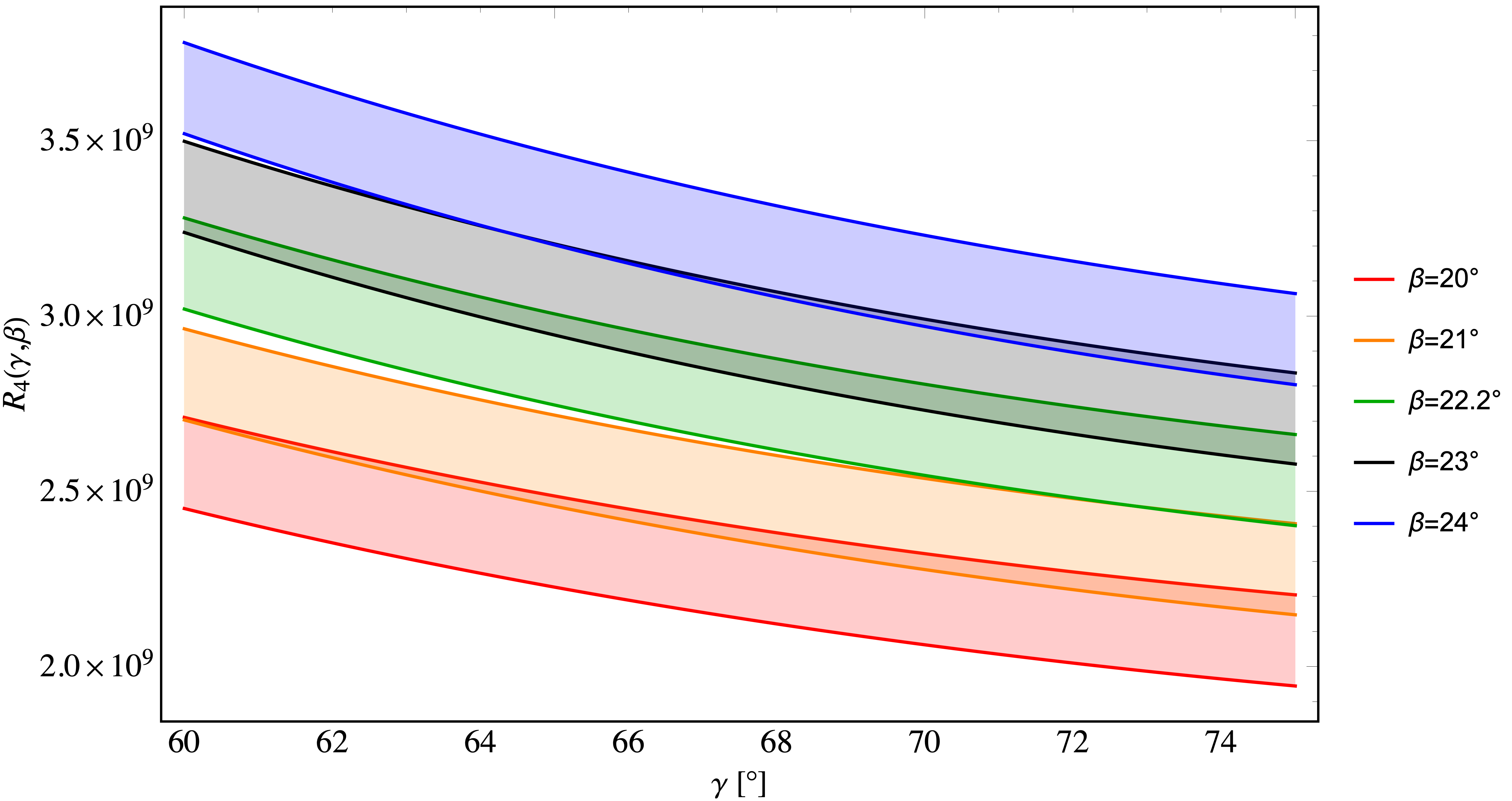}\\
%\includegraphics[width=0.48\textwidth]{Plot5c.png}%
%\hfill%
%\includegraphics[width=0.48\textwidth]{Plot5d.png}%
\caption{\it The ratios {$R_3$ and $R_4$} as a functions of $\gamma$ for different values of $\beta$. {The coloured bands represent the uncertainties associated to parameters different from $\gamma$ and $\beta$.}}
\label{fig:4}
\end{figure}

In Fig.~\ref{fig:4} we
show the ratios {$R_3$ and $R_4$} as  functions of $\gamma$ for different values of $\beta$. 
These plots  have been obtained using the exact formulae for all branching ratios involved but almost the same results would we obtained by using the approximate expressions given above. In particular, the independence of these ratios from $\vcb$ is exact. {One can notice that the uncertainty is dominated by the error on the $\gamma$ and $\beta$ angles, while the uncertainties associated with other parameters, depicted with coloured bands, are around one order of magnitude smaller.}

\boldmath
\subsection{Correlations with $B\to K^*\nu\bar\nu$ and $B\to K\nu\bar\nu$}\label{sec:2b}
\unboldmath
The most recent SM estimate of the branching ratios for these decays,
based on the formulae in \cite{Buras:2014fpa} and 
the form factors in the case of $B\to K^*\nu\bar\nu$ from \cite{Straub:2015ica} and {those for} $B\to K\nu\bar\nu$ from  \cite{Gubernari:2018wyi} read\footnote{Unpublished 2019 analysis of David Straub.} 
\begin{align}
{\mathcal{B}}(B^+\to K^+\nu\bar\nu)_{\rm SM} &=(4.18\pm 0.56)\times 10^{-6}\left|\frac{V_{ts}V_{tb}^*}{0.0402}\right|^2,\label{BKp}\\
{\mathcal{B}}(B^0\to K^{0*}\nu\bar\nu)_{\rm SM} &= (9.08\pm 0.85)\times 10^{-6}\left|\frac{V_{ts}V_{tb}^*}{0.0402}\right|^2,\label{BK0}
\end{align}
which update those in \cite{Buras:2014fpa}.

  Again the largest uncertainties in  these branching ratios originate  in the value of $\vcb$ 
  which cancels out in the ratios
\be\label{R56}
\boxed{R_5(\beta,\gamma)=\frac{\mathcal{B}(\kpn)}{\left[\mathcal{B}(B^+\to K^+\nu\bar\nu)\right]^{1.4}},\qquad
R_6(\beta,\gamma)=\frac{\mathcal{B}(\kpn)}{\left[\mathcal{B}(B^0\to K^{0*}\nu\bar\nu)\right]^{1.4}}.}
\ee
We find then
{\be\label{R5}
R_5(\beta,\gamma)={(2.69\pm0.51)}\times 10^{-3}
\left[\frac{\sin\gamma}{\sin 67^\circ}\right]^{1.39}\left[\frac{G(22.2^\circ,67^\circ)}{G(\beta,\gamma)}\right]^{2.8},
\ee
\be\label{R6}
R_6(\beta,\gamma)={(9.07\pm1.23)}\times 10^{-4}
\left[\frac{\sin\gamma}{\sin 67^\circ}\right]^{1.39}\left[\frac{G(22.2^\circ,67^\circ)}{G(\beta,\gamma)}\right]^{2.8}\,,
\ee}
with $G(\beta,\gamma)$ defined in (\ref{vts}).

The ratios $R_5$ and $R_6$ are shown in Fig.~\ref{fig:4B} as  functions of $\gamma$ for different values of $\beta$. We observe that their 
dependence on $\gamma$ and $\beta$ is the same as for $R_1$. {In particular they are both nearly independent of $\beta$.} {The non-parametric uncertainties are fully dominated by formfactor uncertainties that
  should be significantly reduced in the coming years.}
  
{In Fig.~\ref{fig:4Bbis} we show the correlations between $\mathcal{B}(\kpn)$ and $\mathcal{B}(B^+\to K^+\nu\bar\nu)^{1.4}$ and between $\mathcal{B}(\kpn)$ and $\mathcal{B}(B^0\to K^{0*}\nu\bar\nu)^{1.4}$. In both cases, the depicted linear relations are analogous to the one in the left panel of Fig.~\ref{fig:3abis}.}
{Also these correlations are of interest for Belle II, NA62 and
  LHCb.}

  \begin{figure}[t!]
\centering%
%\includegraphics[width=0.48\textwidth]{Plot5a.png}%
%\hfill%
%\includegraphics[width=0.48\textwidth]{Plot5b.png}\\
\includegraphics[width=0.48\textwidth]{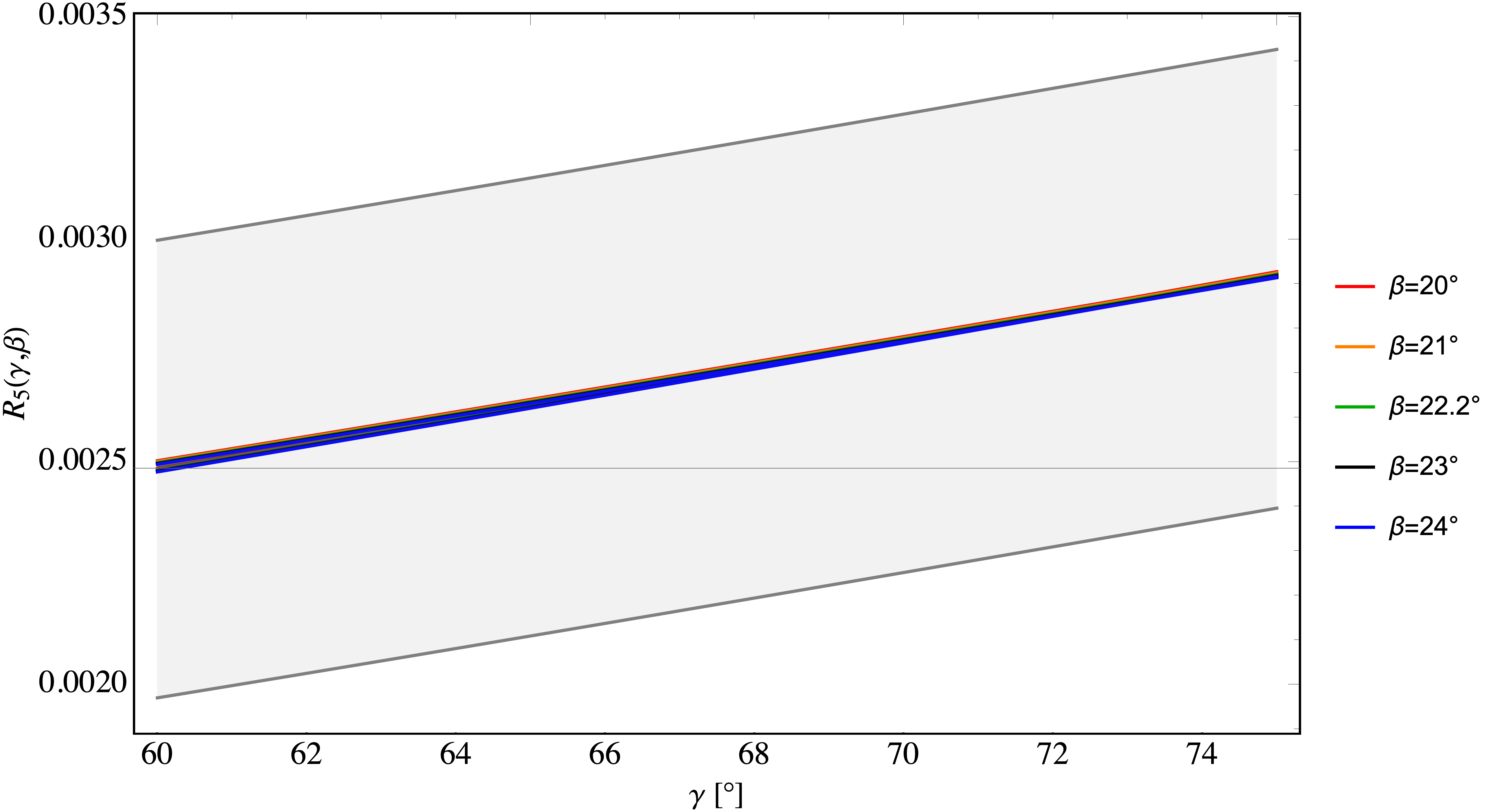}%
\hfill%
\includegraphics[width=0.48\textwidth]{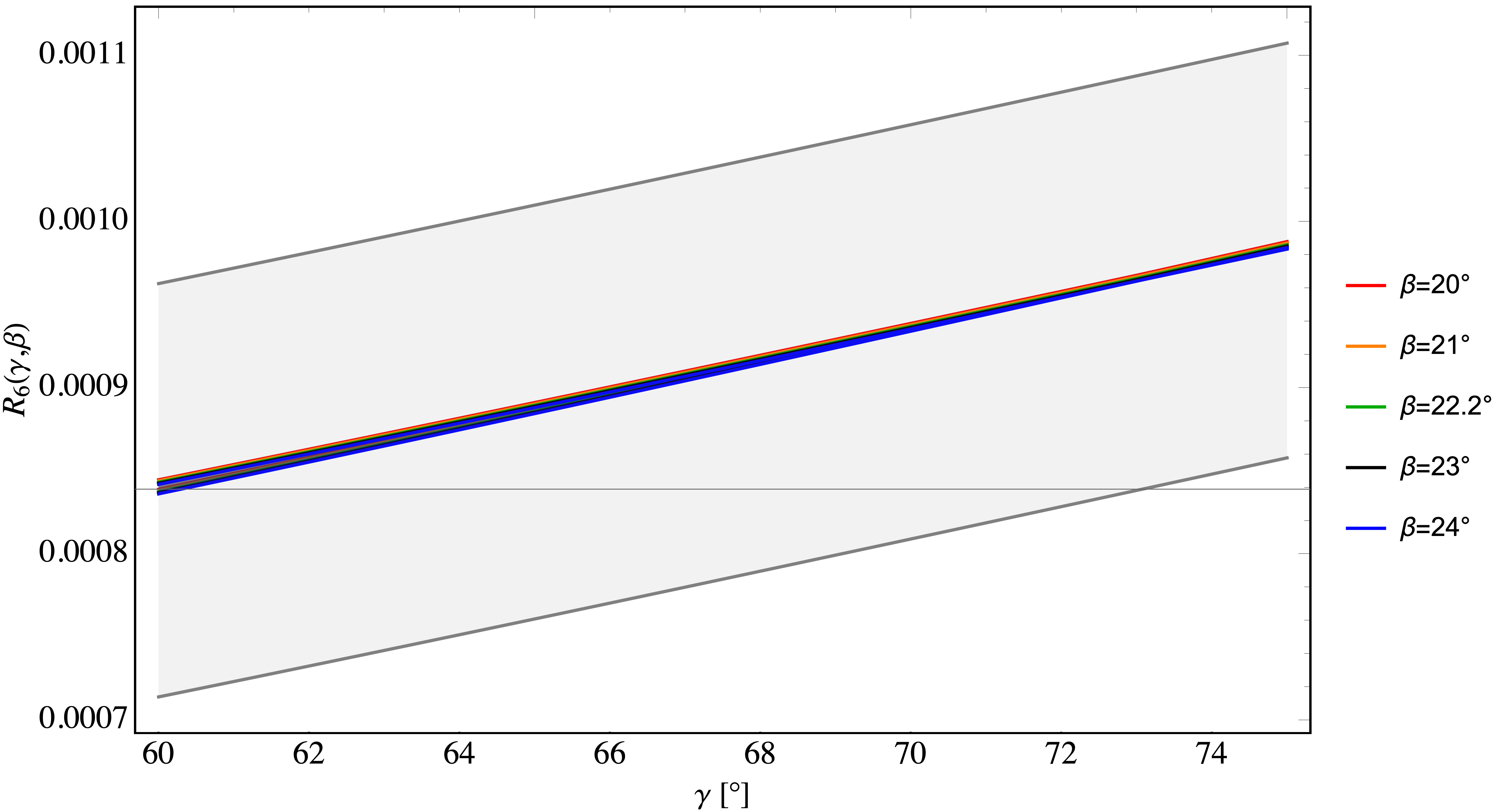}%
\caption{\it {The ratios $R_5$ and $R_6$ as  functions of $\gamma$ for different values of $\beta$. {The coloured bands correspond to $38 < \vcb  \times 10^{3}< 43 $, while the gray ones give the non-parametric uncertainties.}} 
\label{fig:4B}}
\end{figure}

\begin{figure}[t]
\centering%
\includegraphics[width=0.455\textwidth]{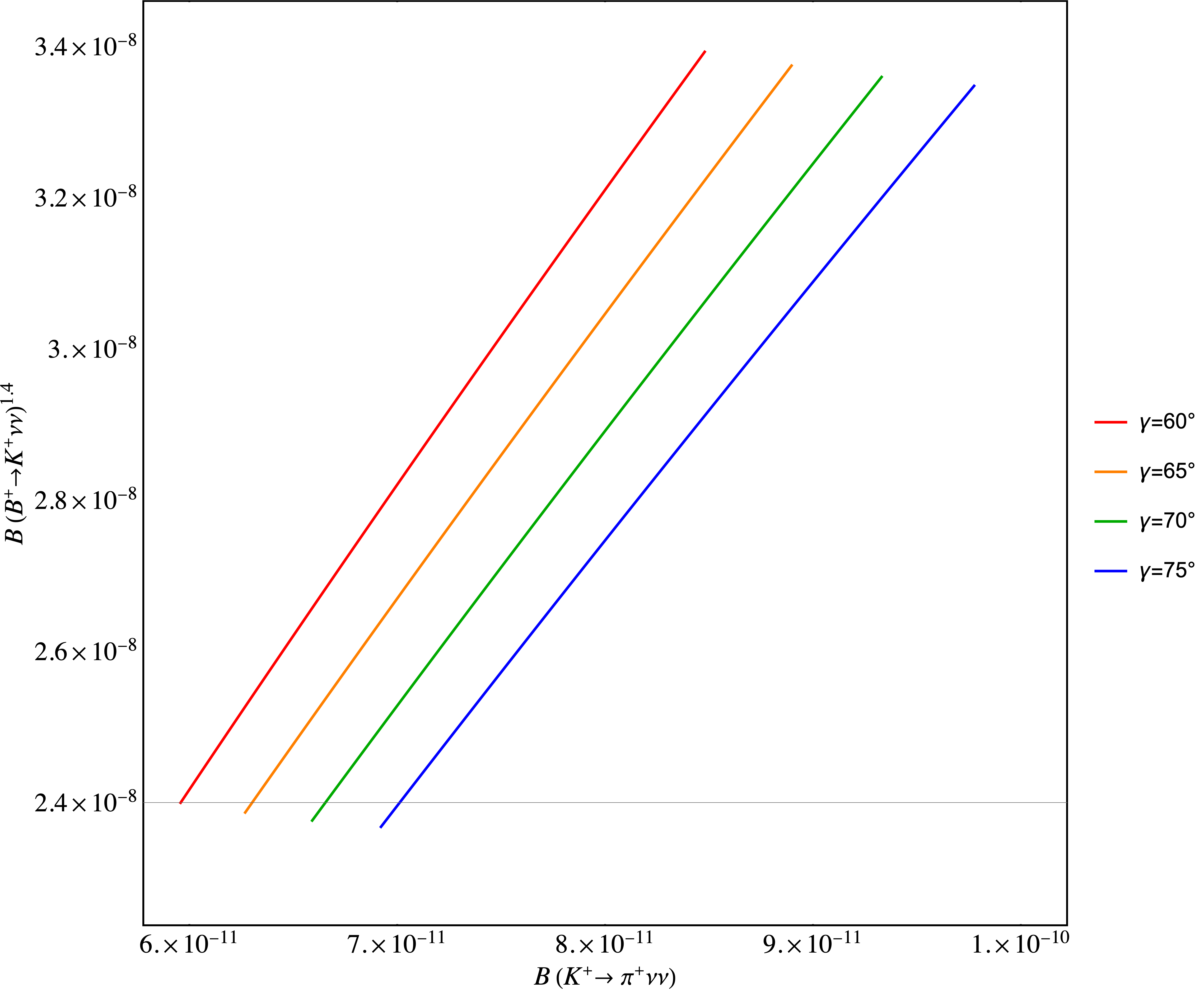}
\includegraphics[width=0.45\textwidth]{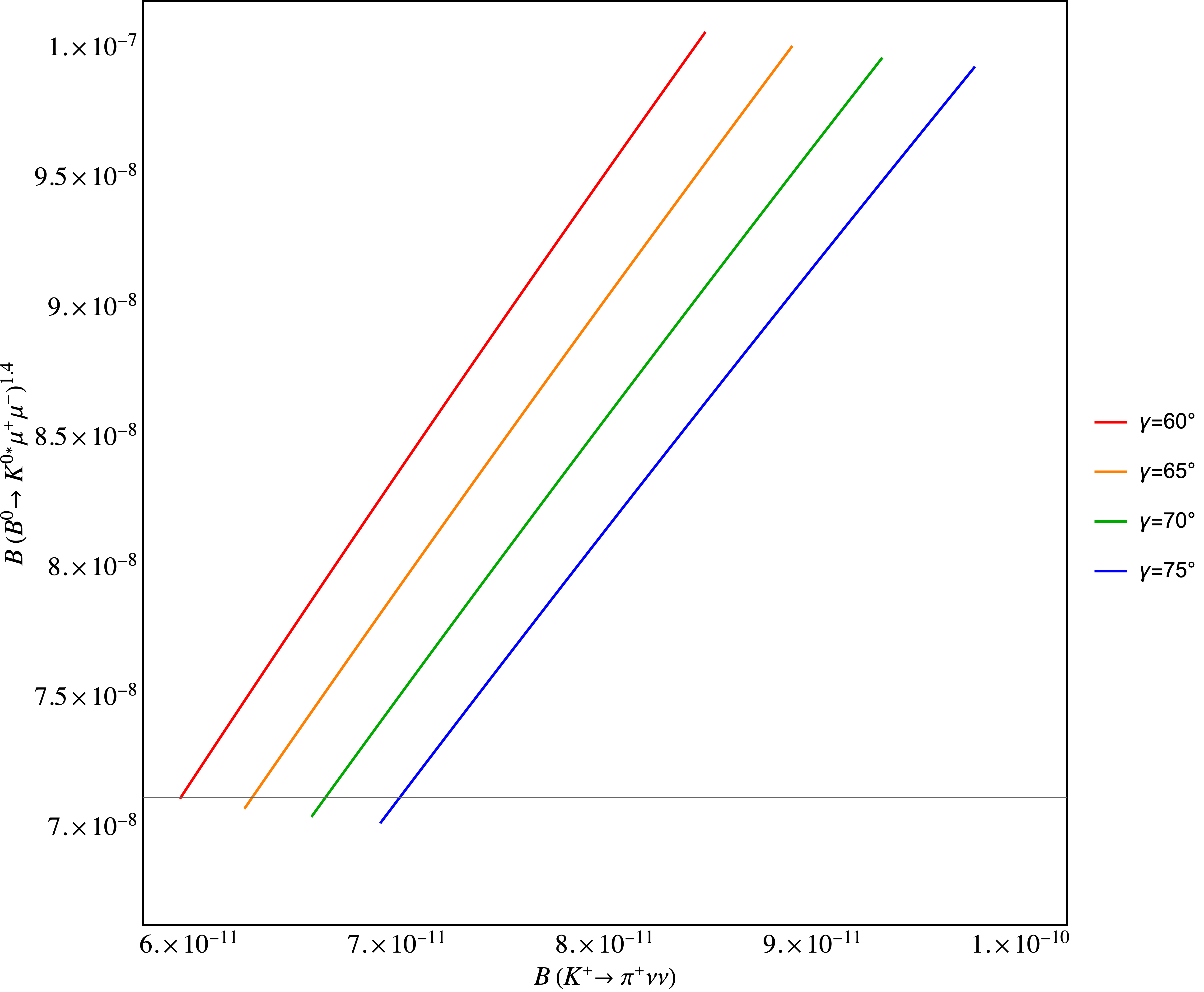}%
\caption{\it {The correlations of $\mathcal{B}(\kpn)$ with $\mathcal{B}(B^+\to K^+\nu\bar\nu)^{1.4}$ (left panel) and with $\mathcal{B}(B^0\to K^{0*}\nu\bar\nu)^{1.4}$ (right panel) as given in (\ref{R5}) and (\ref{R6}), for different  values of $\gamma$ within the SM.
  The ranges of branching ratios correspond to $38 \leq |V_{cb}|\times 10^{3} \leq 43$ {and $20^\circ\leq\beta \leq 24^\circ$.}}} \label{fig:4Bbis}
\end{figure}

Moreover, one has CKM-independent ratios
\be\label{R78}
\boxed{R_7=\frac{\mathcal{B}(B^+\to K^+\nu\bar\nu)}{{\overline{\mathcal{B}}}(B_s\to\mu^+\mu^-)},\qquad
R_8=\frac{\mathcal{B}(B^0\to K^{*0}\nu\bar\nu)}{{\overline{\mathcal{B}}}(B_s\to\mu^+\mu^-)}.}
    \ee
    Using (\ref{BRtheoRpar}) we find 
    \be\label{R7SM}
    (R_7)_\text{SM}={({1.20}\pm 0.17)}\times 10^{3}\frac{1}{\bar R_s}\left(\frac{{230.3}\mev}{F_{B_s}}\right)^2\,,
    \ee
     \be\label{R8SM}
    (R_8)_\text{SM}={({2.62}\pm 0.25)}\times 10^{3}\frac{1}{\bar R_s}\left(\frac{{230.3}\mev}{F_{B_s}}\right)^2\,.
     \ee

    Now the present world average experimental value from LHCb
    \cite{LHCb:2021awg,LHCb:2021vsc}, CMS \cite{CMS:2020rox,Sirunyan:2019xdu} and ATLAS
    \cite{ATLAS:2020acx,Aaboud:2018mst} as given in 
  \cite{Hurth:2021nsi} and the preliminary result from Belle II  \cite{Browder:2021hbl} read respectively
\be
  \label{eq:WAV-Bsa}
    \overline{\mathcal{B}}(B_s\to\mu^+\mu^-)
      = (2.85\; {}^{+0.34}_{-0.31}) \times 10^{-9} \,,\qquad
      \mathcal{B}(B^+\to K^+\nu\bar\nu)= (11\pm4)\times 10^{-6}\,
      \ee
      which implies
             \be
    (R_7)_{\rm EXP}=(3.86\pm 1.48)\times 10^{3}.
       \ee
       The central value is by a factor of 3.2 larger than the SM prediction (\ref{R7SM}) but due to large error in the experimental $B^+\to K^+\nu\bar\nu$ branching ratio the tension  is only at $1.8\sigma$.

       The origin of this significant discrepancy is the fact that whereas the
       central experimental value for $\mathcal{B}(B^+\to K^+\nu\bar\nu)$
       is by a factor of $2.5$ larger than the SM prediction, in the case of
       $\overline{\mathcal{B}}(B_s\to\mu^+\mu^-)$, the data are by a factor
       of 1.3 below its SM value. While these factors correspond to 
       $\vcb=42.0\times 10^{-3}$ and would both change with $\vcb$, the ratio
       $R_7$ is $\vcb$-independent.

              The possible tension of Belle II data with the SM in the case of $\mathcal{B}(B^+\to K^+\nu\bar\nu)$ has been pointed out first in \cite{Browder:2021hbl}
       who found the data by a factor of $2.4\pm0.9$ larger than its 
       SM value estimate from \cite{Buras:2014fpa}  {obtained
         with somewhat different formfactors than used by us here.} However, this
       result corresponds to  $\vcb=42.0\times 10^{-3}$  and the tension would increase for lower exclusive values of $\vcb$. In our {approach} the value of $\vcb$
       does not matter.

\boldmath
\section{$\vcb(\beta,\gamma)$ from $\varepsilon_K$, $\Delta M_s$ and $\Delta M_d$}\label{sec:3}
\unboldmath
\subsection{Preliminaries}
While the  global analyses of the UT \cite{Bona:2007vi,Charles:2004jd} demonstrate good
  consistency of the SM with the data, in this section we want to have first a closer look at  $\varepsilon_K$, $\Delta M_s$ and $\Delta M_d$  with the goal to
  check
  whether within the SM the same value of $\vcb$ allows to obtain for them simultaneously
  good agreement with the data {and for which values of $\beta$ and $\gamma$ this turns out to be possible.}  This is analogous to the expressions
    in (\ref{bklpn1})     and    (\ref{eq:vcb-B3}) but this time $\vcb$
    is expressed in terms of precisely measured $\varepsilon_K$, $\Delta M_s$ and $\Delta M_d$  as opposed to rare decay branching ratios. 

    The motivation for this analysis comes from the fact that in the standard
    analysis of the UT the value of $\vcb$ remains hidden and only
    the angles $\beta$ and $\gamma$ are exposed. In this manner only two
    parameters among the four in (\ref{4CKM}) are visible. While $\vus$
    plays totally subleading role in rare decays, the role of $\vcb$ is even more important than of $\beta$ and $\gamma$.

    Therefore we want to propose here a  test of the {SM} that
    is complementary to the usual UT-analyses. {Namely, we propose} to extract 
    from a given observable the value of $\vcb$ as a function
    of $\beta$ and $\gamma$ for which the SM agrees with the experimental
    data. In what follows we will present this idea using
    $\varepsilon_K$, $\Delta M_s$ and $\Delta M_d$ for which both theory and
    experiment reached good precision but in the future other processes, in
    particular the theoretically clean  rare decays considered by us,
    could also be used for this purpose when the experimental data improve\footnote{An illustration how such an analysis would look like has been recently
      presented in \cite{Buras:2022nfn}.}

    We will find that, while the dependence of $\vcb$ on $\beta$ and $\gamma$
    extracted from  $\varepsilon_K$ is rather rich, the one extracted from
    $\Delta M_d$ involves only  $\gamma$. Finally, $\vcb$ extracted from
    $\Delta M_s$, is practically independent of both   $\beta$ and $\gamma$.
    Our presentation begins therefore with  $\varepsilon_K$ followed by the
    one on $\Delta M_d$ and $\Delta M_s$.

\boldmath
\subsection{$\vcb(\beta,\gamma)$ from $\varepsilon_K$}\label{epsilonK}
\unboldmath
In \cite{Brod:2019rzc} a more accurate formula for $\varepsilon_K$ has been presented. It uses
the unitarity relation $\lambda_c=-\lambda_u-\lambda_t$ instead of
$\lambda_u=-\lambda_c-\lambda_t$ as done in the previous literature. This allows
{to remove significant theoretical uncertainties from charm contribution to $\varepsilon_K$. The new SM expression for  $\varepsilon_K$ reads} \cite{Brod:2019rzc}
\begin{equation}\label{BGS}
    |\epsilon_K|
=  \kappa_\epsilon C_\varepsilon \hat{B}_K
|V_{cb}|^2 \lambda^2 \bar \eta 
 \times \Big[|V_{cb}|^2(1-\bar\rho)
\eta_{tt} S_{tt}(x_t) - \eta_{ut} S_{ut}(x_c, x_t) \Big]\,.
\end{equation}
It replaces the usual phenomenological expression given in \cite{Buras:2008nn}. Here
\be
S_{tt}(x_t)=S_0(x_t)+S_0(x_c)-2 S_0(x_c,x_t),\qquad  S_{ut}(x_c, x_t)=S_0(x_c)-S_0(x_c,x_t),
\ee where
$S_0(x_i)$ and $S_0(x_i,x_j)$ are the standard Inami-Lim functions \cite{Inami:1980fz,Buras:1983ap} with explicit expressions given in Appendix~\ref{App}.

The QCD factors in (\ref{BGS}) at the NLO and NNLO level, respectively, read \cite{Brod:2019rzc}
\be
\eta_{tt}=0.55(2),\qquad \eta_{ut}=0.402(5).
\ee

Next, the kaon bag parameter comprising the hadronic matrix element of
the local $\Delta S = 2$ operators is given by $\hat{B}_K =0.7625(97)$~\cite{Aoki:2019cca}\footnote{{As expected on the basis of Dual QCD approach \cite{Bardeen:1987vg,Gerard:2010jt,Buras:2014maa} $\hat B_K$ will eventually be below $0.75$   which would slightly increase the values of $\vcb$ from $\varepsilon_K$ presented by us.}}. The phenomenological parameter
$\kappa_\epsilon = 0.94(2)$~\cite{Buras:2010pza} comprises
long-distance contributions beyond the lowest order in the operator-product
expansion, which are not included in $\hat B_K$\footnote{As pointed out in
  \cite{Gerard:2010jt} these long-distance contributions could be avoided by considering ${\RE\,\varepsilon_K}$ instead of $|\varepsilon_K|$. But the data
  on ${\RE\varepsilon_K}$ that is extracted from a semi-leptonic asymmetry
  is unfortunately much less accurate than it is  on $|\varepsilon_K|$. Progress
  on the reduction of the error on $\kappa_\epsilon$ is also expected from
  LQCD \cite{Laiho:2009eu}.}.
  
Finally, 
\be
C_\varepsilon =
\frac{G_{F}^2 F_K^2 m_{K^0} M_W^2}{6\sqrt{2}\pi^2(\Delta M_K)_{\rm exp}}={3.635\times 10^4.}
\ee

The new expression in (\ref{BGS}) is significantly more accurate than the old one as far as theoretical uncertainties are concerned but  it is still subject
to large uncertainty due to $\vcb$.

Defining then
\be\label{r1r2}
r_1=\frac{1}{(1-\bar\rho)}\frac{\eta_{ut} S_{ut}(x_c, x_t)}{\eta_{tt} S_{tt}(x_t)},
\quad r_2=\kappa_\epsilon C_\varepsilon \hat{B}_K\lambda^2 \bar \eta(1-\bar\rho)\eta_{tt} S_{tt}(x_t),
\ee
we find
\be\label{vcb2}
\boxed{\vcb^2=\frac{1}{2}\left[r_1+\sqrt{r^2_1+4\frac{ |\epsilon_K|}{r_2}}\right]}
\ee
with $\bar\rho$ and $\bar\eta$ being through (\ref{S4}) functions of $\beta$ and $\gamma$. In this manner  one of the basic parameters in (\ref{4CKM}), {namely $\vcb$, has been traded
  for $ |\epsilon_K|$ that is very precisely measured.}
{Surprisingly, to our knowledge this expression for $\vcb$ in terms of
  $\beta$, $\gamma$ and $|\epsilon_K|$ has never been presented in the literature.

  {Choosing the central experimental value for  $|\epsilon_K|$  in Table~\ref{tab:input} and  the reference values in (\ref{REFCKM})
    ($\lambda=0.225$, $\gamma=67^\circ$ and $\beta=22.2^\circ$) we find
    $\vcb=42.2\times 10^{-3}$ that favours the inclusive determinations of this parameter. The full $\beta$ and $\gamma$ dependence of $\vcb$ following from $(\ref{vcb2})$ is shown in  Fig.~\ref{fig:5}.
    {The green band corresponds to the $1\sigma$ range for $\beta$ determined from $S_{\psi K_S}= 0.699(17)$:
\be\label{betarange}
21.5^\circ\le \beta\le 22.9^\circ\,.
\ee
We will see soon that this is an important constraint.}
    Before describing this result in detail let us extract
 first $\vcb$ from $\Delta M_d$ and $\Delta M_s$.}

 \begin{figure}[t!]
\centering%
\includegraphics[width=0.70\textwidth]{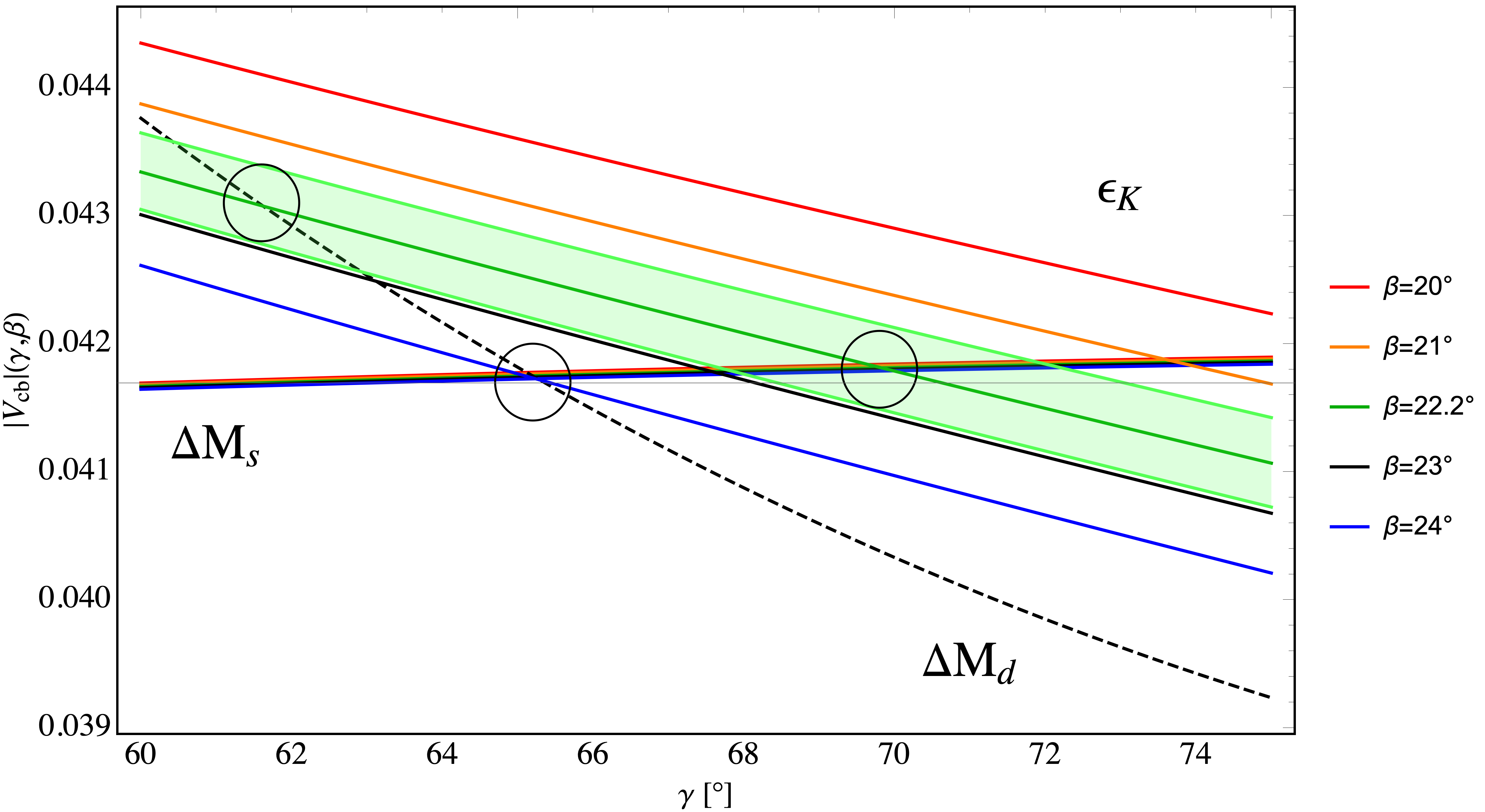}%
%\hfill%
%\includegraphics[width=0.48\textwidth]{Plot6b.png}\\
%\includegraphics[width=0.48\textwidth]{Plot6c.png}%
%\hfill%%
\caption{\it {The values of $\vcb$ extracted from $\varepsilon_K$, $\Delta M_d$ and  $\Delta M_s$ as functions of $\gamma$ for different values of $\beta$. {$\vcb$ extracted from $\Delta M_d$ is independent of  $\beta$.}} 
\label{fig:5}}
\end{figure}

 \boldmath
\subsection{$\vcb(\beta,\gamma)$ from $\Delta M_d$ and $\Delta M_s$}
\unboldmath

{Updating the two very accurate formulae from \cite{Blanke:2016bhf} we have}:
\bea
\label{DMD}
\Delta M_d&=&
0.5065/{\rm ps}\times\left[ 
\frac{\sqrt{\hat B_{B_d}}F_{B_d}}{214.0\mev}\right]^2
\left[\frac{S_0(x_t)}{2.307}\right]
\left[\frac{\vtd}{8.53\times10^{-3}} \right]^2 
\left[\frac{\eta_B}{0.5521}\right]\,,\\ 
\label{DMS}
\Delta M_{s}&=&
%{17.232}/{\rm ps}\cdot\left[ %for 40.3
17.749/{\rm ps}\times\left[
\frac{\sqrt{\hat B_{B_s}}F_{B_s}}{261.7\mev}\right]^2
\left[\frac{S_0(x_t)}{2.307}\right]
\left[\frac{\vts}{41.0\times 10^{-3}} \right]^2
\left[\frac{\eta_B}{0.5521}\right] \,.
\eea

The value {$2.307$} in the normalization of $S_0(x_t)$ is its SM value for 
{$m_t(m_t)=162.83\GeV$.} {The central values of $\vtd$ and $\vts$ exposed here are chosen to make the overall factors in these formulae to be equal
  to the experimental values of the two observables. One can check that, for $\gamma=67^\circ$ and $\beta=22.2^\circ$, these values of $\vtd$ and $\vts$  correspond to
  {$\vcb=41.2\times 10^{-3}$ and  $\vcb=41.8\times 10^{-3}$}, respectively.}
 See Table~\ref{tab:input} for other parameters.

The measurement of  $\Delta M_d$
together with $\Delta M_s$  allows to determine $R_t$ or equivalently $\gamma$ without any dependence  on $m_t$ and $|V_{cb}|$. To an excelent approximation
one finds \cite{Blanke:2016bhf} 
\be\label{Rt}
\frac{\vtd}{\vts}=\xi\sqrt{\frac{m_{B_s}}{m_{B_d}}}\sqrt{\frac{\Delta M_d}{\Delta M_s}}\quad \Longrightarrow \quad  \sin\gamma={0.750}\, \xi, \qquad
\xi=\frac{\sqrt{\hat B_{B_s}}F_{B_s}}{\sqrt{\hat B_{B_d}}F_{B_d}}
\ee
where $\beta=22.2^\circ$, corresponding to $S_{\psi K_S}= 0.699$ has been used.
But the dependence on $\beta$ is very weak as one can check using the expressions in (\ref{vtdvub}) and (\ref{vts}) {so that  this relation
  is an excellent approximation for the full range of $\beta$ used by us.}

This determination of $\gamma$ can be confronted with the tree-level determination of
$\gamma$ with the help of non-leptonic two-body decays as mentioned before.
As the mass differences $\Delta M_{s,d}$ are very precisely measured, the $\gamma$ following from
their ratio depends as seen in  (\ref{Rt}) to an excellent approximation solely on $\xi$. This dependence is shown in Fig.~3 of \cite{Blanke:2016bhf}.

 {Now}, various LQCD collaborations contributed to the evaluation of $\xi$. In particular Fermilab-MILC collaboration \cite{Bazavov:2016nty} with
$\xi=1.206\pm0.019$ and   RBC-UKQCD collaboration with  $\xi = 1.1853\pm 0.0054^{+0.0116}_{-0.0156}$ \cite{Boyle:2018knm}. Similar value has been obtained from
HQET sum rules:  $\xi = 1.2014^{+0.0065}_{-0.0072}$ \cite{King:2019lal}. With
these values $\gamma$ is found in the range $60^\circ\le\gamma \le 65^\circ$.
In particular the present value of $\gamma$ corresponding to the FLAG average for $\xi$ reads \cite{Aoki:2019cca} 
{
\be\label{xiLQCD}
\xi=1.206(17),\qquad  \gamma=64.7(16)^\circ.
\ee
}
Until recently the central values for $\gamma$ from the LHCb collaboration
  were in the ballpark of $74^\circ$ implying in particular some tension
  between the FLAG value and the one from non-leptonic decays pointed
  out   in \cite{Blanke:2016bhf}. 
  However the most recent LHCb value in (\ref{betagamma}) is 
in a good agreement with (\ref{xiLQCD}).

{But as the error on $\gamma$ from  LHCb is still large let us have a closer look and determine  within the SM  $\vcb$  independently from $\Delta M_d$ and $\Delta M_s$. Using (\ref{DMD}) and  (\ref{DMS}) together with
  (\ref{vtdvub}) and (\ref{vts}) we find}
\be\label{VCBd}
  \vcb=\frac{{8.53}\times 10^{-3}}{\lambda \sin\gamma}
  \left[\frac{214.0\mev}{\sqrt{\hat B_{B_d}}F_{B_d}}\right], \qquad (\Delta M_d),
  \ee
  \be
  \vcb=\frac{{41.0}\times 10^{-3}}{G(\beta,\gamma)}
  \left[\frac{261.7\mev}{\sqrt{\hat B_{B_s}}F_{B_s}}\right], \qquad (\Delta M_s).
  \ee
    with $G(\beta,\gamma)$ defined in (\ref{vts}). {For the reference values in (\ref{REFCKM}) ($\lambda=0.225$, $\gamma=67^\circ$ and $\beta=22.2^\circ$) we find $\vcb=41.2\times 10^{-3}$ and 
    $\vcb=41.8\times 10^{-3}$, respectively.}
{While the experimental errors coming from the $\Delta M_d$ and $\Delta M_s$ measurements can be safely neglected, the theoretical errors associated to the uncertainties on the $\sqrt{\hat B_{B_{d,s}}}F_{B_{d,s}}$ factors induce errors respectively of $1.8\%$ and of $1.5\%$ on the corresponding $\vcb$ determinations. These uncertainties are not shown in the figures, for the sake of readability.}
  
  In Fig.~\ref{fig:5} we show $\vcb$ as given above as function of $\gamma$ for different values of $\beta$ and compare it with the one obtained from
  $\varepsilon_K$.

  We observe that
  \begin{itemize}
  \item
    $\vcb$ extracted from $\varepsilon_K$ shows significant dependence on both
    $\beta$ and $\gamma$.
    \item
        $\vcb$ extracted from $\Delta M_d$ is independent of $\beta$ but shows
        a significant dependence on $\gamma$ which is evident from the
        expression in (\ref{VCBd}).
      \item
        $\vcb$ extracted from $\Delta M_s$ is practically independent of both $\beta$
        and $\gamma$ because the element $\vts$ governing $\Delta M_s$
        differs from $\vtd$ only by the function $G(\beta,\gamma)$ which
        is weakly dependent on both angles of the UT.
\end{itemize}

        {Therefore the last finding implies that just
          on the basis of $\Delta M_s$ alone a rather precise value
          of $\vcb$ can be obtained. Considering the full range of $\gamma$
          and $\beta$ and the error on $\sqrt{\hat B_{B_s}}F_{B_s}$ in
            Table~\ref{tab:input} we find
            \be\label{vcbbest}
         \boxed{\vcb={41.78(62)}\times 10^{-3}, \qquad (\Delta M_s)\, } 
         \ee
            {where the largest contribution to the error, ${0.61} \times 10^{-3}$, is due to the uncertainty on $\sqrt{\hat B_{B_s}}F_{B_s}$.} This result agrees very well with the one obtained using HQET sum rules \cite{King:2019rvk}.

        In view of the strong dependence of $\varepsilon_K$ and $\Delta M_d$ on
        $\gamma$ which is presently not precisely known and the persistent
        tension between inclusive and exclusive determinations of $\vcb$
        we point out that presently the result in (\ref{vcbbest}) is the most
        precise determination of this CKM element based on a single quantity.
        {We impose, then, the constraint on $\beta$ coming from $S_{\psi K_S}$, i.e. $\beta=22.2(7)^\circ$, and on $\gamma$ coming from $\varepsilon_K$, that together with $\Delta M_s$ implies {$\gamma=69.8(26)^\circ$ }for $\beta$ varying in its $S_{\psi K_S}$ range; the obtained $\vcb$ becomes}
  \be\label{vcbbest1}
         \boxed{\vcb=41.{81(61)}\times 10^{-3}, \qquad (\Delta M_s\,,~|\varepsilon_K|,~ S_{\psi K_S}).}
            \ee
}
              {The question then arises what happens when also $\Delta M_d$
              is taken into account and what these results imply for $\gamma$.
              To this end let us    note the following pattern implied by
    Fig.~\ref{fig:5}.}
             
  \begin{itemize}
  \item
    The {values} of $\beta$, $\gamma$ and $\vcb$  following simultaneously from  $\varepsilon_K$, $\Delta M_s$ and $\Delta M_d$ are
 {   \be
\boxed{\beta=24.0(14)^\circ,\qquad \gamma=65.2(28)^\circ,\qquad \vcb=41.7(6)\times 10^{-3}}\,.
\ee}
While the value of $\gamma$ is consistent with the FLAG determination
in (\ref{xiLQCD}), the value of $\beta$ is outside the 1$\sigma$ range in
(\ref{betagamma}) and would imply $S_{\psi K_S}=0.743$. This {tension}
could be cured by a NP phase $\phi_{\rm NP}$ in the ballpark of $-2^\circ$
\be
S_{\psi K_S}=\sin(2\beta+2\phi_{\rm NP}) \, .
\ee
{The error on $\beta$ determined in this manner, being twice as large as
  the one from  $S_{\psi K_S}$, demonstrates clearly the virtue of the determination of $\beta$ by means of the latter asymmetry.}
\item
  Imposing then {$\beta=22.2(7)^\circ$, from the $S_{\psi K_S}$ measurement,} we find from $\varepsilon_K$ and $\Delta M_d$
{     \be
\boxed{\beta=22.2{(7)}^\circ,\qquad \gamma=61.6(16)^\circ,\qquad \vcb=43.{1(10)}\times 10^{-3}}\,. 
\ee}
This time $\gamma$ is  below FLAG determination {and} $\vcb$ is even larger
than its inclusive determinations.
\item
  On the other hand from $\varepsilon_K$ and $\Delta M_s$  we find
     \be
\boxed{\beta=22.2{(7)}^\circ,\qquad \gamma=69.{8(26)}^\circ,\qquad \vcb=41.8{(6)}\times 10^{-3}}\,.
\ee
{This is precisely the case leading to (\ref{vcbbest1}); the corresponding
 value of $\gamma$ is this time}  significantly larger than the one from
FLAG in (\ref{xiLQCD}) but consistent with the LHCb value in (\ref{betagamma})
due to its large error.
  \end{itemize}
  These three cases are nicely depicted by three circles in Fig.~\ref{fig:5}.
  There is no question about that there are some tensions between these three determinations visible in the plot but this requires simultaneous determination of
  $\beta$ from $S_{\psi K_S}$\footnote{On the other hand as shown in \cite{Buras:2022wpw} these tensions disappear if only 2+1+1 hadronic matrix elements in
    $\Delta M_{s,d}$ from HPQCD collaboration \cite{Dowdall:2019bea} are used and not the average of     2+1+1 and 2+1 LQCD data as done here.}

    . It is evident from this
  plot how important will be future determinations of $\gamma$ from LHCb and Belle II.    
}

{The values of $\vcb$ in these three cases can be compared with the values quoted by PDG
  obtained using CKMfit and UTfit prescriptions. They are
  $\vcb=40.5(8)\times 10^{-3}$ and $\vcb=42.0(6)\times 10^{-3}$, respectively.} 
{As both collaborations determine practically the same central value of $\gamma=68.7^\circ$, the resulting central values of the branching ratios for $\kpn$ being $7.9\times 10^{-11}$ and  $8.7\times 10^{-11}$ exhibit the $\vcb$-problem
  in question that affects global fits.}

%\ajb{Should we update (81).This would require changing gray band in Fig. 13.}

{Having the results for $\vcb$ in Fig.~\ref{fig:5} {at hand} we can use
  the relation in (\ref{vtdvub}) to obtain analogous results for $\vub$. They are shown in Fig.~ \ref{fig:7}. They should be compared with \cite{Aoki:2019cca}
\be\label{FLAGVUB}
\vub=3.73(14)\times 10^{-3}, \qquad {(\rm FLAG)},
\ee
which are indicated by {gray} bands in the three plots in this figure.
We observe that $\vub$ extracted from any of the three observables has almost the same dependence on $\beta$, dominated by the $\sin \beta$ factor in the ratio between $\vub$ and $\vcb$. It is stronger with respect to the dependence on $\gamma$, especially in the case of $\epsilon_K$ and $\Delta M_s$, where $\vub$ is almost $\gamma$ independent. But the main message from the plot is the
agreement with the FLAG value above for $\beta$ from $S_{\psi K_S}$. In this manner,
the inclusive determinations of $\vub$ with central values being above $0.0040$ are practically excluded.}

In view of this result it is tempting to calculate $\vub$ corresponding
to $\vcb$ in (\ref{vcbbest1}). We find
\be\label{vubbest1}
         \boxed{\vub=3.65(12)\times 10^{-3}, \qquad (\Delta M_s\,,~|\varepsilon_K|,~ S_{\psi K_S}),}
         \ee
         in good agreement with the FLAG's value in (\ref{FLAGVUB}) and
         also  with $\vub=3.77(15)\times 10^{-3}$ from light-cone sum rules \cite{Leljak:2021vte}.

\begin{figure}[t!]
\centering%
\includegraphics[width=0.48\textwidth]{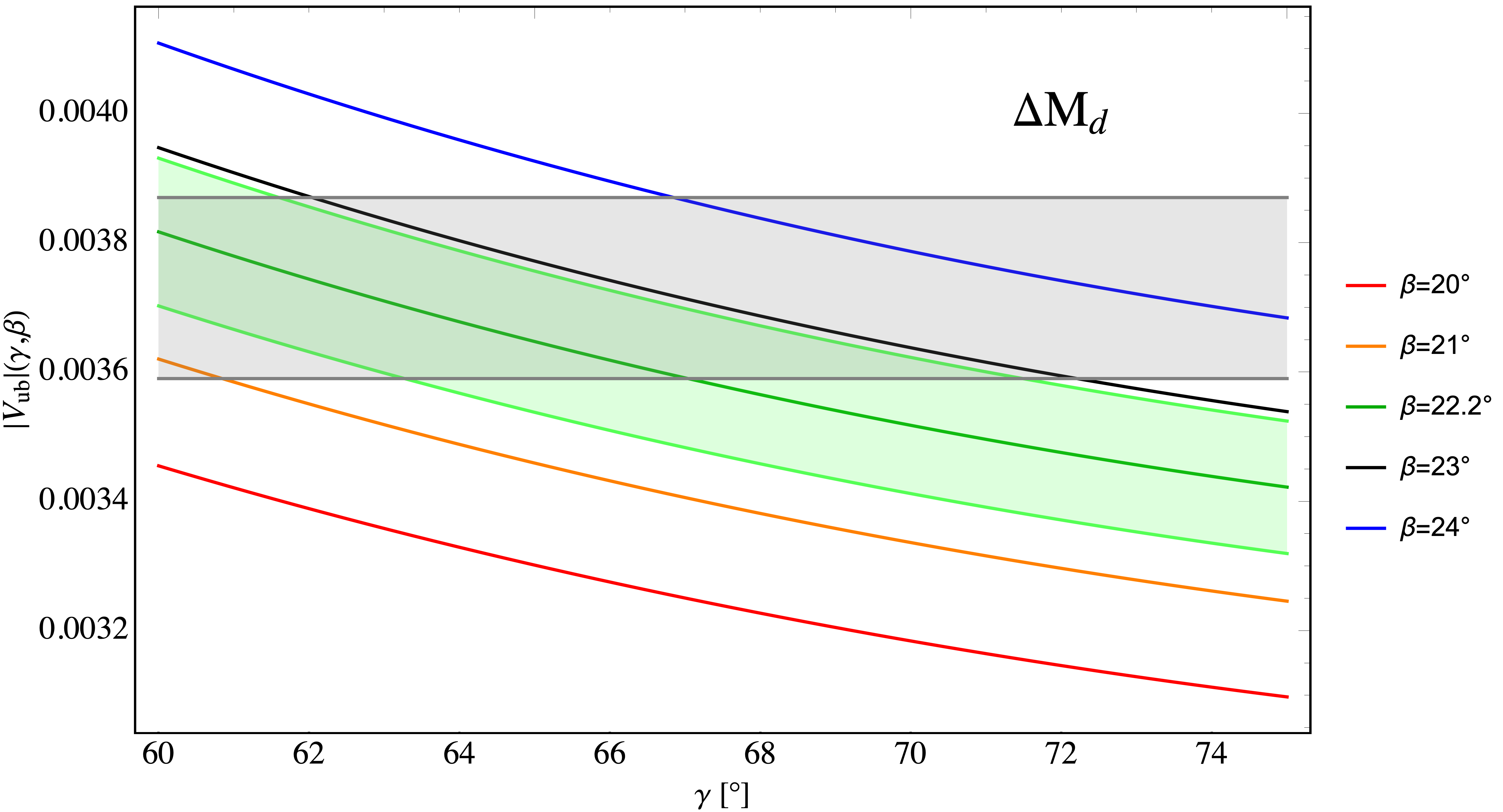}%
\hfill%
\includegraphics[width=0.48\textwidth]{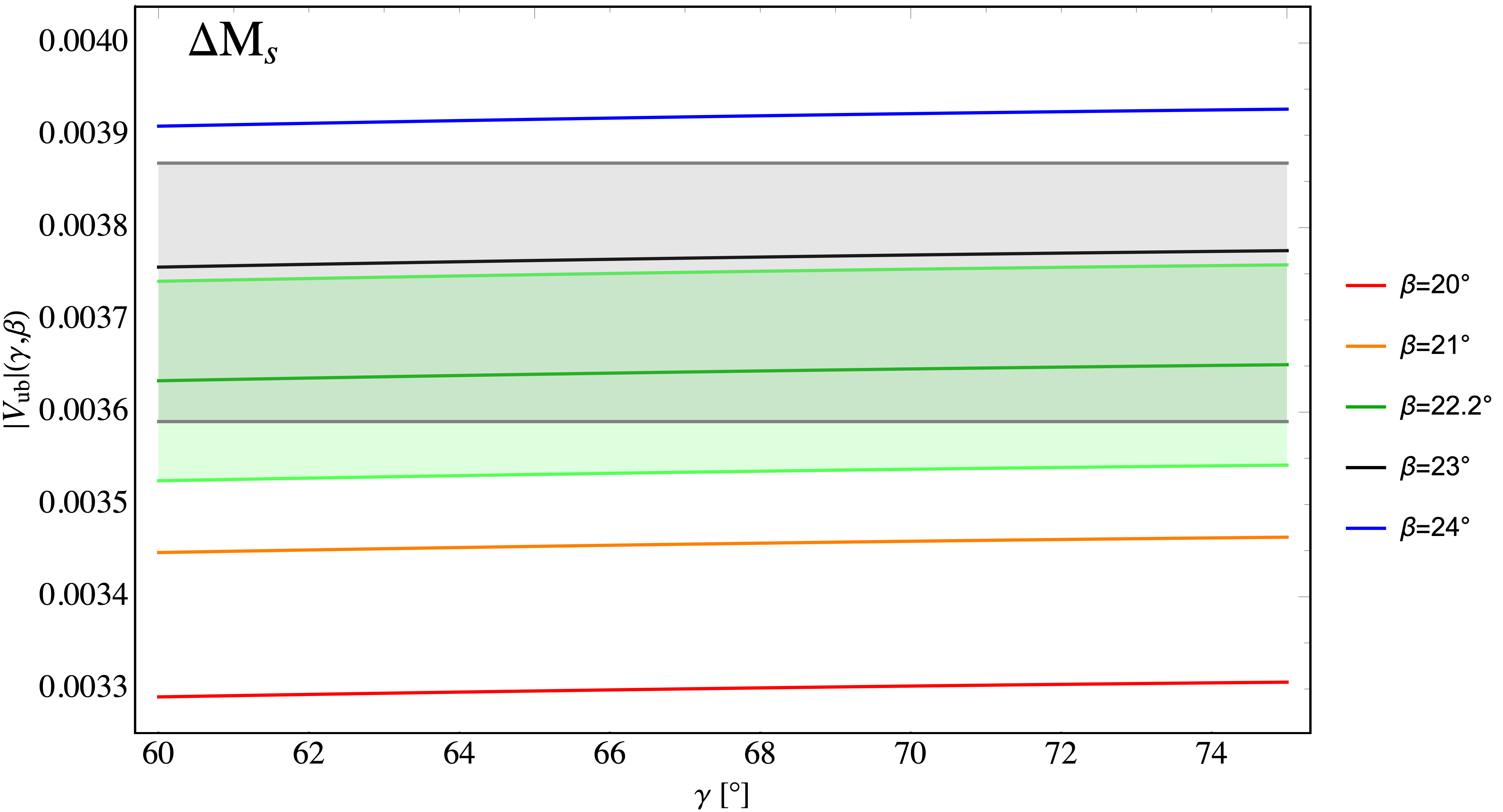}\\
\includegraphics[width=0.48\textwidth]{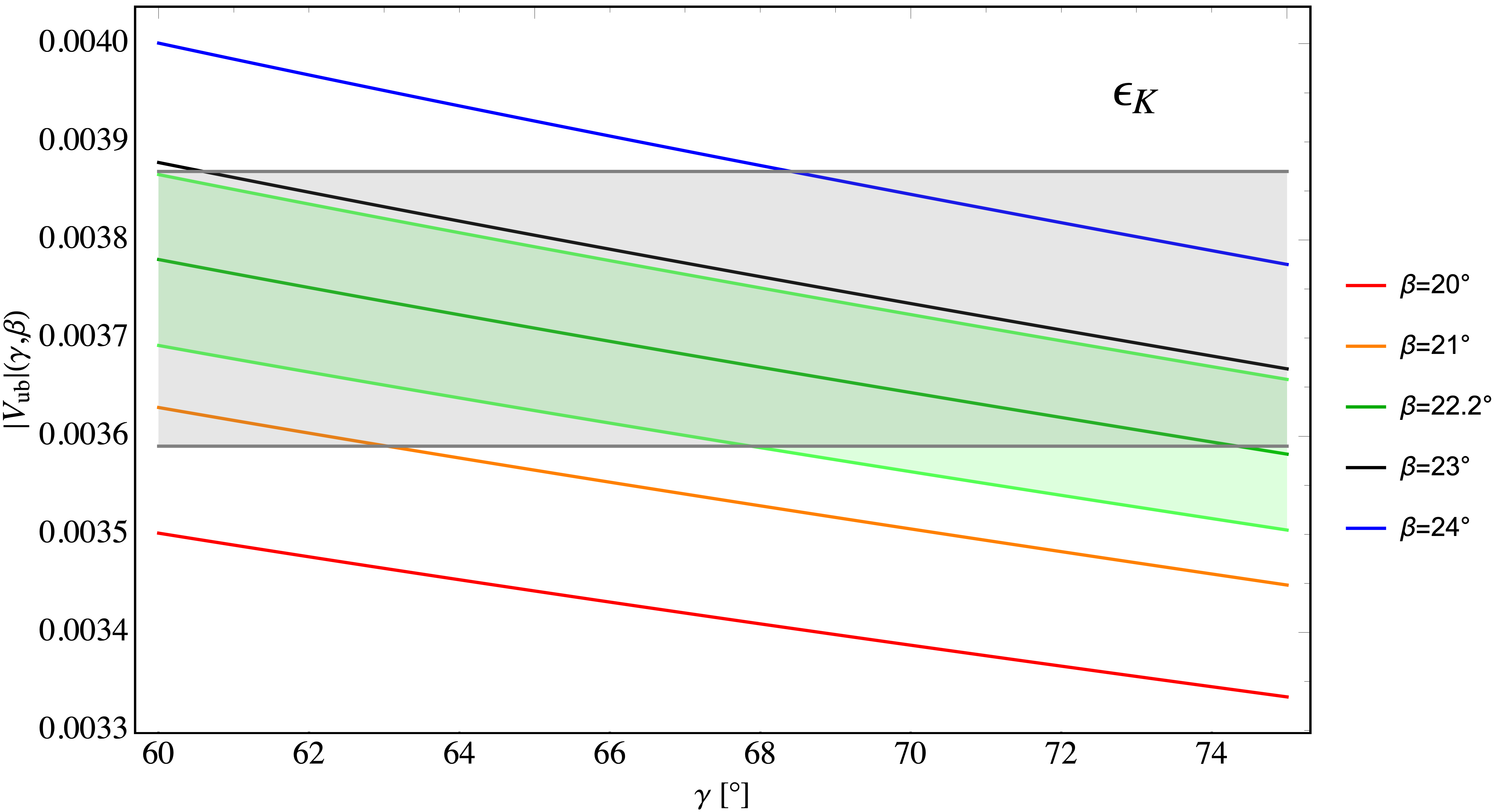}%
\hfill%%
\caption{\it {The values of $\vub$ extracted from $\Delta M_s$, $\Delta M_d$ and $\varepsilon_K$ as functions of $\gamma$ for different values of $\beta$.} 
\label{fig:7} {The gray bands represent the FLAG measurement.}}
\end{figure}

{While the interplay between $\varepsilon_K$, $\Delta M_d$, $\Delta M_s$,
  $\beta$, $\gamma$, $\vub$ and $\vcb$ has been discussed already in a number
  of papers in the past \cite{Lunghi:2008aa,Buras:2008nn,Blanke:2016bhf,Blanke:2018cya}, our presentations in Figs.~\ref{fig:5} and \ref{fig:7} are new.
  But also in these papers some tension between these observables within the SM
  have been found.
  In particular in \cite{Blanke:2018cya} it has been pointed out that
  with $\gamma$ in the ballpark of $74^\circ$, as signalled by the LHCb collaboration in 2018,  and $\vcb$ determined from
  inclusive decays the SM value for $\Delta M_d$ was significantly
  above the data while $\varepsilon_K$ within  the SM was  consistent
  with the data. These finding can be confirmed by inspecting our plots and setting $\gamma=74^\circ$. In particular in this case FLAG and LHCb values for $\gamma$ disagree with each other.

  In the meantime the most recent LHCb value in (\ref{betagamma})
  seems to agree well with the FLAG value in (\ref{xiLQCD}) and the situation
  changed relative to the one addressed in  \cite{Blanke:2018cya}.
  Yet, as we have seen, some tensions remained but in contrast to the latter
  paper we want to address the possible
  tension between $\varepsilon_K$ and $\Delta M_d$ and even $\Delta M_s$
  in the spirit of the present paper, that is in the $\vcb$-independent manner.

  To this end we cast the formula (\ref{BGS}) in a form analogous to
  (\ref{kplusApprox}) to find
  \be\label{eKapp}
  |\varepsilon_K|= {2.015 \times 10^{-3} \times \left( \frac{V_{cb}}{41.0 \times 10^{-3}}\right)^{3.4} \left(\frac{\sin\gamma}{\sin 67^\circ}\right)^{1.67}\left(\frac{\sin \beta}{\sin 22.2^\circ}\right)^{0.87}}
  \ee
  {where the central value has been evaluated with (\ref{BGS}), using the numerical values given in Section~\ref{epsilonK}. The expression above provides an approximation of (\ref{BGS}) with an accuracy of 1.5\%, in the ranges $38< \vcb \times 10^3< 43$, $60^\circ<\gamma<75^\circ$, $20^\circ<\beta<24^\circ$.}

  We consequently define two $\vcb$-independent ratios
  \be
  \boxed{R_9(\beta,\gamma)=\frac{ |\varepsilon_K|}{(\Delta M_d)^{{1.7}}},\qquad
  R_{10}(\beta,\gamma)=\frac{ |\varepsilon_K|}{(\Delta M_s)^{{1.7}}}\, .}
  \ee
  The explicit expressions for them read
  \be\label{R9}
  {R_9(\beta,\gamma)=6.405\times 10^{-3}\, {\rm ps^{1.7}}\left(\frac{\sin 67^\circ}{\sin \gamma}\right)^{1.73}\left(\frac{\sin \beta}{\sin 22.2^\circ}\right)^{0.87}
\bar{R}^\epsilon_d },
    \ee
  \be\label{R10}
  {R_{10}(\beta,\gamma)=1.516\times 10^{-5}\, {\rm ps^{1.7}}\left(\frac{\sin \gamma}{\sin 67^\circ}\right)^{1.67}\left(\frac{\sin \beta}{\sin 22.2^\circ}\right)^{0.87}\left(\frac{G(22.2^\circ,67^\circ)}{G(\beta,\gamma)}\right)^{3.4}
\bar{R}^\epsilon_s }
  \ee
  where
  {\be
  \bar{R}^\epsilon_d=\left( 
\frac{214.0\mev}{\sqrt{\hat B_{B_d}}F_{B_d}}\right)^{3.4}
\left(\frac{2.307}{S_0(x_t)}\right)^{1.7} 
\left(\frac{0.5521}{\eta_B}\right)^{1.7}\, ,
  \ee
  \be
   \bar{R}^\epsilon_s=\left( 
\frac{261.7\mev}{\sqrt{\hat B_{B_s}}F_{B_s}}\right)^{3.4}
\left(\frac{2.307}{S_0(x_t)}\right)^{1.7} 
\left(\frac{0.5521}{\eta_B}\right)^{1.7} \, .
  \ee}

  The ratios $R_9$ and $R_{10}$ are shown in Fig.~\ref{fig:8} as  functions of $\gamma$ for different values of $\beta$. {One can see that, indeed, the dependence of these two ratios on $\vcb$ is very mild, with a variation only of $\lesssim 0.5\%$ for $38< \vcb \times 10^3 < 43$.} On the other hand their experimental values are very precise and read
  \be\label{R9R10}
  R_9=7.082(42)\times 10^{-3}{\rm ps^{1.7}},\qquad
  R_{10}=1.676(8)\times 10^{-5}{\rm ps^{1.7}}\,.
  \ee
  They are shown in  Fig.~\ref{fig:8} as very thin horizontal lines. The circles
  correspond to the ones in  Fig.~\ref{fig:5}. But now the dependence
  on $\vcb$ disappeared and the tensions identified already in Fig.~\ref{fig:5}
  can now be formulated in a different manner:
  \begin{itemize}
  \item
    Imposing the data on $\varepsilon_K$, $\Delta M_d$ and {$S_{\psi K_S}$} implies
    $\gamma=61.6^\circ$.
  \item
    On the other hand imposing  the data on $\varepsilon_K$, $\Delta M_s$ and {$S_{\psi K_S}$} implies
    $\gamma=69.8^\circ$.
  \item
    The agreement on $\gamma$, in fact very close to the central value from LHCb
    in (\ref{betagamma}), can only be obtained for $\beta=24^\circ$, that
    is outside the range obtained from $S_{\psi K_S}$.
  \end{itemize}

  We conclude therefore that it is not possible to obtain simultaneous good agreement
  between the data on $\varepsilon_K$, $\Delta M_d$, $\Delta M_s$ and
  $S_{\psi K_S}$ within the SM independently of the value of $\vcb$ and $\gamma$.
  
}

 \begin{figure}[t!]
\centering%
%\includegraphics[width=0.48\textwidth]{Plot5a.png}%
%\hfill%
%\includegraphics[width=0.48\textwidth]{Plot5b.png}\\
\includegraphics[width=0.48\textwidth]{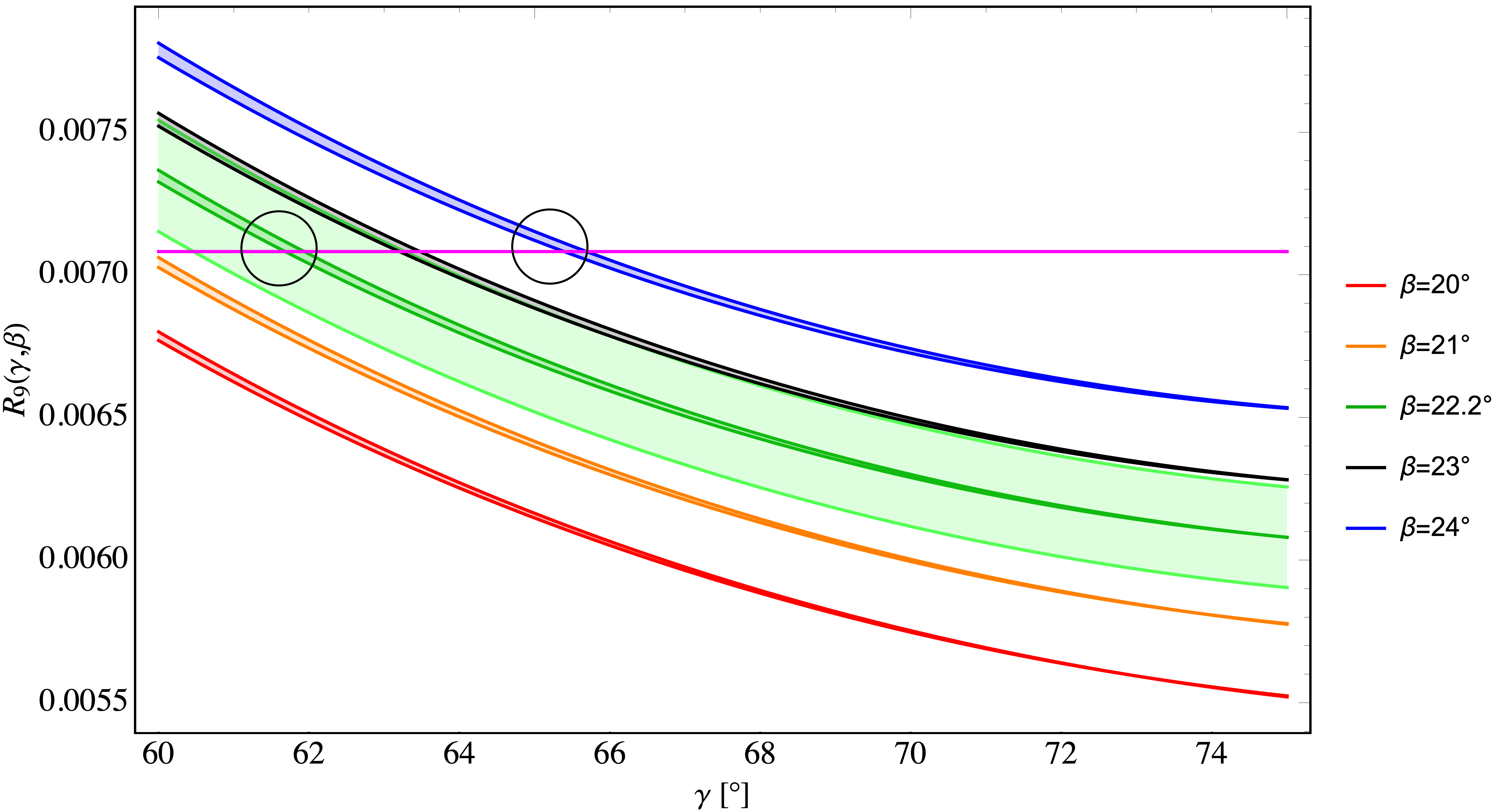}%
\hfill%
\includegraphics[width=0.48\textwidth]{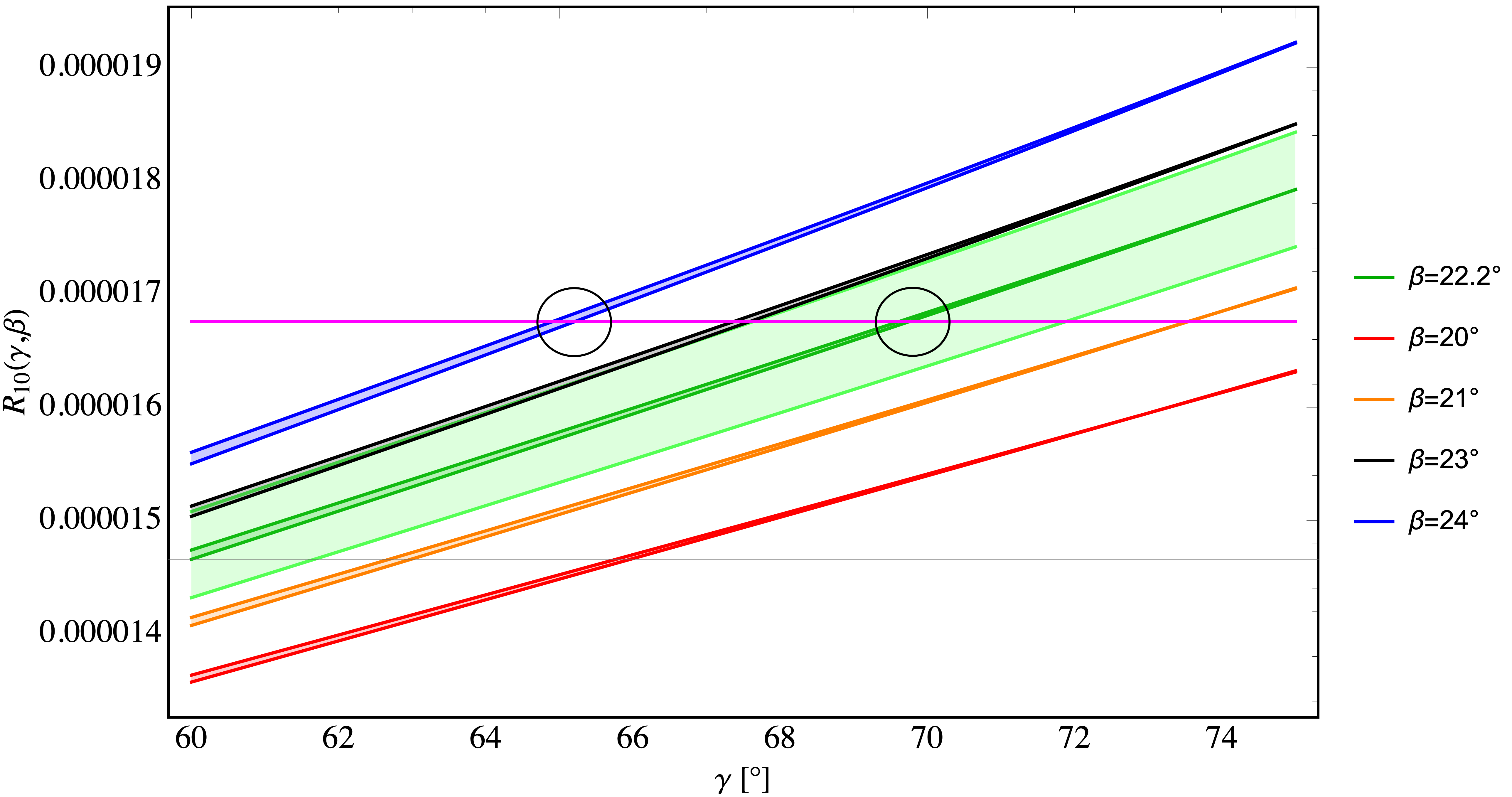}%
\caption{\it {The ratios $R_9$ and $R_{10}$ as functions of $\gamma$ for different values of $\beta$. The coloured bands correspond to $38 < \vcb \times 10^{3}< 43 $. {The horizontal thin bands represent the experimental values as given in
(\ref{R9R10}).}} {The light green band corresponds to the range $21.5^\circ<\beta<22.9^\circ$, with $38\times 10^{-3} < \vcb < 43 \times 10^{-3}$.}
      \label{fig:8}}
\end{figure}

{While, in view of large experimental errors,
 it would be premature to  include presently the rare decays 
in this analysis, we will next  use the
results for $\vcb$ that are extracted from $\varepsilon_K$, $\Delta M_s$ and
$\Delta M_d$  for rare decays.}

\boldmath
\subsection{{Improved SM Predictions for rare $K$ and $B$ Decays}}
\unboldmath
{Beginning with $\kpn$ and $\klpn$, the authors of \cite{Brod:2021hsj} obtained the results in (\ref{Brod}) by performing first a careful analysis of theoretical uncertainties in the evaluation
of both branching ratios in question and subsequently using the CKM
parameters listed in PDG \cite{Zyla:2020zbs} that strictly speaking come from
CKMfitters with $\vcb$ in the ballpark of its exclusive determinations. If they used UTfitters result, also listed there, they would
get significantly larger values for both branching ratios because
in the latter analysis the value of $\vcb$ closer to its inclusive determinations has been used.

The virtue of the strategies presented in our paper is that we can avoid
all such uncertainties by simply inserting the formula (\ref{vcb2}) into
the formulae for  $\kpn$ and $\klpn$  decay branching ratios and
study their dependence on  $\gamma$, $\beta$ and $\vus$ in a $\vcb$-independent manner. To our knowledge such a $\vcb$-independent analysis of rare $K$ decay branching ratios has not  been presented in the literature. As we will
now demonstrate this allows to determine both branching ratio not only
independently of the value of $\vcb$ but practically also independently of the
value of the angle $\gamma$. The main parametric uncertainty comes
then from $\beta$ which, as seen in (\ref{betagamma}),
is already known rather precisely from the measurement of $S_{\psi K_S}$.
The theoretical uncertainties are as in  \cite{Brod:2021hsj} but the fact
that the CKM uncertainties have been practically reduced to the one in $\beta$
allows to obtain results even more accurate than listed in (\ref{Brod})
without any worries about $\vcb$ and $\gamma$. However, in addition to the
experimental error on $\varepsilon_K$ we have to take the theoretical ones in $r_1$ and $r_2$ in (\ref{r1r2}), in particular the ones due to $\kappa_\epsilon$ and $\hat{B}_K$. But these uncertainties have only small impact on the final errors.

We proceed then as follows.
 From
  (\ref{kplusApprox}),  (\ref{k0Approx}) and (\ref{eKapp}) we find approximate formulae
  \be\label{R11}
  \boxed{R_{11}=\frac{\mathcal{B}(\kpn)}{|\varepsilon_K|^{0.82}}=(1.31\pm0.05)\times 10^{-8}{\left(\frac{\sin\gamma}{\sin 67^\circ}\right)^{0.015}\left(\frac{\sin 22.2^\circ}{\sin \beta}\right)^{0.71},  }            }
  \ee
  \be\label{R12a}
\boxed{R_{12}=\frac{\mathcal{B}(\klpn)}{|\varepsilon_K|^{1.18}}=(3.87\pm0.06)\times 10^{-8}
    {\left(\frac{\sin\gamma}{\sin 67^\circ}\right)^{0.03}\left(\frac{\sin\beta}{\sin 22.2^\circ}\right)^{0.9{8}}}}
  \ee
  and
  \be
  \frac{\mathcal{B}(\klpn)}{\mathcal{B}(\kpn)}={(2.95\pm 0.12)}\,|\varepsilon_K|^{0.36}
  {\left(\frac{\sin\gamma}{\sin 67^\circ}\right)^{0.015} \left(\frac{\sin\beta}{\sin 22.2^\circ}\right)^{1.6{9}}\,.}
  \ee
  The first two of these formulae express explicitly the fact that combining
  on the one hand $\kpn$ and $\varepsilon_K$ and on the other hand
  $\klpn$ and  $\varepsilon_K$ allows within the SM to determine
  to a very good approximation the angle $\beta$ independently of the value
  of $\vcb$ and $\gamma$. The last one just follows from them. Indeed
  the dependence on $\gamma$ is very weak.

But these formulae are only approximate and 
  therefore, in what follows, we use exact formulae.
 In Fig.~\ref{fig:6}
we present in the upper panels $\kpn$ and $\klpn$  branching ratios within the SM as functions of
$\gamma$ for different values of $\beta$, {once the $\vcb$ dependence has been eliminated through (\ref{vcb2})}. The dependence on $\vus$ is very weak.
In lower panels we show  $\kpn$ and $\klpn$  branching  as functions of
$\beta$ for different values of $\gamma$
We make the following observations.
\begin{itemize}
\item The $\gamma$ dependence is extremely weak in both cases, while
  the $\beta$ dependence is significant. This is in accordance with
    (\ref{R11}) and (\ref{R12a}).
\item
  In particular, the $\gamma$ dependence of the $\kpn$ branching ratio
  is practically absent when the $\varepsilon_K$ is used\footnote{This independence of $\gamma$ is exact if (\ref{AIACD}) is used instead of the exact expression.}. {In fact, one can notice that the $\gamma$ dependence in (\ref{R11}) is even weaker than in (\ref{R12a}).} {This is related to the decrease of $\vcb$ with increasing $\gamma$ as seen in Fig.~\ref{fig:5} and these two effects of respectively suppressing 
and enhancing the $\kpn$ branching ratio compensate each other.}
  With a precise
  value of experimental $\kpn$ branching ratio the value of $\beta$
  could be determined without any involvement of $\klpn$.
\item
  On the other hand the dependence on $\beta$ in the case of $\klpn$
  is stronger than for $\kpn$, again in agreement with (\ref{R11}) and
  (\ref{R12a}). Therefore this time $\beta$ could be determined from $\klpn$ without
  any involment of $\kpn$.
\end{itemize}

In studying these plots one should again keep in mind that on the basis
of the measurement of $S_{\psi K_S}$ the $1\sigma$
range for $\beta$ is given in (\ref{betarange}).
We show this range as green  bands in the plots in Fig.~\ref{fig:6}.} Therefore in the final
step we impose this constraint.
{Performing the error analysis we find then from $\varepsilon_K$ and
  $S_{\psi K_S}$ with $60^\circ\le\gamma\le 75^\circ$ the result in (\ref{BV}).
To this end we have taken into account the experimental and theoretical uncertainties in  $\varepsilon_K$, experimental ones in the
$S_{\psi K_S}$ measurements  and the non-parametric errors of the considered branching ratios like $P_c$ in the case of $\kpn$. The dominant source of uncertainty in $\kpn$ are $P_c$ and $\beta$, while in the case of $\klpn$ it is
$\beta$.

We observe that
both branching ratios have significantly smaller errors than the ones
in (\ref{KSM}) and (\ref{Brod}). They supersede the usual quoted values in
(\ref{KSM}).}
 \begin{table}
\centering
\renewcommand{\arraystretch}{1.4}
\resizebox{\columnwidth}{!}{
\begin{tabular}{|ll||ll|}
\hline
Decay 
& Branching Ratio
& Decay
&  Branching Ratio
\\
\hline \hline
 $\kpn$ & $(8.60\pm 0.42)\times 10^{-11}$ & $B_s\to\mu^+\mu^-$ & $(3.62^{+ 0.15}_{-0.10})\times 10^{-9}$
\\
 $\klpn$ & $(2.94\pm 0.15)\times 10^{-11}$ & $B_d\to\mu^+\mu^-$ & $(0.99^{+ 0.05}_{-0.03})\ \times 10^{-10}$
\\
$\ksm$ & {$(1.85\pm 0.10)\times 10^{-13}$} &  $B^+\to K^+\nu\bar\nu$ & $(4.45\pm 0.62)\times 10^{-6}$
\\
& &  $B^0\to K^{0*}\nu\bar\nu$ &$(9.70\pm 0.92)\times 10^{-6}$
\\
\hline
\end{tabular}
}
\renewcommand{\arraystretch}{1.0}
\caption{\label{tab:SMBRs}
  \small
  Present most accurate $\vcb$-independent SM estimates  of the branching ratios considered in the paper. The $\gamma$ dependence is either very small or absent. See the text for details.
}
\end{table}

\begin{figure}[t!]
\centering%
\includegraphics[width=0.48\textwidth]{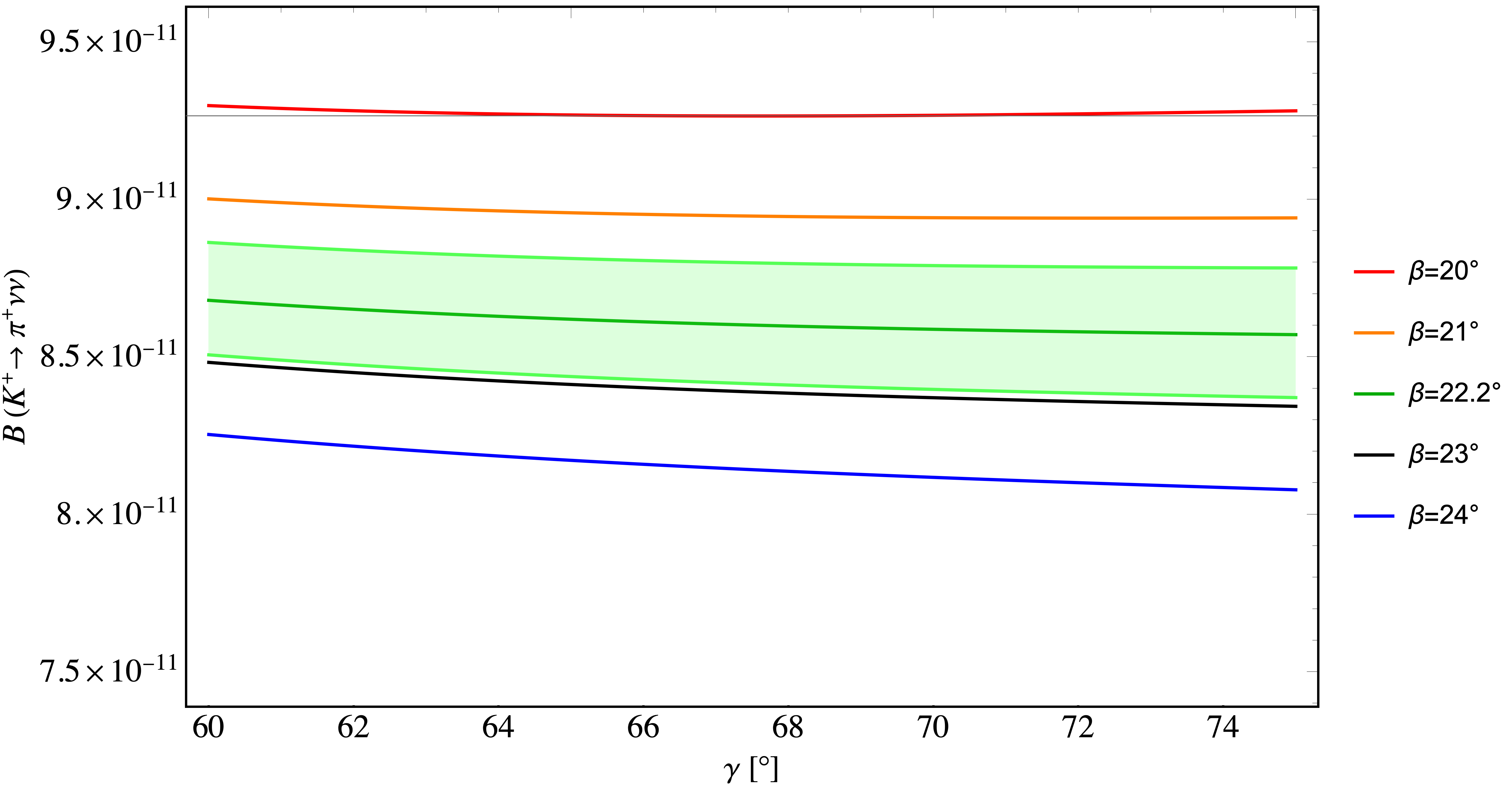}%
\hfill%
\includegraphics[width=0.48\textwidth]{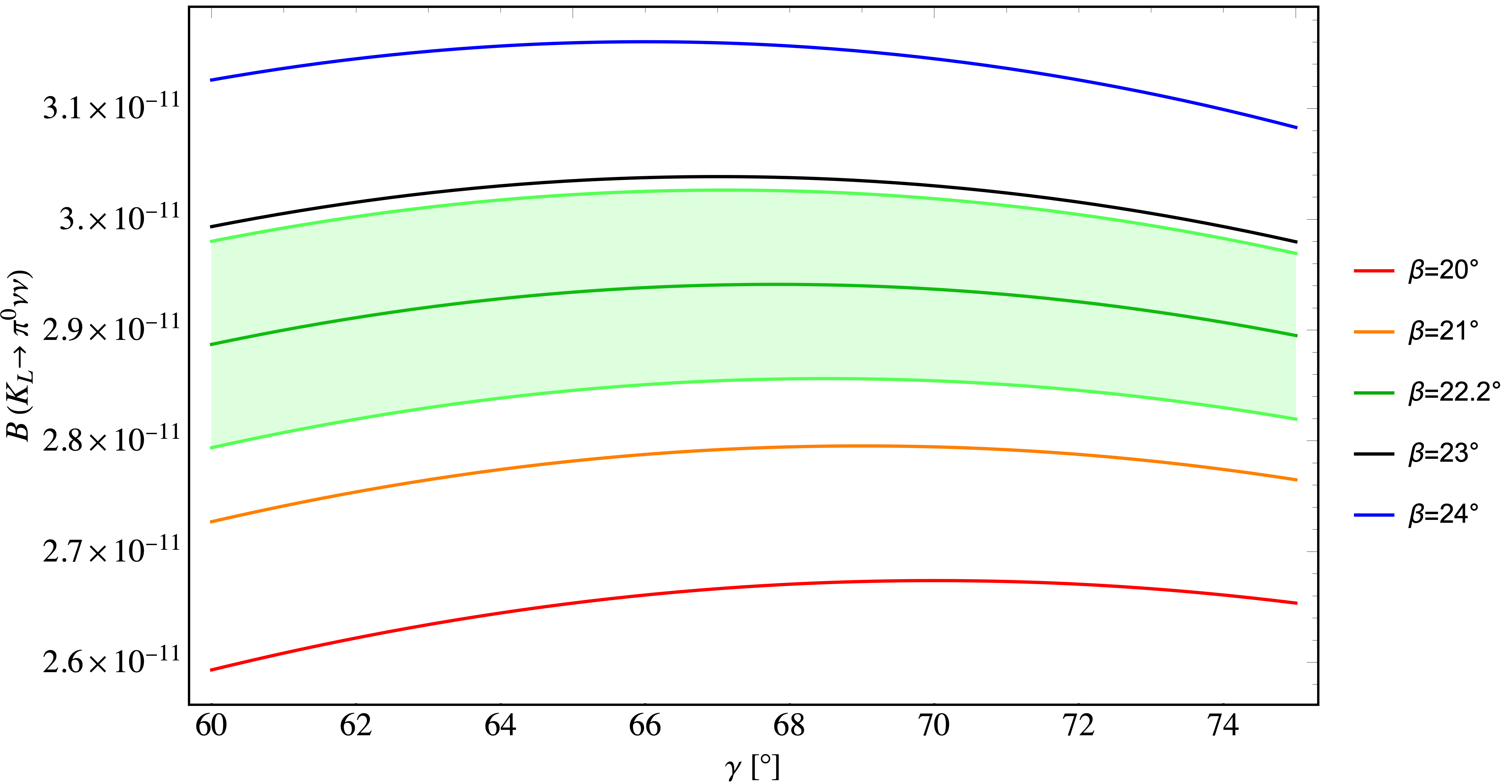}\\
\includegraphics[width=0.48\textwidth]{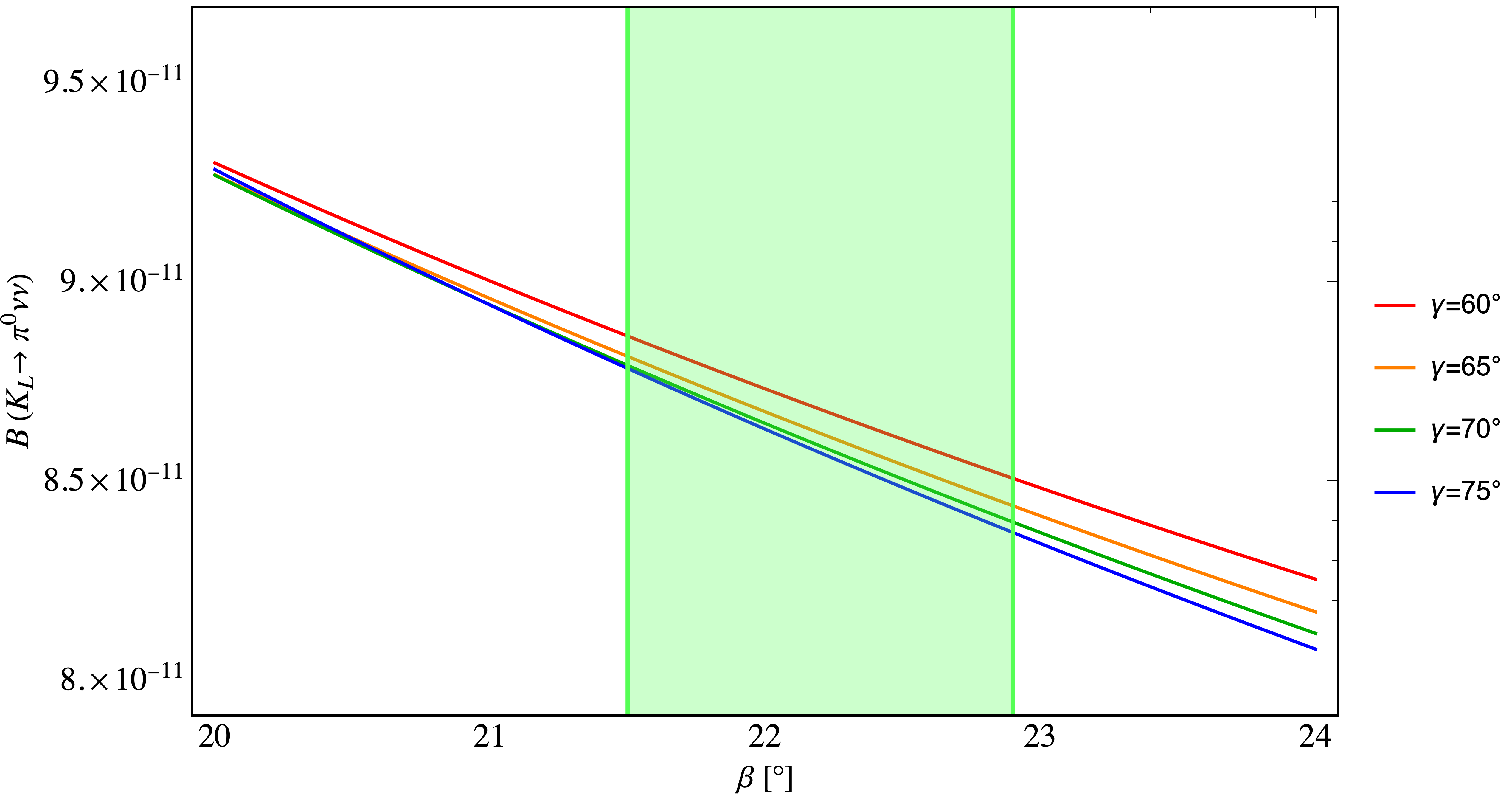}%
\hfill%
\includegraphics[width=0.48\textwidth]{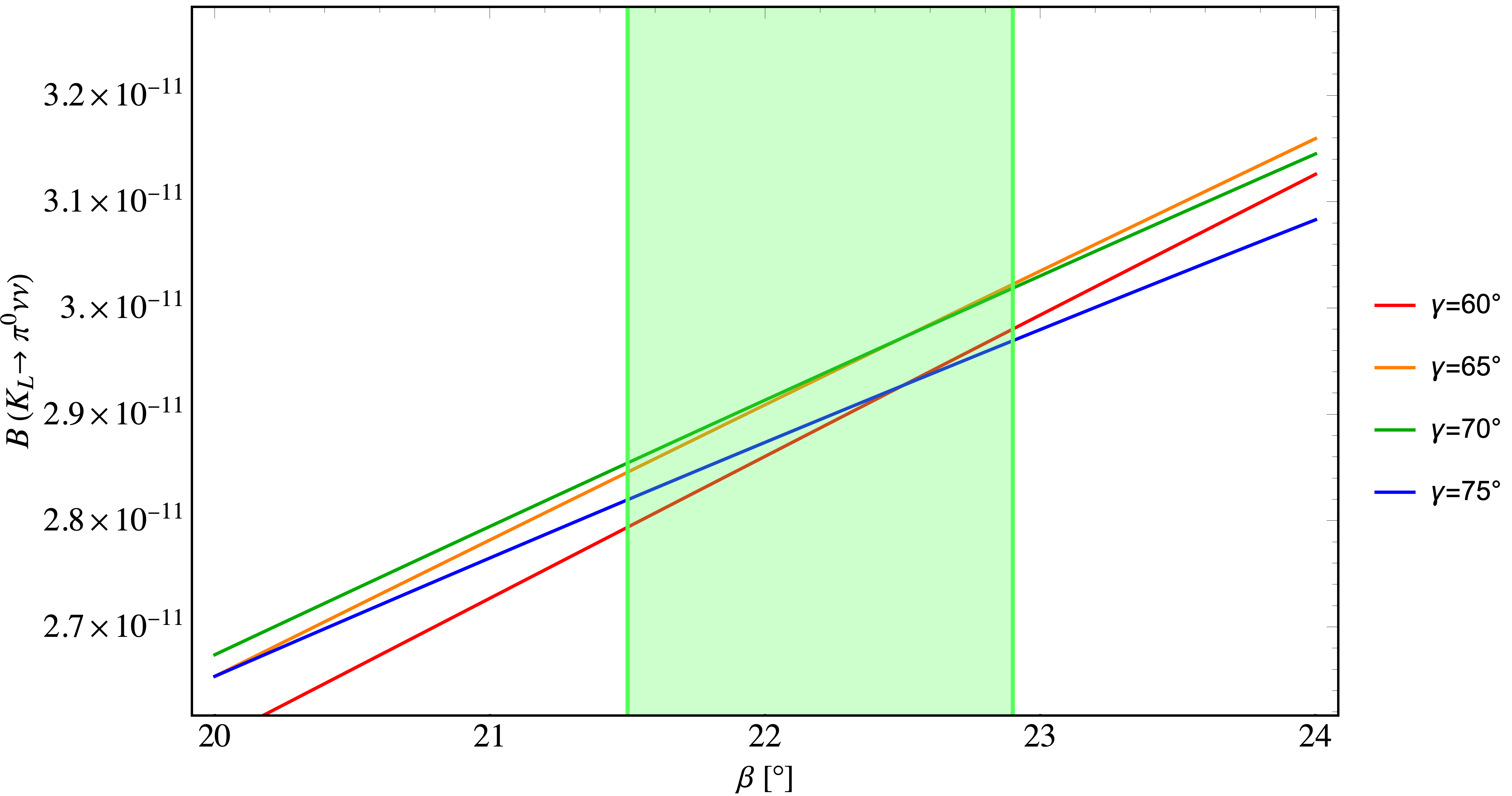}%
\caption{\it {The dependence of the branching ratios $\mathcal{B}(\kpn)$ (left panels) and $\mathcal{B}(\klpn)$ (right panels) on  $\gamma$ for different values of   $\beta=20.0^\circ,21.0^\circ, 22.0^\circ, 23.0^\circ, 24.0^\circ$ and on $\beta$
 for different values of $\gamma=60.0^\circ, 65.0^\circ, 70.0^\circ, 75^\circ$.  
$\varepsilon_K$ has been fixed to its experimental value.} 
\label{fig:6}}
\end{figure}

For $B_{s,d}\to\mu^+\mu^-$ decays analogous use of $\Delta M_{s,d}$ eliminates the CKM dependence  from branching ratios as already demonstrated in \cite{Buras:2003td,Bobeth:2021cxm} {with the result given in (\ref{LHCbTH}).
 Inserting next the results in (\ref{BV})
 and (\ref{LHCbTH}) into the $\vcb$-independent ratios $R_7$, $R_8$ and $R_{\rm SL}$ allows in turn to obtain  the most precise estimate of the
 branching ratios of  $\ksm$, $B^+\to K^+\nu\bar\nu$ and
 $B^0\to K^{0*}\nu\bar\nu$ as well. The results for all branching ratios are
 summarized in Table~\ref{tab:SMBRs} {and their implications in the case of
 $\kpn$ and $B_s\to\mu^+\mu^-$ in   Fig.~\ref{fig:3abis}.}

  It should be emphasized that to obtain precise SM predictions like the
  ones in (\ref{BV}) and (\ref{LHCbTH}) it is crucial to choose the proper
  pairs of observables. For instance combining $\kpn$ with $\Delta M_s$ or
  $B_s\to\mu^+\mu^-$ with $\varepsilon_K$ would not allow us precise predictions
  for $\kpn$ and $B_s\to\mu^+\mu^-$ even after the elimination of the $\vcb$ because of the left-over $\gamma$ dependence in both cases.} {Moreover
  selecting a subset of optimal observables for a given SM prediction
  avoids the assumption of the absence of NP in other observables which would
  be questionable in view of the inconsistencies between various
  determinations of $\vcb$ identified by us.}

{The accuracy of our predictions could certainly be improved through
  a better determination of $\beta$ with the help of $S_{\psi K_s}$ or a better
  determination of $\vub$. As an example we refer to a recent determination
  of $\vub$ with the help of exclusive semi-leptonic $B$ decays with
  $\vub=3.68(5)\times 10^{-3}$ \cite{Gonzalez-Solis:2021awb}, that is more accurate than what we obtained in (\ref{vubbest1}).  Through (\ref{vtdvub}) using this result a more
  accurate $\beta=22.4(5)^\circ$ can be determined. Adding this additional constraint we find
\be\label{BV2}
{\mathcal{B}(\kpn)_\text{SM}= {(8.55\pm 0.37)}\times 10^{-11}\,,\quad
\mathcal{B}(\klpn)_\text{SM}={(2.97\pm 0.12)}\times 10^{-11}\,,}
\ee
that is slightly more accurate than our results in (\ref{BV}). However, we keep
(\ref{BV}) as our official result until the impressive determination of $\vub$
in  \cite{Gonzalez-Solis:2021awb} will be confirmed by other experts on
exclusive semi-leptonic $B$ decays.}

\boldmath
\section{A Guide to $\vcb$-independent Relations}\label{sec:4B}
\unboldmath
In the course of our analysis we have introduced various ratios and relations
between different observables of which some were independent of $\vcb$ and the
rest practically independent of it, that is with the dependence on $\vcb$
significantly below $1\%$ in the full range of $\vcb$ considered by us.
We have also checked that the dependence on $\vus=\lambda$ of these relations and  ratios
was either absent or totally negligible. Consequently the only relevant dependences which were left were only on the angles $\beta$ and $\gamma$.

{The basis for the derivation of these correlations were approximate
  but accurate formulae with the general power-like structure
  \be\label{criticalform}
  {\mathcal{B}(\text{Decay}_i)=C_i\vcb^{r_1}[\sin\gamma]^{r_2}[\sin\beta]^{r_3},}
  \ee
with the coefficients $C_i$ either being constants or being very weakly dependent on $\beta$ and $\gamma$ like the function $G(\beta,\gamma)$.   
The powers $r_i$ can be regarded as {\em critical exponents} of flavour physics.
In Table~\ref{tab:critical} we collect these critical exponents for
all observables considered by us and give references to the corresponding expressions
with the power-like structure. Expressing then $\vcb$ in terms of a given branching ratio and inserting it into a power-like formula for another branching ratio
allows to derive analytically all correlations presented in our paper.
One finds then that  some of the ratios are independent of $\beta$ and $\gamma$, some are dependent only on $\gamma$, some only on $\beta$ and 
some on both $\beta$ and $\gamma$.}

\begin{table}
\centering
\renewcommand{\arraystretch}{1.0}
\resizebox{\columnwidth}{!}{
\begin{tabular}{|l|lll|l|}
\hline
\tiny  Observable
& \tiny $r_1$
&  \tiny $r_2$
& \tiny $r_3$
& \tiny Formula
\\
\hline \hline
  \tiny $\kpn$, &\tiny $2.8$ &\tiny $1.39$ & \tiny$0.0$ & \tiny(\ref{kplusApprox})
\\
  \tiny$\klpn$, & \tiny$4.0$ & \tiny$2.0$ & \tiny$2.0$ &\tiny (\ref{k0Approx})
\\
  \tiny$\ksm$, & \tiny$4.0$ & \tiny$2.0$ &\tiny $2.0$ & \tiny(\ref{eq:ksm-br-SDAppr})
\\
  \tiny$|\varepsilon_K|$, &\tiny $3.4$ & \tiny$1.67$ & \tiny$0.87$ & \tiny(\ref{eKapp})
\\
\hline
 \tiny$B_s\to\mu^+\mu^-$ & \tiny$4.0$ & \tiny$0.0$ & \tiny$0.0$ & \tiny(\ref{BRtheoRpar})
\\
 \tiny$B_d\to\mu^+\mu^-$ & \tiny$2.0$ & \tiny$2.0$ & \tiny$0.0$ & \tiny(\ref{BRtheoRpard})
\\
 \tiny$B^+\to K^+\nu\bar\nu$ & \tiny$2.0$ & \tiny$0.0$ & \tiny$0.0$ & \tiny(\ref{BKp})
\\
\tiny$B^0\to K^{0*}\nu\bar\nu$ & \tiny$2.0$ & \tiny$0.0$ & \tiny$0.0$ & \tiny(\ref{BK0})
\\
\hline
\tiny$\Delta M_d$ & \tiny$2.0$ & \tiny$2.0$ &\tiny $0.0$ & \tiny(\ref{DMD})
\\
\tiny$\Delta M_s$ & \tiny$2.0$ & \tiny$0.0$ & \tiny$0.0$ &\tiny (\ref{DMS})
\\
\hline
\end{tabular}
}
\renewcommand{\arraystretch}{1.0}
\caption{\label{tab:critical}
  \small
  Critical exponents of flavour physics entering the formula (\ref{criticalform}). In the last column the reference to the given power-like formula is given.
}
\end{table}

  In Table~\ref{tab:guide} we indicate 
which of the relations found by us has weak, strong or none dependence
on $\beta$ and $\gamma$, {where "none" {stands} also for a dependence that induces a ratio variation of less than $0.5\%$, in the considered ranges}. This table allows to find in no time the
analytic expressions for each relation in the text and the corresponding 
plot as a function of $\gamma$ for different values of $\beta$.
The ratios $R_q$ proposed in \cite{Buras:2003td} and
analyzed recently in \cite{Bobeth:2021cxm} 
are defined through
\be\label{CMFV6}
R_q=\frac{\mathcal{B}(B_q\to\mu^+\mu^-)}{\Delta M_q}= C \frac{\tau_{B_q}}{\hat B_q}\frac{(Y_0(x_t))^2}{S_0(x_t)},\qquad q=d,s\,,
\ee
with
\be
C = 6\pi \frac{1}{\eta_B^2}\left(\frac{\alpha}{4\pi \sin^2\theta_W}\right)^2\frac{m_\mu^2}{M_W^2}= 4.291\times 10^{-10}\,.
\ee
They are CKM independent {and have been used to obtain the results in (\ref{LHCbTH}).}

 We would like to emphasize that among {sixteen} ratios listed in Table~\ref{tab:guide} only $R_7$, $R_9$, $R_{10}$ and $R_s$ could be confronted until now with the data.
 As $R_9$ and $R_{10}$ depend strongly on $\gamma$ no definite conclusion on them
 could be reached but as we discussed in the previous section and seen in
 Fig.~\ref{fig:8} it is not possible to obtain simultaneous agreement on both with very precise  data for any value of $\gamma$ if the constraint from $S_{\psi K_S}$ is  taken into account. On the other hand $R_7$ and $R_s$ are independent
 of CKM parameters and it was interesting to find that in both cases some tensions  with the data have been identified. It will be interesting to compare one
 day all {sixteen} ratios with future data. In particular when the value of $\gamma$ will be known with high precision we will be able to check if all these
 ratios agree with the improved data. If this will not be the case the pattern of possible
 deviations from the SM predictions calculated by us may give some hints for
 the particular NP at work. Therefore we are looking forward to the day
 on which the angle $\gamma$ will be precisely known and all branching
 ratios analysed by us accurately measured. Then we will be able to replace
 the last two columns  in Table~\ref{tab:guide} by SM predictions for these ratios and the corresponding data.
 
\begin{table}
\centering
\renewcommand{\arraystretch}{1.4}
\resizebox{\columnwidth}{!}{
\begin{tabular}{|l|l|ll|ll|}
\hline
  Ratio
& Observables
& $\beta$
&  $\gamma$
& Formula
& Figure
\\
\hline \hline
 $R_0$                & $\kpn$, $\klpn$ & strong
& {none}           & (\ref{B1vsB2})  & \ref{fig:3}
\\
  $R_1$                & $\kpn$, $B_s\to\mu^+\mu^-$ & {none}
& strong            & (\ref{master1})  & \ref{fig:3a}
\\
 $R_2$                & $\kpn$, $B_d\to\mu^+\mu^-$ & {none}
& strong             & (\ref{master2})   & \ref{fig:3a}
\\
  $R_3$                & $\klpn$, $B_s\to\mu^+\mu^-$ & strong
& strong            & (\ref{R3})  & \ref{fig:4}
\\
 $R_4$                & $\klpn$, $B_d\to\mu^+\mu^-$ & strong
& strong             & (\ref{R4})  & \ref{fig:4}
\\
  $R_5$                & $\kpn$, $B^+\to K^+\nu\bar\nu$ & {none}
& strong            &  (\ref{R5})   & \ref{fig:4B}
\\
 $R_6$                & $\kpn$, $B^0\to K^{0*}\nu\bar\nu$   & {none}
& strong             &  (\ref{R6})   & \ref{fig:4B}
\\
  $R_7$                & $B_s\to\mu^+\mu^-$, $B^+\to K^+\nu\bar\nu$ & none
& none           & (\ref{R7SM}) & $-$
\\
 $R_8$                & $B_s\to\mu^+\mu^-$, $B^0\to K^{0*}\nu\bar\nu$   & none
& none            & (\ref{R8SM})  & $-$
\\
  $R_9$                & $|\varepsilon_K|$, $\Delta M_d$ & strong
& strong          & (\ref{R9})  &  \ref{fig:8}
\\
 $R_{10}$                &  $|\varepsilon_K|$, $\Delta M_s$ & strong
& strong          &(\ref{R10})  & \ref{fig:8}
\\
$R_{11}$                &  $\kpn$, $|\varepsilon_K|$ & strong
& none         &(\ref{R11})  & \ref{fig:6}
\\
$R_{12}$                & $\klpn$, $|\varepsilon_K|$, & strong
& none        &(\ref{R12a})  & \ref{fig:6}
\\
  $R_{d}$                & $B_d\to\mu^+\mu^-$, $\Delta M_d$ & none
& none       & (\ref{CMFV6})  & $-$
\\
 $R_{s}$                &  $B_s\to\mu^+\mu^-$, $\Delta M_s$ & none
& none         & (\ref{CMFV6})  & $-$
\\
 $R_{\rm SL}$                &  $\klpn$, $\ksm$ & none
& none         & (\ref{SR1})  & $-$
\\
\hline
\end{tabular}
}
\renewcommand{\arraystretch}{1.0}
\caption{\label{tab:guide}
  \small
  Guide to the relations presented in the paper. We distinguish between
  strong, weak and none dependences on $\beta$ and $\gamma$. In the last
  two columns the reference to the formula for a given ratio and to the corresponding plot are given.
}
\end{table}

\section{Conclusions}\label{sec:4}
In the present paper, following and extending significantly the strategies of
\cite{Buchalla:1994tr,Buras:1994rj,Buchalla:1996fp,Buras:2002yj,Buras:2015qea,Buras:2003td,Bobeth:2021cxm,Blanke:2018cya}, we have proposed to search for NP in rare Kaon and $B$-meson decays without the necessity
  to choose the values of the CKM elements $\vcb$ and $\vub$, that introduce
  presently large parametric uncertainties in the otherwise theoretically
  clean decays $\kpn$, $\klpn$, $\ksm$, $B_{s,d}\to\mu^+\mu^-$, $B\to K(K^*)\nu\bar\nu$, in the parameter $\varepsilon_K$ and $\Delta M_{s,d}$.
 {Table~\ref{tab:guide} is a useful  guide to $\vcb$-independent relations and corresponding figures obtained in our paper.}

  In addition to various updates and to stressing the usefulness of the
  set (\ref{4CKM}) \cite{Buras:2002yj,Blanke:2018cya}
  the main results of our present paper are as follows.
  
  \begin{itemize}
  \item
    We reemphasized that despite the strong dependence of the branching ratios for $\kpn$ and $\klpn$ on $\vcb$ the correlation between
    them within the SM and models with CMFV, as given analytically in (\ref{B1vsB2}) and numerically in Fig.~\ref{fig:3}, is practically independent of $\vcb$ and $\gamma$ \cite{Buchalla:1994tr}. {The dependence of this correlation on $\beta$ can
    be in principle used for the extraction of $\beta$  \cite{Buchalla:1994tr}.}
  \item
    {We discussed} the correlation between the branching ratios for $\kpn$ and $B_s\to\mu^+\mu^-$ in (\ref{master1}), already proposed in \cite{Buras:2015qea}, but here
    presented in a different manner.
    In particular 
    we introduced the ratios $R_1$ and $R_2$, in (\ref{R12}), of
    the branching ratio for $\kpn$ to  the branching
    ratios for $B_s\to\mu^+\mu^-$ and $B_d\to\mu^+\mu^-$ raised both to the power of $1.4$. {While, as seen in Figs.~\ref{fig:CKMdependence} and 
      \ref{fig:CKMdependenceB}, all these branching ratios depend strongly on 
      $\vcb$, 
    $R_1$ and $R_2$
    are to an excellent approximation $\vcb$-independent.} They exhibit then
    only
    a sizable dependence on $\gamma$ and a very weak one on $\beta$ as seen
    in Fig.~\ref{fig:3a}. {Even more interesting will be the plots in {Fig.~\ref{fig:3abis} }
      when the precision on both branching ratios and $\gamma$ will be increased
      in the coming years.}
  \item
    {We presented an improved, relative to \cite{Buras:2015qea},  triple correlation between the branching ratios
      for $\kpn$, $B_s\to\mu^+\mu^-$ and $B_d\to\mu^+\mu^-$ in (\ref{master3})
      that practically  does not depend     on the CKM parameters within the SM .}
  \item
    {We discussed} the correlations between the branching ratios for $\kpn$ and $\klpn$
 with the branching ratios for $B_{s,d}\to\mu^+\mu^-$ and $B\to K(K^*)\nu\bar\nu$
 represented by the $\vcb$-independent ratios $R_3-R_6$
   in (\ref{R34}) and (\ref{R56}).
   {The $\gamma$ and $\beta$ dependence of $R_3$ and $R_4$ has been shown in Fig.~\ref{fig:4} and the one of $R_5$ and $R_6$ in Fig.~\ref{fig:4B}.}
\item
  On the other hand the ratios of $B\to K(K^*)\nu\bar\nu$ branching
    ratios to the $B_s\to\mu^+\mu^-$ one as well as 
    the ratio of the short distance contribution to the $\ksm$ branching ratio and the one for $\klpn$ in the SM are independent of CKM parameters except for
    $\lambda$.
  These correlations can be found in (\ref{R7SM}), (\ref{R8SM}) and in (\ref{SR1}).   In this context we have pointed out that the ratio of the branching ratios for $B^+\to K^+\nu\bar\nu$ and
 $B_s\to\mu^+\mu^-$ from Belle II and LHCb signals a $1.8\sigma$ tension with
 its SM value.
  \item
    In the context of the determination of $\gamma$ by means of the ratio $\Delta M_d/\Delta M_s$  we have emphasized
    that the tension between this determination and the one from tree-level $B$-decays pointed out in 2016 in \cite{Blanke:2016bhf} {diminished. The final verdict will be given by future precise measurements of $\gamma$ by the  LHCb
      and Belle II collaborations \cite{Krishnan:2018gdn,Kou:2018nap} that could reach the precision of $\pm1^\circ$.}
  \item
    We have proposed  a  test of the SM that
      is complementary to the usual UT-analyses. It
exhibits the parameter $\vcb$ hidden in the latter analyses. To this end
we proposed to extract 
    from a given observable the value of $\vcb$ as a function
    of $\beta$ and $\gamma$ for which the SM agrees with the experimental
    data. We have illustrated this idea considering 
$\varepsilon_K$, $\Delta M_s$ and $\Delta M_d$ for which both theory and
    experiment reached already good precision. The result is presented in
    Fig.~\ref{fig:5}. We find that, when $\varepsilon_K$ and $\Delta M_d$
    are considered simultaneously and the data on $S_{\psi K_S}$ are taken into account, {$\gamma=61.3(1.6)^\circ$ and $\vcb=43.1(1.0)\times 10^{-3}$} are obtained.
    But when $\varepsilon_K$ and $\Delta M_s$ are simultaneously considered we find 
 {$\gamma=70.3(2.6)^\circ$ and $\vcb=41.7(6)\times 10^{-3}$}.
    In order to exhibit this tension in a $\vcb$-independent manner we
    considered suitable ratios of $\varepsilon_K$ and $\Delta M_{d,s}$ that do not depend  on $\vcb$. The result is presented in Fig.~\ref{fig:8}.
    As seen  in Figs.~\ref{fig:5} and \ref{fig:8} the agreement between these two determinations can only be obtained for {a} value of $\beta$ that differs by
    $2\sigma$ from the one obtained from $S_{\psi K_S}$.
  \item
    We conclude therefore that it is not possible to obtain full
    agreement
  between the data on $\varepsilon_K$, $\Delta M_d$, $\Delta M_s$ and
  $S_{\psi K_S}$ within the SM independently of the value of $\vcb$ and $\gamma$.
  While this tension is still moderate it hints for some NP at work. In this
  context a precise measurement of $\gamma$ will be important. If {its} value
  will turn out to be in the ballpark of $66^\circ$ as signalled
  by the most recent LHCb result, a new CP-violating phase will be required
  to obtain the agreement with the data on  $S_{\psi K_S}$.
\item
  In the context of these investigations we have pointed out that $\Delta M_s$ offers  presently the best estimate of $\vcb$ in the SM if only one quantity is considered. The result is {shown in Fig.~\ref{fig:5} and} given in (\ref{vcbbest}) and in (\ref{vcbbest1}), {where in the latter also the $\epsilon_K$ and $S_{\psi K_S}$ measurements are taken into account}. 
  {Including $S_{\psi K_S}$ allows to determine $\vub$ as well so that
    in summary we find
    \be
    \boxed{ |V_{cb}|=41.8(6)\times 10^{-3}, \qquad |V_{ub}|=3.65(12)\times 10^{-3}.}
    \ee
    {This value of $\vub$ is} in perfect agreement with the FLAG's value in (\ref{FLAGVUB}) while the one for $\vcb$ agrees well with the UTfit determination     $\vcb=42.0(6)\times 10^{-3}$. {On the other hand it
    is} significantly higher than the CKMfit one, i.e. $\vcb=40.5(8)\times 10^{-3}$.}
 % \item
 %   Having the results for $\vcb$ in Fig.~\ref{fig:5} it is straightforward to find the corresponding results for $\vub$. We show them in Fig.~\ref{fig:7}. The agreement with FLAG's exclusive determination is found. This is in particular seen in (\ref{vubbest1}) obtained by means of $\Delta M_s$, $|\varepsilon_K|$ and $S_{\psi K_S}$. 
  \item
    We obtained the analytical expression for $\vcb$
    corresponding to the experimental value of $\varepsilon_K$, that is a
    function of $\gamma$ and $\beta$. Having eliminated the $\vcb$ dependence
this allowed to study transparently  the $\gamma$ and $\beta$
    dependences of $\kpn$ and $\klpn$ branching ratios. The results for $\kpn$ and $\klpn$ are  presented in Fig.~\ref{fig:6}.
       For $B_{s,d}$ decays it is more convenient to use $\Delta M_{s,d}$ for this purpose as this eliminates the CKM dependence  from branching ratios as already demonstrated in \cite{Buras:2003td,Bobeth:2021cxm}.
 \end{itemize}

  However, at present the most important results of our paper appear
    to be precise predictions for $K\to\pi\nu\bar\nu$ branching ratios in (\ref{BV}) and for $B_{s,d}\to\mu^+\mu^-$ in (\ref{LHCbTH}) that
    supersede the results quoted in the literature. {Beyond the $\vcb$-independence their virtues are as follows.
      \begin{itemize}
      \item
        In the case of $\kpn$ and $\klpn$ the $\gamma$ dependence is very weak
        as seen in (\ref{R11}) and (\ref{R12a}) and the only relevant CKM
        dependence comes from the angle $\beta$ that is already accurately
        determined  by $S_{\psi K_S}$. Both uncertainties in $\beta$ and $\gamma$
        have been included in the errors. In particular the one from $\gamma$
        corresponds to the range $60^\circ\le\gamma\le 75^\circ$ that is significantly larger than the one resulting from global CKM fits.
      \item
        In the case of  $B_{s,d}\to\mu^+\mu^-$, as seen in (\ref{CMFV6})
        no CKM dependence is involved so that the result is independent
        of any CKM global fits.
        \end{itemize}
}
       
Having these results
    allowed us to obtain the best estimate of the branching ratios
 of  $\ksm$, $B^+\to K^+\nu\bar\nu$ and $B^0\to K^{0*}\nu\bar\nu$ as well. The results for all branching ratios are
      summarized in Table~\ref{tab:SMBRs}  {and their implications in the case of
 $\kpn$ and $B_s\to\mu^+\mu^-$ in   Fig.~\ref{fig:3abis}.}

  {The future of our strategies will depend on the precision on the determination of $\gamma$ and $\beta$ in those tree-level decays in which it can be
    demonstrated that NP contributes in a negligible way. In this context
    LHCb and Belle II experiments aiming at $1^\circ$ accuracy for $\gamma$
  could contribute in an important manner.}
 
  However, similar to what was stated in \cite{Bobeth:2021cxm},
  we also emphasize that taking ratios of observables cancels not
  only parametric, theoretical and experimental uncertainties. It can in principle cancel also {some} NP effects {that could be} present, in our case, in $\kpn$, $\klpn$, $\ksm$,
 $B_{s,d}\to\mu^+\mu^-$, $B\to K(K^*)\nu\bar\nu$, {$|\varepsilon_K|$}
  and
  in the ratio of the mass differences $\Delta M_{s,d}$. 
  %Also NP could enter in  $|\varepsilon_K|$.
  Therefore the complete search for NP
  must also  consider eventually all  observables separately that brings back CKM   uncertainties and in certain cases also hadronic uncertainties. Yet, the strategies presented here allow to test the mutual
  consistency of various correlations predicted by the SM without the worry about the values of $\vcb$ and $\vub$.
  This simplifies the search for NP through the violation of these SM correlations. Indeed, it was already possible to
  conclude in \cite{Bobeth:2021cxm} that
   some NP is hinted in the SM correlation between
   $\overline{\mathcal{B}}(B_{s}\to\mu^+\mu^-)$ and $\Delta M_{s}$. {Similar
   tension seems to be hinted in the correlation between $\overline{\mathcal{B}}(B_{s}\to\mu^+\mu^-)$ and $B^ +\to K^+\nu\bar\nu$ pointed out
   here.}

   In this context we would like to refer to  a series of papers by Fleischer and collaborators \cite{Banelli:2018fnx,Fleischer:2019wlx,Fleischer:2021yjo} in
   which, in the spirit of the proposal in \cite{Buras:2003td}, ratios
   of processes governed by the transition $b\to ul\bar\nu_l$ have been considered in order to cancel $\vub$ dependence. Already these ratios, when compared with their SM
   values can, as in \cite{Bobeth:2021cxm} and in the present paper, signal
   the presence of NP at work. Considering then these ratios in the context
   of different NP scenarios that can be distinguished through different
   operators like vector, scalar, pseudoscalar and tensor ones, the authors of
    \cite{Banelli:2018fnx,Fleischer:2019wlx,Fleischer:2021yjo} extracted
   from the present data the regions for the Wilson coefficients of these
   operators independently of the value of $\vub$. Subsequently, using
   these coefficients they determined the values of $\vub$ for each NP
   scenario. While this strategy is interesting in itself, it shows that
   at the end the value of $\vub$ extracted from the data will depend on
   NP involved and taking the ratios of various observables will not avoid it.
   Even in specific NP scenarios considered in  \cite{Banelli:2018fnx,Fleischer:2019wlx,Fleischer:2021yjo} it was crucial to assume that NP did not affect
   the coefficient of the SM left-handed operator. Without this assumption
   one could always rescale all Wilson coefficients by an arbitrary number
   which would cancel in the ratio. In turn Wilson coefficients of NP
   operators would not be determined and this would also be the case of $\vub$.
   Imposing no NP contribution to the SM operator does not allow such
   rescaling and the strategy for $\vub$ determination in the presence of NP proposed in these papers could be executed.
   Whether this interesting strategy is more powerful than the global fits remains to be
   seen when the data improves. {But in any case it provides an additional insight
in a possible NP at work.}

   In the context
  of the present paper we are  looking
  forward to improved results for $\kpn$, $\klpn$, $\ksm$,  $B_{s,d}\to\mu^+\mu^-$, $B\to K(K^*)\nu\bar\nu$ 
  and also $\gamma$ which would allow to test several of the $\vcb$-independent
  correlations found by us. 

\bigskip
  
{\bf Acknowledgements}

\noindent
{We would like to thank Pietro Baratella, Jean-Marc G{\'e}rard and Peter Stangl for discussions and comments on the manuscript.}
A.J.B acknowledges financial support from the Excellence Cluster ORIGINS,
funded by the Deutsche Forschungsgemeinschaft (DFG, German Research Foundation), 
Excellence Strategy, EXC-2094, 390783311. E.V. has been partially funded by the Deutsche Forschungs-gemeinschaft (DFG, German Research Foundation) under Germany’s Excellence Strategy- EXC-2094 - 390783311, by the Collaborative Research Center SFB1258 and the BMBFgrant  05H18WOCA1  and  thanks  the  Munich  Institute  for  Astro-  and  Particle  Physics(MIAPP) for hospitality.

\appendix
\section{Basic Functions}\label{App}
The basic functions that enter various formulae are given as follows.

For rare decays without the inclusion of QCD corrections one has
\begin{equation}\label{X0}
X_0(x_t)={\frac{x_t}{8}}\;\left[{\frac{x_t+2}{x_t-1}} 
+ {\frac{3 x_t-6}{(x_t -1)^2}}\; \ln x_t\right]\,,
\end{equation}
\begin{equation}\label{Y0}
Y_0(x_t)={\frac{x_t}{8}}\; \left[{\frac{x_t -4}{x_t-1}} 
+ {\frac{3 x_t}{(x_t -1)^2}} \ln x_t\right]\,,
\end{equation}
where $x_t=m^2_t/M_W^2$.

The inclusion of NLO 
QCD corrections \cite{Buchalla:1998ba,Misiak:1999yg} and NLO electroweak
corrections \cite{Brod:2010hi},in the case of $X_0(x_t)$ {gives with the
  most recent value of $m_t(m_t)$ in Table~\ref{tab:input}
  \cite{Brod:2021hsj}
\be
X(x_t)=1.462\pm0.017_{\rm QCD}\pm0.002_{\rm EW}=1.462\pm0.017\,
\ee
which updates the value $X(x_t)=1.481\pm0.009$ used in \cite{Buras:2015qea}
that is also quoted in \cite{Buras:2020xsm}.}

The inclusion of NLO and NNLO 
QCD corrections and NLO electroweak correction in the case of $Y_0(x_t)$
results in \cite{Bobeth:2013uxa}
\begin{equation}\label{yyx}
Y(x_t) = \eta_\text{eff}\,Y_0(x_t)\,, \qquad  \eta_\text{eff}=0.9882\pm 0.0024\,.
\end{equation}

For $\Delta M_{s,d}$ and $\varepsilon_K$ the relevant functions without QCD
corrections are
\begin{eqnarray}
\label{S0}
S_0(x_t)&=&\frac{4x_t-11x^2_t+x^3_t}{4(1-x_t)^2}-
 \frac{3x^3_t \ln x_t}{2(1-x_t)^3},\\
\label{BFF1}
S_0(x_c)&=&x_c,\\
\label{BFF12}
S_0(x_c, x_t)&=&x_c\left[\ln\frac{x_t}{x_c}-\frac{3x_t}{4(1-x_t)}-
 \frac{3 x^2_t\ln x_t}{4(1-x_t)^2}\right].
\end{eqnarray}
In the last two expressions we have kept only linear terms in $x_c\ll 1$, 
but of course all orders in $x_t$. The last function generalizes $S_0(x_t)$ 
in (\ref{S0}) to include box diagrams with simultaneous top-quark and
charm-quark exchanges.
QCD corrections are included in the main text with the help of $\eta_{ij}$ and
$\eta_B$ factors.

For numerical calculations one should of course use the exact expressions given above but to get an idea on the size of these functions and for derivation of 
analytical formulae one can use the following approximate but simple expressions \begin{equation}\label{PBE1}
~X_0(x_t)=0.660~x_t^{0.575},  \qquad
 Y_0(x_t)=0.315~x_t^{0.78},\qquad S_0(x_t)=0.784~x_t^{0.76}\,.
\end{equation}
In the range $160\GeV \le m_t(m_t) \le 165\GeV$ 
these approximations reproduce the
exact expressions to an accuracy better than $0.2\%$ which is sufficient 
for deriving analytic formulae. Then
\be
X_0(x_t)=1.49~\left(\frac{m_t(m_t)}{163\GeV}\right)^{1.15},
\quad\quad
Y_0(x_t)=0.95~\left(\frac{m_t(m_t)}{163\GeV}\right)^{1.56},
\ee
\begin{equation}
 S_0(x_t)=2.31~\left(\frac{m_t(m_t)}{163\GeV}\right)^{1.52}\,.
\ee

%--------+---------+---------+---------+---------+---------+---------+---------+
%
%
%
%--------+---------+---------+---------+---------+---------+---------+---------+
\renewcommand{\refname}{R\lowercase{eferences}}

\addcontentsline{toc}{section}{References}

\bibliographystyle{JHEP}

\small

\bibliography{Bookallrefs}

\end{document}